\documentclass{article}[11pt, a4paper]
\pdfoutput=1
\usepackage[backend=biber]{biblatex}
\usepackage{amsmath,amssymb,amsfonts}
\usepackage{algorithmic}
\usepackage{graphicx, float}
\usepackage{textcomp}
\usepackage{caption}
\usepackage{subcaption}
\usepackage{url}
\usepackage{array}
\usepackage{supertabular}
\usepackage{multirow}
\usepackage{pdflscape}
\usepackage{acro}
\usepackage{authblk}
\usepackage[margin=2.5cm]{geometry}
\usepackage{wrapfig}
\usepackage{setspace}

\usepackage{fontawesome5}

\usepackage{enumitem}

\DeclareAcronym{ACS}{
	short={ACS},
	long={Anonymous Communication System}
}

\DeclareAcronym{ANDOS}{
	short={ANDOS},
	long={all or nothing disclosure of secrets}
}

\DeclareAcronym{AOT}{
	short={AOT},
	long={Anonymization by Oblivious Transfer}
}

\DeclareAcronym{BAR}{
	short={BAR},
	long={Broadcast Anonymous Routing}
}

\DeclareAcronym{DC}{
	short={DC},
	long={Dining Cryptographers}
}

\DeclareAcronym{DCN}{
	short={DCN},
	long={Dining Cryptographers Network}
}

\DeclareAcronym{Dissent}{
	short={Dissent},
	long={Dining-cryptographers Shuffled-Send Network}
}

\DeclareAcronym{ECC}{
	short={ECC},
	long={Elliptic Curve Cryptography}
}

\DeclareAcronym{EU}{
	short={EU},
	long={European Union}
}

\DeclareAcronym{HORNET}{
	short={HORNET},
	long={High-speed Onion Routing at the Network Layer}
}

\DeclareAcronym{HTTP}{
	short={HTTP},
	long={Hypertext Transfer Protocol}
}

\DeclareAcronym{I2P}{
	short={I2P},
	long={Invinsible Internet Project}
}

\DeclareAcronym{IoT}{
	short={IoT},
	long={Internet of Things}
}

\DeclareAcronym{IP}{
	short={IP},
	long={Internet Protocol}
}

\DeclareAcronym{ISO}{
	short={ISO},
	long={International Organization for Standardization}
}

\DeclareAcronym{ITU}{
	short={ITU},
	long={International Telecommunication Union}
}

\DeclareAcronym{LAN}{
	short={LAN},
	long={Local Area Network}
}

\DeclareAcronym{LTE}{
	short={LTE},
	long={Long-Term Evolution}
}

\DeclareAcronym{M2}{
	short={M2},
	long={Multicasting Mixes for Efficient and Anonymous Communication}
}

\DeclareAcronym{MACAddress}{
	short={MAC},
	long={Media Access Control Address},
	plural={s},
	long-plural={es},
}

\DeclareAcronym{MAM}{
	short={MAM},
	long={Mutual Anonymous Multicast}
}

\DeclareAcronym{MPC}{
	short={MPC},
	long={Multi-Party Computation}
}

\DeclareAcronym{MPSaaS}{
	short={MPSaas},
	long={MPC as a system-as-a-service}
}

\DeclareAcronym{NAT}{
	short={NAT},
	long={Network Address Translation}
}

\DeclareAcronym{NIAR}{
	short={NIAR},
	long={Non-interactive Anonymous Router}
}

\DeclareAcronym{NSA}{
	short={NSA},
	long={National Security Agency}
}

\DeclareAcronym{OT}{
	short={OT},
	long={Oblivious Transfer}
}

\DeclareAcronym{P3} {
	short={P3},
	long={Private Keyword-Based Push and Pull}
}

\DeclareAcronym{P5}{
	short={P5},
	long={Peer-to-Peer Personal Privacy Protocol}
}

\DeclareAcronym{PANORAMIX}{
	short={PANORAMIX},
	long={Privacy and Accountability in Networks via Optimized Randomized Mix-nets}
}

\DeclareAcronym{PIR}{
	short={PIR},
	long={Private Information Retrieval}
}

\DeclareAcronym{PRG}{
	short={PRG},
	long={Pseudorandom Generator}
}

\DeclareAcronym{RAC}{
	short={RAC},
	long={Freerider-Resilient Scalable Anonymous Communication Protocol}
}

\DeclareAcronym{RACE}{
	short={RACE},
	long={Resilient Anonymous Communication for Everyone}
}

\DeclareAcronym{SSH}{
	short={SSH},
	long={Secure Shell}
}

\DeclareAcronym{TCP}{
	short={TCP},
	long={Transmission Control Protocol}
}

\DeclareAcronym{Tor}{
	short={Tor},
	long={The Onion Router}
}

\DeclareAcronym{UDP}{
	short={UDP},
	long={User Datagram Protocol}
}

\DeclareAcronym{US}{
	short={US},
	long={United States}
}

\DeclareAcronym{VPN}{
	short={VPN},
	long={Virtual Private Network}
}

\DeclareAcronym{WLAN}{
	short={WLAN},
	long={Wireless Local Area Network}
}

\DeclareAcronym{XOR}{
	short={XOR},
	long={exclusive or}
}

\DeclareAcronym{XPIR}{
	short={XPIR},
	long={Private Information Retrieval for Everyone}
}

\DeclareAcronym{ZKP}{
	short={ZKP},
	long={Zero-Knowledge Proof}
}

\DeclareAcronym{SICTA}{
	short={SICTA},
	long={Successive Inference Cancellation Tree Algorithm}
}

\DeclareAcronym{SOCKS}{
	short={SOCKS},
	long={Socket Secure}
}

\graphicspath{{figures/}}
\def\BibTeX{{\rm B\kern-.05em{\sc i\kern-.025em b}\kern-.08em
    T\kern-.1667em\lower.7ex\hbox{E}\kern-.125emX}}
\begin{document}

\newcommand{\thetitle}{A Survey on Anonymous Communication Systems with a Focus on Dining Cryptographers Networks}

\title{\thetitle}

\author[1]{Mohsen Shirali}
\author[2]{Tobias Tefke}
\author[2]{Ralf C. Staudemeyer}
\author[3]{Henrich C. P{\"o}hls}

\affil[1]{Computer Science and Engineering, Shahid Beheshti University, Tehran 19839-63113, Iran (e-mail: \texttt{m\_shirali@sbu.ac.ir})}
\affil[2]{Schmalkalden University of Applied Sciences, Blechhammer D-98574, Schmalkalden, Germany (e-mail: {\texttt{t.tefke@stud.fh-sm.de}, \texttt{r.staudemeyer@hs-sm.de}})}
\affil[3]{Chair of IT-Security, University of Passau, Passau D-94036, Germany (email: \texttt{hp@poehls.com})}


\newcommand{\fix}[2]{{\bf FIX}\footnote{{\bf{Fix #1:} }#2}}
\newcommand{\chk}[2]{{\bf CHECK}\footnote{{\bf{CHECK #1:} }#2}}
\newcommand{\comment}[1]{}

\maketitle

{\small{Corresponding author: Mohsen Shirali (\texttt{m\_shirali@sbu.ac.ir})}}

\begin{abstract}
	Traffic analysis attacks can counteract end-to-end encryption and use leaked communication metadata to reveal information about communicating parties.
With an ever-increasing amount of traffic by an ever-increasing number of networked devices, communication privacy is undermined.
Therefore, \acp{ACS} are proposed to hide the relationship between transmitted messages and their senders and receivers, providing privacy properties known as anonymity, unlinkability, and unobservability.
This article aims to review research in the \acp{ACS} field, focusing on \acfp{DCN}.
The \ac{DCN}-based methods are information-theoretically secure and thus provide unconditional unobservability guarantees.
Their adoption for anonymous communications was initially hindered because their computational and communication overhead was deemed significant at that time, and scalability problems occurred.
However, more recent contributions, such as the possibility to transmit messages of arbitrary length, efficient disruption handling and overhead improvements, have made the integration of modern \ac{DCN}-based methods more realistic.
In addition, the literature does not follow a common definition for privacy properties, making it hard to compare the approaches’ gains.
Therefore, this survey contributes to introducing a harmonized terminology for \ac{ACS} privacy properties, then presents an overview of the underlying principles of \acp{ACS}, in particular, \ac{DCN}-based methods, and finally, investigates their alignment with the new harmonized privacy terminologies.
Previous surveys did not cover the most recent research advances in the \ac{ACS} area or focus on \ac{DCN}-based methods.
Our comprehensive investigation closes this gap by providing visual maps to highlight privacy properties and discussing the most promising ideas for making \acp{DCN} applicable in resource-constrained environments.
\end{abstract}

{\textbf{Keywords:}}
Privacy-Preservation, Anonymity, \acf{ACS}, \acf{DCN}, Unobservability.

\acresetall

\section{Introduction}
\label{sec:introduction}

The continuously growing collection of data by pervasive computing techniques and great advances in communication during the current information age are bringing many benefits to society.
This encompasses transformative changes and opportunities created in many aspects of daily life, e.g. healthcare, transportation, education and social interaction~\cite{cranor2016towards,mendes2017privacy}.
However, much of these collected data might be sensitive or contain personal information.
Therefore, their collection and transmission poses serious privacy concerns.
These could prevent a wider incorporation of new technologies into daily lives~\cite{pohls2014rerum,cook2018using}.

Furthermore, people might desire strong communications privacy and anonymity on the Internet in many situations.
These include circumstances in which people need to report information they may have on unlawful activities without fear of retribution or punishment.
Moreover, people who live under regimes that try to limit what their citizens can say and do on the Internet need solutions to circumvent censorship and restrictions concerning the freedom of speech.
Additionally, even private citizens may want to be able to freely browse the web, without third-parties collecting statistics on their browsing habits and selling that personal information to other companies~\cite{edman2009anonymity}.

End-to-end encryption is often used to combat this situation and to ensure the confidentiality of transmitted messages over intermediate links.
Then, only the intended recipient can read the message~\cite{tragos2015securing}.
However, encryption only hinders third parties from reading transmitted information.
It cannot hide the fact that the message exchange is taking place and parties are communicating~\cite{ren2010survey,Syverson1997}.
Even over encrypted channels, an observer on the network might be able to gather the so-called metadata, which includes information like communication endpoints, the sheer size of exchanged packets~\cite{rupprecht2019breaking}, the frequency and timing of packets in correlation to other packets, events\footnote{Events occur when the state of an object within a communication changes significantly~\cite{Aalst2016}.} and location details~\cite{staudemeyer2018road}.
This information, when extracted and combined with a priori knowledge, statistics and processed (e.g. by machine learning algorithms) can be rich enough to even bypass end-to-end encryption~\cite{staudemeyer2019takes,barman2020prifi}.
The attacks on the users' privacy, which work without the attacker having knowledge of the communications' contents, are called traffic analysis attacks~\cite{raymond2001traffic,danezis2007introducing}.
While the mere knowledge that devices exchange information might not be interesting on its own, the whole situation changes dramatically, when we can map devices to locations and device types, or we can find out the used services from the communication patterns (The interested reader will find more information about traffic analysis and de-anonymization attacks as well as implemented examples in Appendix~\ref{sec:background:attacks}).

In order to counter such traffic analysis attacks and to minimise any kind of information disclosure occurring as part of the exchange of metadata, more than end-to-end encryption is necessary.
For providing privacy to metadata, an additional layer of privacy protection running on top of the existing communication protocols is needed.
This layer must hide the fact that communication takes place~\cite{edman2009anonymity,staudemeyer2018road}.
Different protocols and mechanisms, that are generally known as \acp{ACS}, have been proposed to this aim.
\acp{ACS} hide the relation between transmitted messages and their senders and/or receivers.
Thereby, they allow their users to communicate privately within a network environment~\cite{kotzanikolaou2017broadcast}.
For instance, in (wireless) networks, anonymous communication features could prevent traffic analysis by making the real network traffic indistinguishable from random noise~\cite{barman2020prifi}.

The \acp{ACS} can offer different levels of protection, e.g. anonymity, unlinkability and unobservability (these communication properties are defined in Section~\ref{sec:background:privacy-terms} in detail).
However, providing different levels of guarantee for privacy comes at a price of performance, scalability limitations and increases practical deployment complexity.
So that, all protocols and techniques proposed to enable anonymous communication have to choose a trade-off between efficiency in terms of throughput, latency, and scalability on the one hand and security and privacy guarantees on the other hand.

The basic building blocks of nearly all widely known developed \acp{ACS} are two concepts proposed by Chaum: Mix networks in 1981~\cite{chaum1981untraceable} and \acf{DCN} in 1988~\cite{chaum1988dining}.
Mix networks (also called Mixes) take a bunch of messages and then scramble, delay and re-encode them.
In this way, an eavesdropper can no longer easily correlate incoming with outgoing messages~\cite{staudemeyer2018road}.
The strength of Mix techniques relies on multiple transmission and routing of messages.
However, this introduces delays between the time a message is sent and the time it arrives at the intended recipient~\cite{edman2009anonymity}.
Besides, \ac{DCN} is a broadcast round-based protocol wherein only one member can publish one \textit{l}-bit message per round.
The privacy of \ac{DCN} relies on information coding, and \ac{DCN} provides unconditional secure unobservable communication~\cite{barman2020prifi}.

The \acfp{VPN}, proxies, Mix-based solutions and onion-based routings like \ac{Tor}~\cite{dingledine2004tor} are popular \acp{ACS} and have been widely adopted in practice due to their support for most network protocols~\cite{barman2020prifi}. Nevertheless, they mostly offer limited anonymity protection and an observer who traces packets can still mount traffic analysis attacks to break anonymity guarantees~\cite{nguyen2003breaking,pfitzmann1994breaking}.
On the contrary, with \ac{DCN}-based \acp{ACS} an adversary monitoring the users is unable to distinguish messages carrying actual content from random noise.
However, these solutions have their own challenges, such as providing round (or slot)-reservation techniques and dealing with disruptions.
Moreover, the initial \ac{DCN}-based \acp{ACS} suffered from high computational and communication overheads and lack of scalability~\cite{staudemeyer2019takes}.
For this reason, even though the history of \ac{DCN} dates back to almost three decades ago, they were rarely implemented in real-world anonymous communications until solutions were proposed to improve efficiency and make them more realistic~\cite{edman2009anonymity}.

Due to the increasing importance of privacy and the variety of proposed \acp{ACS} and their applications, several surveys have been conducted on private and anonymous communication so far.
Some of these surveys will be mentioned in Section~\ref{subsec:Surveys}.
However, despite a large number of these articles, there are still issues that need to be addressed.
For instance, despite traffic-analysis resistance provided by \ac{DCN} protocols, their recent improvements or implementations in constrained environments (like~\cite{staudemeyer2019takes,bauer2017dining} and~\cite{Franck2021FastECC}) have not been addressed in survey articles so far.
The proposed methods for protection against disruption and contributions that have been made to reduce high computation and latency overheads of \ac{DCN} approaches in order to make them more practical and efficient to be used in constrained environments are less discussed.
Additionally, to the best of our knowledge, there has not been a recent survey article that included newly published \acp{ACS}' research and projects, such as cMix~\cite{chaum2017cmix} and Nym~\cite{diaz2021nym} or \ac{DCN}-based solutions like PriFi~\cite{barman2020prifi}, Arbitrary length k-anonymous~\cite{modinger2021arbitrary} and Shared-Dining~\cite{modinger2021shared}.

The challenge is a lack of a comprehensive study to present recent innovations for anonymous communication, which motivates us to conduct a panoramic review of novel \acp{ACS} focusing on DCN-based systems.
Our contribution is as follows: Firstly, we evaluate the most recent progress and developments in the area of \acp{ACS}, in particular, we analyse \ac{DCN}-based methods from the past to the present because of their information-theoretic privacy features.

Beyond looking at \acp{ACS} and grouping them into families according to their key design decisions, secondly, we look at the \acp{ACS} from the privacy perspective. 
We analysed them regarding their achievements in offering various privacy properties, which is a challenging task without having a common understanding of privacy. This article fills this gap:
We observe that the properties to describe anonymous communications do not follow a common definition, which leads to confusion and challenges in comparing different methods. 
Hence, we define the main commonly used terms based on the literature. 
With a harmonized terminology as a common ground, we then investigate the reviewed \acp{ACS} based on the privacy properties they offer.

Thirdly, the article contributes to focus on the fitness of \acp{ACS} when used in everyday applications on the Internet or within networks with more resource-limited nodes, like in the Internet of Things.Finally, this survey concludes with a discussion of the most promising ideas developed by different protocols to mitigate the current \acp{ACS} challenges.

The remainder of this article is structured as follows: Section~\ref{sec:background:privacy-terms} reviews and harmonizes the definitions of privacy properties.
Section~\ref{sec:GENERAL OVERVIEW OF ACS} briefly overviews general \acp{ACS} methods such as Mix networks, Onion Routing based solutions and \acf{MPC} and offers our analysis of the privacy properties achieved by the different \acp{ACS} proposals (see the map of privacy properties of \acp{ACS}s in Section~\ref{sec:acs_privacy_map}).
Then, the original \ac{DCN} protocol with its main challenges and \acp{ACS} based on \acp{DCN} are studied comprehensively in Section~\ref{sec:DCN-BASED ACS} and Section~\ref{subsec:DCN-4-methods}, respectively. 
Again we offer an overview map of the privacy properties achievable with \acp{DCN} methods in Section~\ref{subsec:DCN-5-discussion}.
Finally, Section~\ref{sec:conclusion} concludes this contribution.
Supplementary materials are provided in the Appendices including the abbreviations list in Appendix~\ref{Appendix-A-Terminology} and the extensions of the terminology (network and security), adversarial model and traffic-analysis attacks in Appendix~\ref{sec:BACKGROUND KNOWLEDGE} to provide common background knowledge for interested readers.
Finally, a detailed review of all the main \acp{ACS} which are discussed is provided in Appendix~\ref{AppendixC:Detailed_Overview} to wrap up this survey.
\section{Privacy Terminology}
\label{sec:background:privacy-terms}
A complete body of terminology for talking about privacy has been proposed by Pfitzmann and Hansen in 2000~\cite{Pfitzmann2000}.
Since then, their definitions have been cited highly in the anonymous communication publications, and they have become the reference terminology.
In this survey, we largely follow the most recent version of their terminology~\cite{pfitzmann2010terminology} and definitions based on it in the literature.

\textbf{Anonymity.} “The state of being not identifiable within a set of subjects, which is called the \textit{anonymity set}”.
“\textbf{Not identifiable} within the anonymity set” means that only using the information the attacker has at his discretion, the subject is “not uniquely characterized within the anonymity set” or the subject is “not distinguishable from the other subjects within the anonymity set”\footnote{Direct quotes show where the exact wording of~\cite{pfitzmann2010terminology} is used in the definitions}.

In simple words, anonymity is provided when multiple subjects form a set, for instance in message transmission, it cannot be distinguished who sends or receives the message.
Hence, the anonymity set is the set of all possible subjects or actors within a system.
Identically in network communication, all the nodes that could have been involved will form the anonymity set.
The anonymity property can be refined further based on the role that a specific subject has.
Regarding a specific message exchange; the subject can be either the \textit{sender} or \textit{recipient} of a message.
The set of subjects which could have sent a specific message is called the \textit{sender anonymity set}.
Similarly, all subjects who could have received a particular message form the \textit{recipient anonymity set}~\cite{edman2009anonymity}.

In some cases, sender and recipients want to identify each other while achieving \textit{third-party anonymity}, meaning that they want to be sure that they are interacting with the intended party – while not wanting any other external party to be able to determine that they are communicating with each other~\cite{diaz2021nym}.

The probability that a verifier can successfully determine the real subject is exactly $\frac{1}{n}$, where $n$ is the number of members in the anonymity set~\cite{ren2010survey}.
As this definition implies, an \ac{ACS} must consist of at least two subjects in order to provide anonymity property, so the anonymity set should always have more than one member.

Reiter and Rubin~\cite{reiter1998crowds} widened the term of anonymity by adding the degree of anonymity.
The degree of anonymity is an indicator telling how exposed a sender and/or receiver is on the spectrum between having absolute privacy and being provably exposed.
Later, Shields and Levine~\cite{Shields2000} refined this indicator by adding specific mathematical definitions to the degrees of anonymity.

Identifiability means that the attacker can sufficiently identify the subject within a set of subjects and is the opposite of anonymity~\cite{pfitzmann2010terminology}.
Hence, in the literature, \textit{unidentifiability} is sometimes used as an equivalent of anonymity.
Moreover, \textit{undetectability} improves the unidentifiability property by making it impossible for an attacker to figure out whether a specific subject exists~\cite{staudemeyer2019takes,pfitzmann2010terminology}.

\textbf{Unlinkability.} A user may make multiple uses of resources or services; however, others are unable to determine whether the same user caused certain specific operations in the system.~\cite{isoiec27551}.
In an abstract sense, \textit{unlinkability} refers to the inability to determine which pieces of data available at different parts of a system may or may not be related to each other~\cite{diaz2021nym}.
A network providing unlinkability ensures that neither messages nor network nodes can be correlated.
Therefore, the probability of finding relations between senders, recipients and messages stays the same before and after eavesdropping on the traffic~\cite{erdin2012anonymous}.

\textit{Unlinkability} relates to anonymity as follows: \textit{Sender anonymity} means that a particular message is not linkable to any sender, and no sender is linkable to any message.
Respectively, \textit{recipient anonymity} means that a particular message is not linkable to any recipient, and no recipient is linkable to any message.
Pfitzmann and Hansen further defined the unlinkability between senders and recipients in an anonymous system as \textit{relationship anonymity}\cite{pfitzmann2010terminology}.

\textit{Relationship anonymity} means that each message is unlinkable to each potentially communicating pair of subjects and is a \textbf{weaker} property than each of sender anonymity and recipient anonymity (sender anonymity or recipient anonymity each alone implies relationship anonymity)~\cite{pfitzmann2010terminology}.

If the unlinkability property holds, an adversary observing senders and recipients in the network is not able to discern any relationship between communicating nodes and cannot distinguish who is communicating with whom~\cite{edman2009anonymity}.  

\textbf{Unobservability.} Ensures that the communication pattern between senders and recipients remains hidden from the adversary.
It conceals the activities of users and adds idle users to the anonymity set~\cite{diaz2021nym}.
Thus, unobservability hides the fact that a subject is sending or receiving a message, and it  is achieved through the use of \textit{"cover"} (or \textit{"dummy"}) traffic~\cite{edman2009anonymity}.

Unobservability is \textbf{stronger} privacy feature than unlinkability and anonymity~\cite{danezis2008survey}.
Unobservability always reveals only a subset of the information that anonymity reveals~\cite{pfitzmann2010terminology} (with respect to the same attacker, when we have unobservability, we have anonymity as well).
In a network with anonymity; when a user sends a message, the adversary cannot identify which of the observed output message corresponds to the user; while unobservability means that the adversary cannot even determine whether the user is sending any message at all, or whether it is just being idle~\cite{danezis2008survey,diaz2021nym}.

Similar to anonymity sets, we have unobservability sets that describe the unobservability for a set of subjects considering the subjects' role in communication.
\textit{Sender unobservability} or \textit{Sender online unobservability} means that it is impossible to tell whether a sender within the unobservability set currently transmits a message.
In other words, sender unobservability is the inability of an adversary to decide whether a specific sender (for any concurrently online sender of the adversary's choice) is communicating with any potential or not~\cite{piotrowska2017loopix}.
Sender unobservability directly implies the notion of sender anonymity where the adversary tries to distinguish between two possible senders communicating with a target recipient.

Likewise, \textit{recipient unobservability} means that it is impossible to tell if a recipient within the unobservability set currently receives a message.
Therefore, it is defined as the inability of an adversary to decide whether any sender is communicating with a specific recipient or not, for any recipient of the adversary’s choice~\cite{piotrowska2017loopix}.

\textit{Relationship unobservability} then means that it is impossible to figure out whether anything is sent out of a set of could-be senders to a set of could-be recipients~\cite{danezis2008survey}.
In other words, it is not noticeable if a message is transmitted by any of all possible sender-recipient pairs within the relationship unobservability set~\cite{ren2010survey}.

In summary, unobservability ensures that all activities between network nodes remain unnoticeable to eavesdroppers.
The messages carrying actual information are not distinguishable from random noise messages and cannot be correlated~\cite{staudemeyer2018road}.
As a result, unobservability does not only hide communicating parties.
It also hides which subjects exchanged messages during a period of observation~\cite{edman2009anonymity}.
Unobservability ensures unlinkability and unidentifiability, under the assumption of a continuous flow of dummy traffic.
 
Anonymous communication systems seek all, but provide at least a subset of these privacy properties~\cite{edman2009anonymity}.
However, they are mostly intended and recommended ensuring the highest level: \textbf{unobservable communication}.
Unobservability is beneficial to any application and data as it frustrates traffic analysis by an attacker who observes local traffic~\cite{staudemeyer2019takes}.

\textbf{Perfect preservation.} The privacy properties can be evaluated in terms of their change over time. The \textit{perfect preservation} of a property means that its value will not decrease with regard to the attacker’s knowledge from the current time relative to the attacker’s background knowledge (the a-priori-knowledge of the attacker). For instance, perfect preservation of a subject’s anonymity means the anonymity property stays the same over time. Indeed, anonymity does not change when it is compared by taking the attacker's observation into account (new knowledge) with the attacker's background knowledge. The change in anonymity over time- which may be reflected in a decrease in the size of the anonymity set-  is called \textit{Anonymity Delta}~\cite{pfitzmann2010terminology}.   

Moreover, it should be noted that the security and privacy properties of a system may be held conditionally or unconditionally.
A conditional secure system only provides security under certain circumstances: 
Its security depends on the hardness of a computational problem or the limitation of the adversary's computational power~\cite{delfs2015introduction}. 
If an \ac{ACS} functionality depends on conditionally secure cryptographic algorithms, it is categorised as conditionally secure or cryptographically secure \ac{ACS}~\cite{chaum1988dining}.
On the other hand, a system is proved to be unconditional if its security will not be broken by an attacker who is computationally unbounded and has an unlimited amount of time~\cite{ren2010survey}.
For instance, an ideal \ac{ACS} which uses one-time random keys that are at least as long as the message is an unconditional secure --but unfortunately impractical-- \ac{ACS}.
In this way, the conditionality/unconditionality of the privacy properties provided by an \ac{ACS} determines its resilience against attackers.

Furthermore, within the \acp{ACS}, the privacy properties are investigated from a \textit{global} perspective; the level of privacy provided by the whole system to all of its users together, and not for each \textit{individual} subject. As an example, global anonymity refers to the anonymity provided by a system to all of its users together, while individual anonymity is the anonymity of one individual subject.
\section{General Overview of Anonymous Communication Systems}
\label{sec:GENERAL OVERVIEW OF ACS}
This article studies \acp{ACS} with a focus on \ac{DCN}-based anonymous communication methods.
In order to provide the readers with the broader concept of \acp{ACS}, their main methods and some important approaches based on them are briefly discussed in this section.
An overview of the offered privacy properties by different \acp{ACS} and their classification are presented to conclude this section.
This overview investigates the alignment of \acp{ACS} with the provided privacy terminology in Section~\ref{sec:background:privacy-terms}.
\subsection{Mix Networks}
\label{subsec:MixNetworks}

Throughout the literature and various categories listed for \acp{ACS} in the publications, Mix networks or Mix-based solutions are one of the main \acp{ACS} concepts.
All Mix-based protocols use a set of mix servers (Mixes) which receive asymmetric encrypted messages from different sources and put them in a queue.
Mixes delete replays, collect and decrypt the received messages and if a certain amount of messages has been queued, they are all pushed out in a rearranged order~\cite{staudemeyer2019takes}.
By doing this, the Mix servers attempt to make the communication paths (including those of the sender and the receiver) ambiguous (see in Figure~\ref{fig:1-a}).
The desired anonymity is reached by relying on statistical properties of background traffic (also called cover traffic)~\cite{ren2010survey}.
In fact, a Mix network is a mean of sender anonymity.

One trustful Mix among the multiple Mixes of the network, which are selected to pass the traffic, is sufficient to ensure anonymity in Mix-based methods.
Hence, the strength of a Mix-based protocol is based on the trust relationship between Mixes, and Mix networks are not capable of providing unconditional anonymity~\cite{ren2010survey}.
However, several stages of encryption and message transmission impose large delays on messages, making Mix-based systems undesirable for real-time communication.
Furthermore, this highlights the need for \acp{ACS} providing low latency message exchange.

Concerning Mixes, the topology of the network strongly influences overhead and anonymity~\cite{diaz2010impact}.
Therefore, several Mix network prototypes have been developed to address various applications' requirements, particularly the need for low-latency communication, which, for instance, is needed for web browsing and online chats.

Loopix~\cite{piotrowska2017loopix} is a Mix network which groups nodes into different layers, where nodes in each layer can communicate with all the nodes in the immediately previous and following layers.
Loopix adds independent delays to incoming messages (Poisson mixing) in order to obfuscate message timing.
cMix~\cite{chaum2017cmix} also is a precomputation-based Mix network with fixed cascade architecture.
It completely eliminates computationally expensive public-key operations during run-time at the senders, recipients and Mix-nodes.
This protocol uses multi-party group homomorphic encryption in order to create a shared secret in the precomputation phase.
The decreased real-time cryptographic latency and lowered computational costs for clients made cMix a well-suited system for low-latency applications with lightweight clients.
Moreover, MiXiM~\cite{guirat2020mixim} provides a flexible simulation framework for Mix networks to evaluate different design options and their trade-offs.
The MiXiM framework allows assessing combinations of Mix network building blocks by running experiments and providing results for metrics including anonymity, end-to-end latency and traffic overhead~\cite{guirat2021mixim}.

In addition to these prototypes, \ac{PANORAMIX}~\cite{Panoramix} and \ac{RACE}~\cite{RACE}, programs funded by \ac{EU} and \ac{US} respectively, along with Nym~\cite{diaz2021nym} and Elixxir~\cite{Elixxir2022}, as two commercial ventures, have strengthened the interest in developing Mix networks during recent years~\cite{diaz2022mix}.
For instance, among the mentioned projects, the Nym network~\cite{diaz2021nym} is proposed to provide a generic infrastructure to be integrated flexibly within myriad services and applications.
Its decentralized and incentivized infrastructure offers stronger privacy guarantees to its users by providing a massive anonymity set.
The Nym is composed of a decentralized Mix network and an anonymous credential cryptosystem.
The anonymous credentials allow users to prove their "right of use" when privately accessing services over the Mix network.
In fact, the credentials are used as Nym tokens to reward the nodes which provision a high-quality service by adequately routing the traffic.
In addition, a blockchain maintained by users decentralizes the operations of the entire Nym network (including the membership and configuration of the Mix network, the issuing of anonymous credentials and the distribution of rewards).
Thus, the Nym network is able to provide a scalable privacy infrastructure to protect network traffic metadata for a broad range of message-based applications and services~\cite{diaz2021nym}.
\subsection{Proxies and VPNs}
\label{subsec:Proxies_and_VPNs}

While Mixes explicitly batch and reorder incoming messages, proxies, as the simplest solutions for anonymous communication, merely forward all incoming traffic (e.g. a \ac{TCP} connection) immediately without any packet reordering~\cite{edman2009anonymity}.
Moreover, \acfp{VPN}  -- which are popular due to their low latency and support for most network protocols on the Internet~\cite{barman2020prifi} -- are also designed to establish a secure tunnel between a client and a \ac{VPN} server.
\acp{VPN} can be used to encrypt and secure the whole network traffic, not just \acs{HTTP} or \acs{SOCKS} requests from a browser (like a proxy server).
In such a way, it is not possible to easily map the incoming and outgoing traffic when a proxy or a \ac{VPN} is used.
However, message frequencies and flows can still be analysed.
Hence, an observer with access to traffic entering and leaving the proxy or network over extended periods can reveal the communication relation.
While \acp{VPN} are convenient, using them offers a very limited degree of anonymity~\cite{staudemeyer2018road,staudemeyer2019takes, barman2020prifi}.
\subsection{Onion Routing based Solutions}
\label{sec:onion-routing}

Later, according to the principle of Chaum’s Mix cascades~\cite{chaum1981untraceable}, onion routing based methods were introduced by Reed, Syverson, and Goldschlag~\cite{goldschlag1996hiding,reed1998anonymous,goldschlag1999onion} as an equivalent of Mix networks within the context of circuit-based routing.
Onion routing differs from Mixes by not routing each packet separately.
Instead, the client chooses a path and then opens a circuit through the network by sending the first message and labelling the chosen path.
In a circuit, each onion router knows its predecessor and successor, but it does not have any information about other nodes in the circuit~\cite{ren2010survey}.
After establishing the circuit, each message having a particular label is routed on this predetermined path.
In the end, a message can be sent to close the path~\cite{danezis2008survey}.

\begin{figure}
	\centering
	\begin{subfigure}[t!]{0.4\textwidth}
		\centering
		\includegraphics[width=\textwidth]{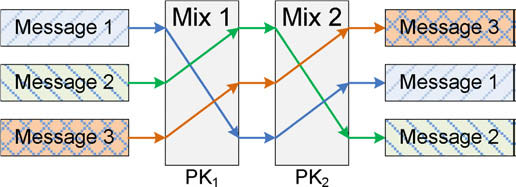}
		\caption{\textit{Chaumian Mix-net}}
		\label{fig:1-a}
	\end{subfigure}
	\hfill
	\begin{subfigure}[t!]{0.5\textwidth}
		\centering
		\includegraphics[width=\textwidth]{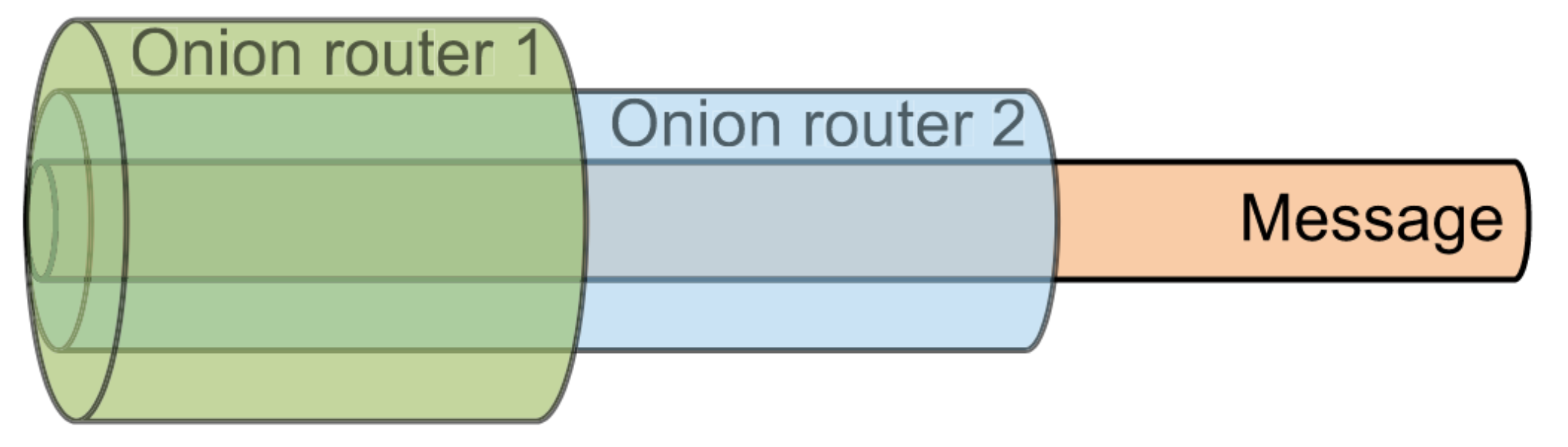}
		\caption{\textit{Onion routing}}
		\label{fig:1-b}
	\end{subfigure}
	\caption{The basic ideas of (a)  Chaumian Mix-nets and (b) Onion routing (derived from~\cite{ren2010survey}).}
	\label{fig:fig1}
\end{figure}

The onion data structure, or simply onion, is composed of layer upon layer of encryption wrapped around the payload (as shown in Figure~\ref{fig:1-b}).
When each onion router receives the fixed-length messages, it performs cryptographic operations on each message and thereby removes a layer of encryption by using its own private key.
This also uncovers the routing instructions for the next onion router in the circuit, then, the message will be forwarded to the next node.
This process is being repeated until the message is delivered to the final onion router.
In this way, intermediary nodes have no knowledge of the origin, destination and content of the message~\cite{ren2010survey}.
Often, the information travelling through each of the labelled circuits is referred to as an anonymous stream~\cite{danezis2008survey}.

The \acp{ACS} based on onion routing provide an application-independent socket connection.
Therefore, they can be easily used by many applications (e.g. web browsing, \acs{SSH} and instant messaging)~\cite{ren2010survey}.
However, onion routing-based \acp{ACS} differ regarding how the onion routers are organized, how encryption algorithms are applied, how the tunnels are established, whether the transport-layer uses \acs{TCP} or \acs{UDP}, or whether the clients relay traffic to other clients~\cite{erdin2012anonymous}.
Thus, a large number of \acp{ACS} based on onion routing as underlying approach have been deployed.
These solutions have attracted millions of users due to the low-latency connections they offer.

\ac{Tor}~\cite{dingledine2004tor,syverson2004tor} is a distributed-trust, circuit-based low-latency anonymous communication network built upon the onion routing design~\cite{conrad2014survey}.
\ac{Tor} is an overlay network (a communication network constructed on top of another network~\cite{Galuba2009Overlay}.
It consists of a set of voluntary servers called onion routers, which are used to build circuits and relay messages~\cite{ren2010survey}.
The \acf{I2P} is another message-oriented system offering anonymization services by using peer-to-peer low-latency communication.
In fact, \ac{I2P} is another overlay network, mainly designed to enable fully anonymous communication between two parties inside the network~\cite{timpanaro2011monitoring,timpanaro2012i2p}.

Crowds~\cite{reiter1998crowds}, Hordes~\cite{levine2002hordes}, LASTor~\cite{akhoondi2012lastor}, Torsk~\cite{mclachlan2009scalable}, \acf{HORNET}~\cite{chen2015hornet}, and \acf{NIAR}~\cite{shi2021non} are based on onion routing as well.
Crowds was specially designed to hide a specific user's action within the actions of many others during web browsing~\cite{reiter1998crowds,lu2019survey,fernandez2012survey}.
Instead of operating a set of onion routers, Crowds' clients relay the traffic of others.
To create an anonymous web request, the representative process of each client (called \textit{jondo}) establishes a random path by choosing another \textit{jondo} from the crowd and forwarding the request to it~\cite{edman2009anonymity}.
Upon receiving the request by the selected \textit{jondo}, it decides to either randomly forward the request again to another \textit{jondo} or forward it to the intended recipient.
The server's response will also be routed in the reverse path through Crowds.

Analogously, Hordes provides a similar degree of anonymity but with a significant performance advantage (in terms of latency in data delivery and the amount of participants’ required work by using multicast communication to anonymously route the reply to the initiator)~\cite{levine2002hordes}.
Another protocol which is known as \acf{RAC}~\cite{mokhtar2013rac} also bases on the principle of onion routing protocols.
\ac{RAC} provides better scalability and resolves the free-riders issue. In this context free-riders are users who have no interest in acting as relay and drop the messages they are supposed to relay~\cite{mokhtar2013rac}.

Furthermore, anonymous broadcast messaging systems based on Mixing and onion routing have also been investigated.
Vuvuzela~\cite{van2015vuvuzela}, Pung~\cite{angel2016unobservable}, Stadium~\cite{tyagi2017stadium} and Karaoke~\cite{lazar2018karaoke} are designed for private message sharing in order to support a large number of users and provide a `cover' for sensitive use cases.
However, these methods also have their own vulnerabilities and impose large delays that prevent them from being accepted.
As an example, Vuvuzela~\cite{van2015vuvuzela} works by routing user messages through a chain of servers and adopts ideas from differential privacy~\cite{dwork2014algorithmic} to prove strong guarantees about the level of privacy provided by dummy traffic.
It is designed for private message sharing in which both sender and receiver pull some information from the system.
Vuvuzela can scale up to two million online users and achieves a throughput of four messages per minute per client with a 37-second end-to-end latency on commodity servers.
But, all messages must have fixed size and the server pads them to the largest message size making its adoption to online storage services inefficient~\cite{akhoondi2015scalable}.
More crucially, it cannot hide the fact that a user is connected to its network~\cite{van2015vuvuzela}.

Moreover, there have been significant works on designing file-sharing systems that allow people to anonymously store, publish, and retrieve data (see survey~\cite{edman2009anonymity} for more information).
Peer-to-peer storage and retrieval systems such as Freenet~\cite{clarke2001freenet}, FreeHaven~\cite{dingledine2001free} and, later, GnuNet~\cite{bennett2003gap,grothoff2003gnunet} provide anonymous persistent data stores.
They use multiple hops to retrieve data associated with a key in a distributed data store.

In summary, solutions based on onion routing employ an application-layer overlay routing and public key cryptography in order to provide sender anonymity.
The most popular protocols of them, such as \ac{Tor}, offer large anonymity sets in the order of millions of users~\cite{kotzanikolaou2017broadcast}.
However, in addition to the latency imposed by the required sequential operations on different servers, the stateful nature of hops or routers makes traffic analysis a serious threat to onion routing approaches~\cite{sergeev2013network}.
Thus, it must be said that onion routing protocols are not designed to protect users against global adversaries~\cite{johnson2013users}.
It has been shown that onion routing protocols are susceptible to a variety of traffic analysis attacks~\cite{wright2004predecessor,murdoch2005low,bauer2007low,hopper2010much,mittal2011stealthy}, even those performed by local adversaries~\cite{cai2012touching,kwon2015circuit,panchenko2011website,wang2013improved}.
For example, in 2014, a study~\cite{chakravarty2014effectiveness} showed that more than 81\% of \ac{Tor} clients can be de-anonymized via traffic analysis.

\subsection{Broadcast/Multicast-based Solutions}
\label{subsec:Broadcast}

Anonymous Communications Systems based on broadcast or multicast methods are designed to provide anonymity, in a scalable manner, through one-to-many communications among hosts~\cite{lu2019survey}.
When using broadcast or multicast communication, the message is sent to all (or a set of) nodes of a network, and it protects the receiver's anonymity.
In this case, instead of using an implicit destination address to enable the only intended recipient to recognise the message, public-key encryption can be utilised.
Every message broadcasts to every participant, then all recipients attempt to decrypt them, whereas only the intended will succeed by using the correct private key.
This also ensures confidentiality, integrity and authenticity~\cite{staudemeyer2017security}.
If broadcast or multicast techniques are used, the senders send their messages to a group of recipients (while these recipients look the same).
According to the definition of anonymity (see Section~\ref{sec:background:privacy-terms}), a higher number of recipients lowers the chance for an attacker to guess who the real receiver is.
Hence, broadcast-/multicast-based systems can increase the anonymity of its participants.

There exist multiple \acp{ACS} using broadcast or multicast-based communication.
Even \ac{DCN}-based methods can be categorised into this type.
However, since \ac{DCN}-based \acp{ACS} also provide sender anonymity and are mainly derived from Chaum's protocols; therefore, it is more preferable to consider them as a main separated category.
\acf{P5}~\cite{srinivasan2002p5}, K-Anonymity~\cite{von2003k}, \acf{M2}~\cite{perng2006m2}, \acf{MAM}~\cite{xiao2006design} and \acf{BAR}~\cite{kotzanikolaou2017broadcast} can be named as broadcast-/multicast-based \acp{ACS}.
For instance, \ac{P5} creates a broadcast hierarchy, in which different hierarchy levels provide different levels of anonymity at the cost of communication bandwidth and reliability~\cite{srinivasan2002p5}.
In \ac{P5}, all messages sent to a certain receiver are transmitted from a single upstream node.
Thus, the receiver does not know the original message sender, also the sender does not know who the receiver is (or which host or address the receiver is using).
\ac{P5} provides individual participants with a trade-off between the degree of anonymity and communication efficiency.
The users always have the flexibility to decrease their level of anonymity in order to increase their performance~\cite{ren2010survey, lu2019survey}.

Another example of broadcast-based \acp{ACS} is \acf{BAR} which proposed a scalable anonymous Internet communication system that combines broadcast features with layered encryption of Mix networks~\cite{kotzanikolaou2017broadcast}.
In this system, a selective broadcast mechanism can provide significantly lower broadcast costs.
Moreover, an efficient filter mechanism allows users to filter out noise traffic and selectively decrypt only those messages intended for them.
Unlike Mix network systems, it provides sender, receiver and unlinkability with forward secrecy.
The system consists of three different parties: users, the \ac{BAR} servers acting as the broadcast servers, and a system coordinator whose role is to publish system parameters and support its operation~\cite{kotzanikolaou2017broadcast}.
The \ac{BAR} design is not distributed, there is a coordinator that acts as a single entity to manage the users, servers and clusters (which must be available in real-time).
Hence, the system performance completely depends on the coordinator plus the number of broadcast servers, since each server can only handle up to some hundreds of users according to the implementation results~\cite{BARgithub2016}.

\subsection{Oblivious Transfer}
\label{subsec:OT}

The first form of \ac{OT} was introduced by Rabin in~\cite{RABIN1981exchange}, which used as a secret exchange protocol between two parties~\cite{gupta2015make}.
An \ac{OT} protocol enables a sender to transfer a record of information from a sequence of records to a receiver, while the sender remains oblivious about which record is selected, also the protocol hides the rest of the records from the receiver~\cite{javani2021privacy, Schoenmakers2005}.
A slightly more advanced form of \ac{OT} is \textit{`chosen one-out-of-two'} \ac{OT}, denoted as $ OT_1^2$~\cite{even1985randomized}, where the sender has two private inputs $ (X_1, X_2)$, and the receiver can choose to get either $X_1$ or $X_2$ and learns nothing about the other input~\cite{Schoenmakers2005,naor2005computationally}.

Similarly, the generalised form of $ OT_1^2$ was introduced by Brassard \emph{et al.}~\cite{brassard1987all} under the name \acf{ANDOS}.
In \textit{1-out-of-n} denoted $ OT_1^n$, the sender has \textit{n} private inputs and the receiver can choose to get one of them on her choice, without learning anything about the other inputs and without knowing the sender which input is transferred~\cite{naor2005computationally,javani2021privacy}.

Indeed, \ac{OT} is a cryptographic primitive that provides the capability for selecting and transferring data between two parties and can be used as a building block in different contexts where there is a requirement to hide or limit the information about data transfer~\cite{javani2021privacy}. This property of \ac{OT} is used to design a delivery mechanism in a novel \ac{ACS} called \ac{AOT}~\cite{javani2021aot}. \ac{AOT} is a protocol that uses \ac{OT} to facilitate anonymous two-way communication and deliver the messages to the recipients. The \ac{AOT} is a system based on Mix network architecture that comprises three levels of nodes, where each one performs a different function. The senders send the encrypted payloads along with their corresponding tags to the network. The tags are derived from secret keys, which are shared in advance between the sender and receiver of each message and will be later published on a public bulletin board. Then in the network, Level-1 nodes strip the sender information of messages and send them to Level-2 nodes in batch, while Level-2 nodes add dummy messages and send the reordered batches of dummy and real messages to Level-3. At the end, Level-3 nodes publish tags associated with messages on the bulletin board. By identifying any tag, a user knows a message is prepared for him and uses \ac{OT} to request the message associated with that tag from a Level-3 node. Using \ac{OT} hides which messages are received by users from a larger set of messages, hence a network adversary cannot link senders and receivers. In summary, the combination of OT and Mix networks increases the anonymity provided by the system, also users do not have to register with the system; any entity who knows the public key of any middle-layer node in the system can send messages with \ac{AOT}~\cite{javani2021privacy,javani2021aot}.

In the same fashion as \ac{OT}, \ac{PIR} protocols allow a client to retrieve (or fetch) an item from a destination in possession of the client, a database, without revealing which record is retrieved~\cite{chor1995private, akhoondi2015scalable, gilad2019metadata}. Riffle~\cite{Kwon2015riffle} uses hybrid mix networks and \ac{PIR} techniques to implement anonymous messaging with an acceptable privacy guarantee, but it cannot handle changes in the network topology~\cite{alexopoulos2017mcmix}. Riposte~\cite{corrigan2015riposte} also uses \ac{PIR} techniques in a system with multiple servers to provide anonymous message broadcasting (Riposte will be explained in detail in Section~\ref{subsec:4.4.5Riposte}). Pung~\cite{angel2016unobservable}, \acf{P3}~\cite{kissner2004private}, and \ac{XPIR}~\cite{aguilar2016xpir} are other anonymous communication systems based on \ac{PIR}, which use a key-value store to allow clients to deposit and retrieve messages without anyone learning the existence of conversation. Although using smart database organization helps these methods to scale to a large number of users, they exhibit substantial client load. As a result, \ac{PIR} techniques require high bandwidth and computation, and they usually provide data anonymity meaning the destination knows the client, but, it does not realise what records are read or written.
\subsection{Multi-Party Computation}
\label{subsec:MPC}

Secure \acf{MPC} is a method which enables a group of distrusting parties $P_1, \ldots , P_n$ to collaboratively compute a function \textit{f}. Hereby, each party $P_i$ can contribute an input $x_i$ to the computation; no other party learns these values, but all learn the result $f(x_1 , \ldots , x_n)$~\cite{von2018management,mueller2000anonymous}. Indeed, the only information each party learns about the inputs of other participants is this result. Therefore, \ac{MPC} can be used by multiple independent data owners, who do not trust each other or any common third party, to carry out a distributed computing task that depends on all of their private inputs in a secure manner~\cite{evans2018pragmatic}. Oblivious Transfer, which is described in the previous item, is one of the cryptographic primitives for building secure multi-party computations to enable the utilisation of data without compromising privacy.

In general, \ac{ACS} solutions which use \ac{MPC} techniques~\cite{kwon2017atom} provide strong anonymity guarantees. Sometimes they use mixing to pass the intermediary results through the network, although the sequential mixing is time-consuming and slows down the protocol. For instance, MCMix casts the problem of anonymous messaging in the setting of \ac{MPC}~\cite{alexopoulos2017mcmix}. The MCMix system enables clients to use a dialling functionality to call other clients and establish a random tag. Subsequently, the dialler and dialee use this tag in the conversation functionality to send messages (even concurrently)~\cite{alexopoulos2017mcmix}.

Furthermore, there have been several efforts to use \ac{MPC} as \acf{MPSaaS}~\cite{barak2018end, lu2019honeybadgermpc}. AsynchroMix is an application of the \ac{MPSaaS} approach to anonymous broadcast communication. In a typical client-server setting, the clients send their confidential messages to server nodes which continuously process those encrypted inputs from the clients. The system selects a subset of clients whose inputs are mixed together before making them public. AsynchroMix employs a \ac{MPC} implementation called HoneyBadgerMPC~\cite{HoneyBadgerMPC2020}, which relies on the pre-processing paradigm. Thus, it features robust online phases along with non-robust but efficient offline phases~\cite{lu2019honeybadgermpc}.
\subsection{Existing Surveys on Anonymous Communication Systems}
\label{subsec:Surveys}

Interested readers who want to study \acp{ACS} in more detail can refer to other surveys such as the following articles, while each one has further elaborated \acp{ACS} with their perspective.

To name a few, Danezis and Diaz in~\cite{danezis2008survey} reviewed briefly the underlying principles, the advantages and disadvantages of various \acp{ACS}. Further, an overview of the research in anonymous communications in terms of their basic definitions, cryptographic primitives, network protocols and their applications is conducted in~\cite{ren2010survey}. Also, a summary of \acp{ACS} is provided by~\cite{kelly2011exploring}. Moreover, Fernández in~\cite{fernandez2012survey} categorised different schemes of anonymous communication according to their architecture, client-server and peer-to-peer. These systems are compared based on their resistance against the most notorious attacks, and several aspects of them are exposed, including scalability, anonymity and unobservability, security and censorship-resistance. Even among more recent studies, a survey on \acp{ACS} in the \acf{IoT} is published in~\cite{wang2019survey} to explore them based on computational offloading and lightweight cryptography. Other than that, different types of anonymous communication based on different anonymous mechanisms such as routing, broadcast, etc. and a formalization of the notion of anonymity for measuring its degree are provided in~\cite{lu2019survey}.

However, one of the most cited studies on anonymity is the one presented by Edman and Yener in 2009~\cite{edman2009anonymity}.
In this paper, major concepts and technologies to design, develop, and deploy systems for enabling private and anonymous communication on the Internet are described.
Anonymous systems are classified into high-latency and low-latency systems depending on their intended application and their latency tolerance.
Considering this classification, high-latency anonymity systems are able to provide strong anonymity, but they impose significant delays between sending and receiving transmitted messages (up to multiple hours and random message drops).
The high-latency systems, also known as message-based systems~\cite{edman2009anonymity}, are typically applicable for non-interactive applications that can tolerate delays of several hours or more.
Hence, they can only be used for asynchronous services which accept notable delays~\cite{Serjantov2003}.
In contrast, low-latency anonymous systems offer better performance and can be adopted for use-cases which require higher levels of interactivity and bidirectional communication channels, e.g. web browsing or \ac{SSH} sessions.
Consequently, low-latency systems are also called connection-based systems~\cite{edman2009anonymity,Serjantov2003}. Furthermore, \ac{DCN}-based systems are introduced as a means to offer unobservability in this survey.

According to the literature review, there is a need for a comprehensive study of \ac{DCN}-based methods. For this reason, the next section of this article is dedicated to a detailed review of these methods. 
\subsection{Overview of Provided Privacy by \acp{ACS}}
\label{sec:acs_privacy_map}

In this section, a comparison of the offered privacy features by different \ac{ACS} methods is prepared in the form of a chart, which we introduce as privacy map.
To draw this map, privacy properties (anonymity, unlinkability, unobservability and their role-based subcategories) are considered in line with the definitions provided in Section~\ref{sec:background:privacy-terms}.
Then, the main \ac{ACS} methods cited in Section~\ref{sec:GENERAL OVERVIEW OF ACS} are investigated in terms of the quality of their offered privacy properties.
In fact, the privacy map visualises the alignment of ACS methods with the privacy terminology.

The size of anonymity/unobservability sets is represented by circles in three sizes to show the maximum number of supported users in the sets (\textit{small, medium or large}).
The circle sizes are determined based on reported evaluation results for each method and the highest number of supported participants at their most efficient setup (acceptable overhead and performance in practice) is represented as the degree of anonymity/unobservability.

\begin{figure}
	\includegraphics[width=\textwidth, height=0.95\textheight,keepaspectratio]{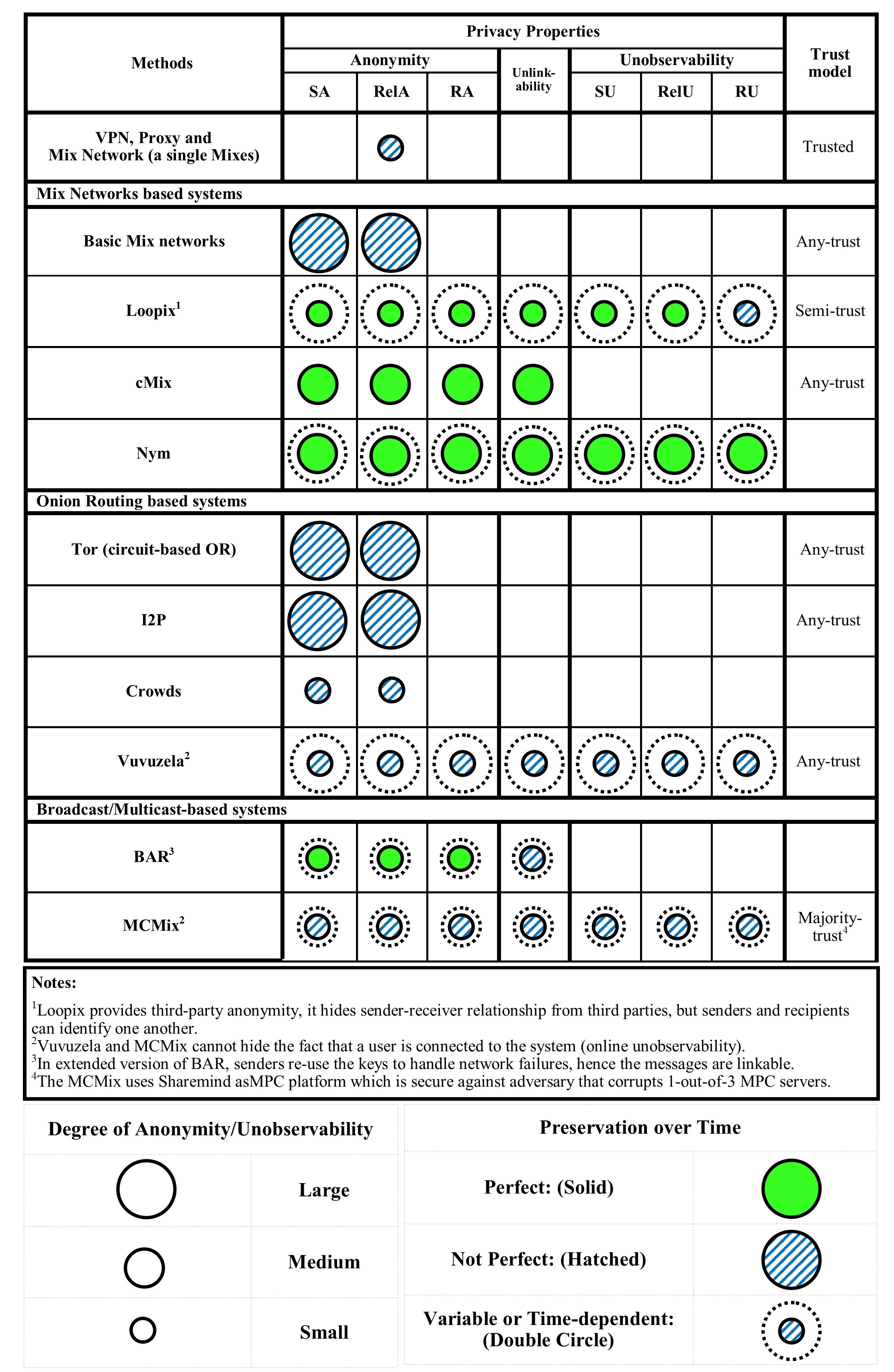}
	\caption{Comparison of main general \acp{ACS} from the privacy properties perspective.\label{fig:privacy-matrix}}
\end{figure}

In addition, the preservation of anonymity/unobservability properties are presented by the circles' style; \textit{solid} colour represents the methods which provide perfect preservation of anonymity/unobservability, while the \textit{hatched} circles are used to show the methods which are not able to provide perfect preservation.
The size of anonymity/unobservability sets of a not-perfect \ac{ACS} changes over time.
Furthermore, \textit{double-circle} icons are used to illustrate the methods which offer variable or time-dependent properties.
In these methods, the size of anonymity/unobservability sets depends on the participation of nodes/users during a specific time interval (i.e. usually known as epochs or time slots). Hence, the anonymity set does not have the same size at all times, and the maximum supported sets are much larger than the actual sets.
The assumptions on the trust model are also mentioned in the privacy map.
Figure~\ref{fig:privacy-matrix} presents the privacy map to compare general \acp{ACS}.

According to the defined privacy properties in Section~\ref{sec:background:privacy-terms}, three different classes can be imagined for the \acp{ACS}, in general;
(1) Anonymity-support group with VPNs and mostly onion routing-based systems.
(2) unlinkability-support group including systems, which uses Mix networks.
(3) And unobservability-support group, which provides the highest protection for the users and is supported by systems like recent implementations of Mix networks like cMix~\cite{chaum2017cmix} and Loopix~\cite{piotrowska2017loopix}.
This classification is also visible in Figure~\ref{fig:privacy-matrix}, in addition, this privacy map clearly highlights the effectiveness of different approaches in providing privacy protection.

Forwarding and redirecting the packets alone in the way that happens in onion routing methods such as Tor~\cite{syverson2004tor} and I2P~\cite{timpanaro2012i2p} is not enough to provide strong privacy protection, and it can only offer anonymity within typically a large anonymity set which should be expected to be degraded over time.
Further, batching, re-ordering messages or imposing random delays during the message transmission is necessary to hide the relationship between messages and their links to the senders/receivers.
The approach taken by Loopix~\cite{piotrowska2017loopix} is good evidence for this claim.
Additionally, when a message can be accessed by multiple subjects at the same time, for instance, by broadcasting the message or publishing it on a  shared board/server, this will lead to receiver anonymity.
BAR~\cite{kotzanikolaou2017broadcast} is an example of this approach.
The last finding is that the only way to make the actual activities of subjects indistinguishable is to hide them by utilising randomly generated dummy traffic.
In this way, the attacker can not understand the difference between real messages and noise, and thus communications will be unobservable.
Indeed, Loopix~\cite{piotrowska2017loopix}, Nym~\cite{diaz2021nym} and MCMix~\cite{alexopoulos2017mcmix} all use random traffic to offer unobservability.

Apart from looking at the \acp{ACS} from the privacy perspective and categorising them based on their offered privacy or grouping the \acp{ACS} based on their main employed idea to protect privacy, it is also possible to classify these methods based on their desired application. 
In this way, \acp{ACS} can be mostly divided into two groups, latency-sensitive and latency-tolerant.

The main criteria for users who want to protect their privacy during latency-sensitive applications such as instant private messaging, web browsing and financial transactions is to experience seamless and (near) real-time communication. 
Therefore, the \acp{ACS}, which impose low or medium latency during the communication phase, can be classified in the latency-sensitive class. 
VPN and Tor~\cite{syverson2004tor}, which are the most widely used and well-known methods, are included in this class, along with the recently proposed Loopix~\cite{piotrowska2017loopix} and cMix~\cite{chaum2017cmix} methods.

On the other hand, the methods, which are designed to provide higher privacy protection by adding extra random delays, such as the delayed version of Loopix~\cite{piotrowska2017loopix} and MCMix~\cite{alexopoulos2017mcmix}, are counted in the latency-tolerant class. 
The methods of this category are for example desirable to be used for posting on blogs, file sharing and email communications. 
Detailed information on the \acp{ACS}' desired applications is provided in Table~\ref{tab:acs-overall-comparision}.

\section{Anonymous Communication Systems based on Dining Cryptographers Networks}
\label{sec:DCN-BASED ACS}
\ac{DCN} is a broadcast round-based protocol, which provides unconditional secure unobservable communication. The name of this protocol, \ac{DCN}, comes from a little story, first introduced by David Chaum in 1988~\cite{chaum1988dining}- Dining Cryptographers’ Network; “Three cryptographers meet for dinner in a restaurant which has paid beforehand. They are curious who paid for the dinner — either one of the cryptographers or it was sponsored by the employer (for instance the \ac{NSA} or the government). In case one of the cryptographers paid, however, they do not want to reveal who exactly.” Chaum came up with the \ac{DCN} protocol, which provides unconditional security for messages’ originators in a closed group to solve this problem~\cite{scholz2007dining, staudemeyer2019takes}. Unconditional secure means that it can be proven that it is impossible to find out who paid~\cite{chaum1988dining}. The seminal protocol only allowed participants to unobservably publish a 1-bit message per round, which is called “superposed sending”~\cite{chaum1988dining,chaum1985security,pfitzmann1987networks}.

In this section, we discuss the \ac{DCN} protocol, its offered level of anonymity and features besides the main challenges, which have caused \ac{DCN} to relatively remain neglected in the first years.
In addition, the comprehensive review of the proposed solutions based on \ac{DCN} protocol is provided in \ref{subsec:DCN-4-methods} to investigate the recent contributions in the development and implementation of \ac{DCN}-based \acp{ACS}.
   
\subsection{The Basic DCN Protocol}
\label{subsec:DCN-1-basic}

In the following, an example taken from~\cite{edman2009anonymity} is used to demonstrate the basic principle behind \ac{DCN} protocol for \textit{n} cryptographers (or players, participants, users). "Let assume \textit{n} cryptographers seated in circle as nodes in an undirected circular graph. Every link in this graph, between two neighbouring cryptographers, represent a one-bit secret shared key between the nodes. Let $x_{i, j}$ be the bit shared between neighbouring cryptographers \textit{i} and \textit{j}. Further, let $s_i$ be cryptographer i’s secret bit indicating whether or not he paid for the meal. Thus, each cryptographer \textit{i} is announcing the result of $z_i = x_{(i-1),i} \oplus x_{i,(i+1)} \oplus  s_i$ to the network. In this way, by receiving a message from each cryptographer, in the end, all the cryptographers can compute the result as shown in Equation~\ref{eq:dcn}.
\begin{align}
	\label{eq:dcn}
	&Z = z_1\oplus z_2 \oplus \ldots \oplus z_n \nonumber \\
	& = (x_{n,1} \oplus x_{1,2} \oplus  s_1)\oplus (x_{1,2} \oplus x_{2,3} \oplus  s_2) \oplus \ldots \nonumber \\
	&\oplus (x_{(n-1),n} \oplus x_{n,1} \oplus  s_n) \nonumber \\
	&= x_{n,1}\oplus x_{n,1} \oplus s_1\oplus x_{1,2}\oplus x_{1,2}\oplus s_2 \oplus \ldots \nonumber \\
	&\oplus x_{(n-1),n} \oplus x_{(n-1),n} \oplus  s_n \nonumber \\
	&= s_1\oplus s_2\oplus \ldots \oplus s_n
\end{align}
where the third step follows from a simple reordering of terms. If no cryptographer paid for the meal, $s_i = 0$ for all \textit{i}. Otherwise, there is precisely one $s_i$ that is non-zero and the result $Z = 1$."

Later, \ac{DCN} were enhanced to support arbitrary message lengths and shared secret sizes. By running this protocol in several rounds and assigning each round to only one user, this user can anonymously publish an \textit{l}-bit message. The following example describes \ac{DCN} for six-bit message transmission.

\begin{figure}
	\centering
	\begin{subfigure}[t!]{0.4\textwidth}
		\centering
		\includegraphics[width=\textwidth]{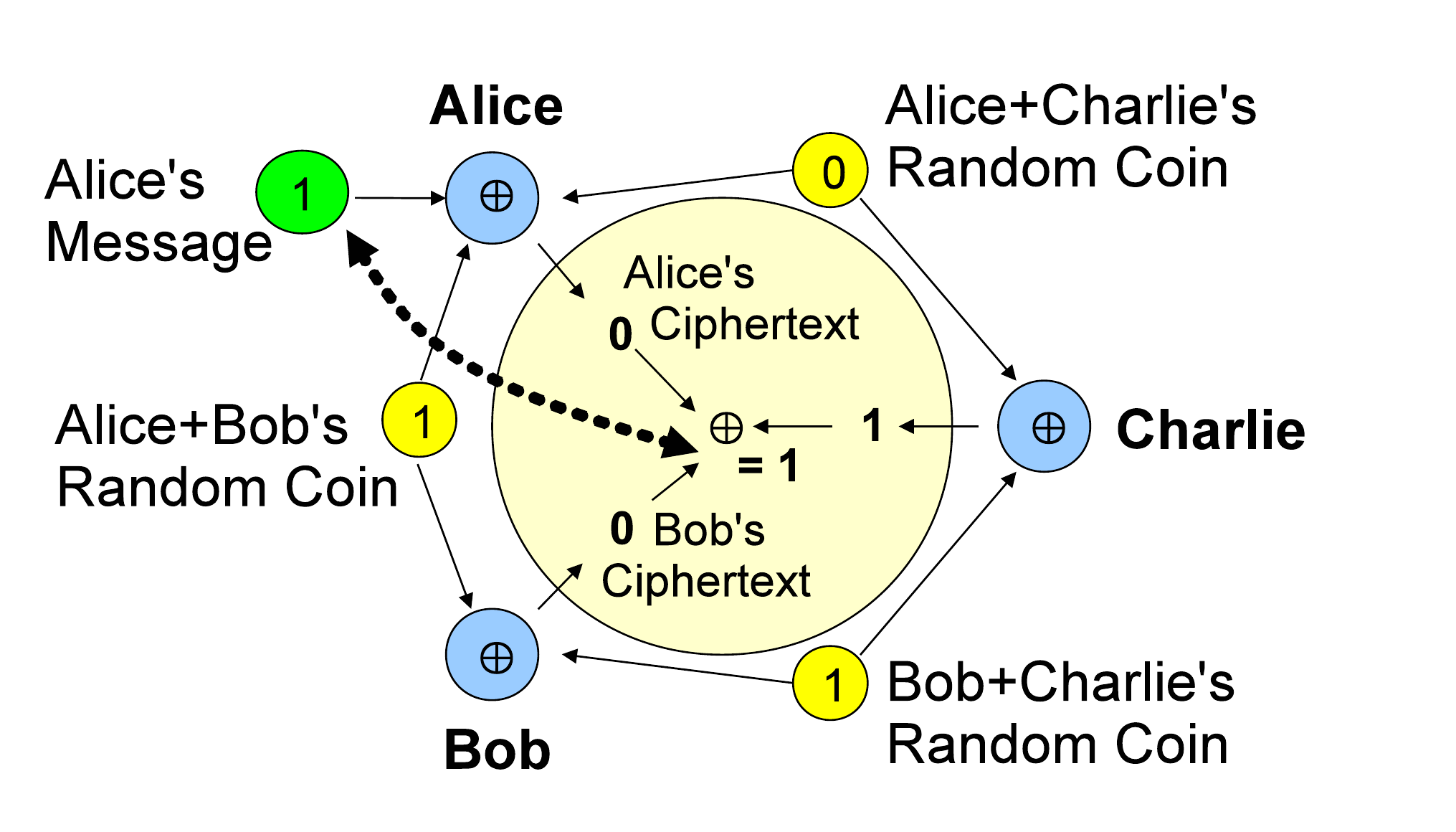}
		\caption{Basic DCN protocol to exchange a one-bit message~\cite{corrigan2013proactively}.}
		\label{fig:DCN-basic-flow}
	\end{subfigure}
	\hfill
	\begin{subfigure}[t!]{0.4\textwidth}
		\centering
		\includegraphics[width=\textwidth]{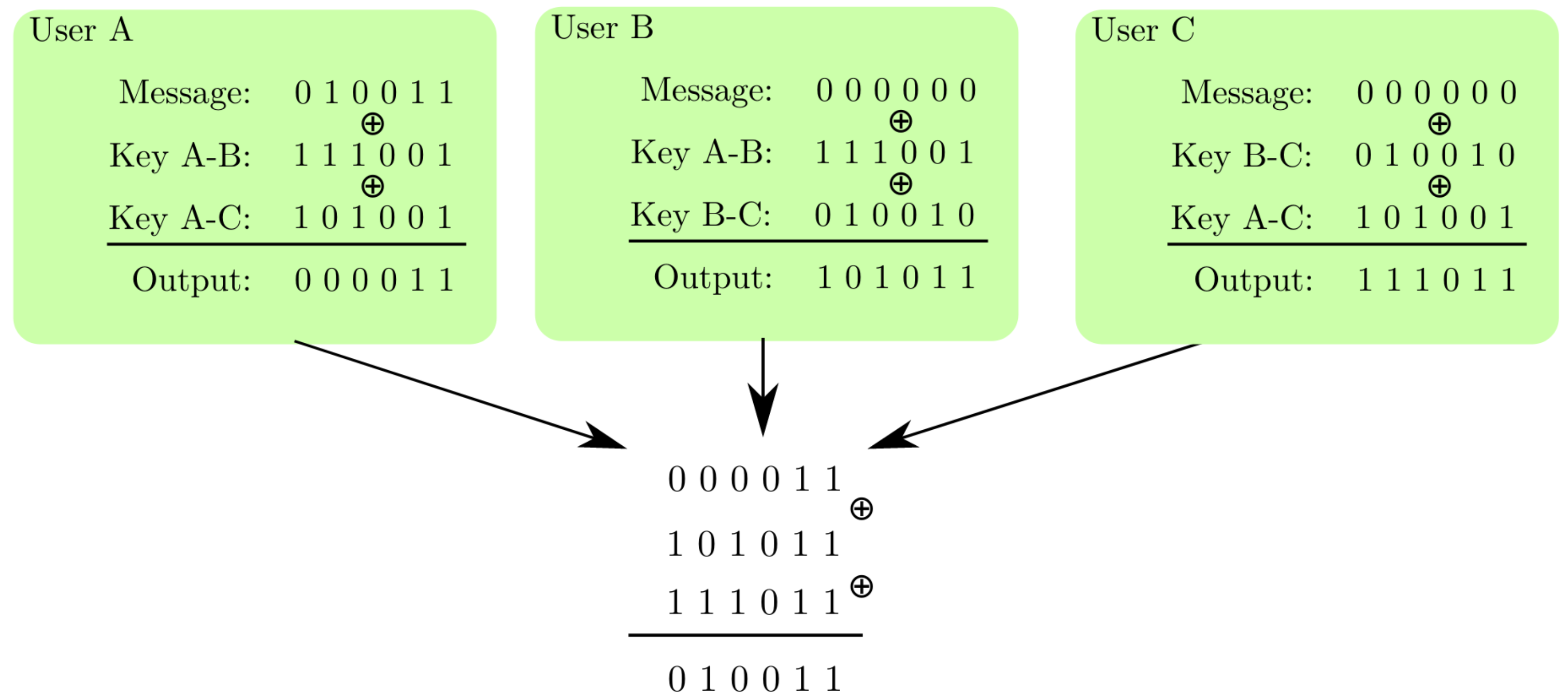}
		\caption{A \ac{DCN} example with three participants to show how this protocol works to transmit a six-bit message per round~\cite{krasnova2016footprint}.}
		\label{DCN-6bit}
	\end{subfigure}
	\caption{The basic DCN protocol flow for three users in a setting in which (a) one-bit and (b) six-bit messages are being transmitted.}
	\label{fig:fig2}
\end{figure}

\textbf{Examples.} In Figure~\ref{fig:DCN-basic-flow}, the \ac{DCN} protocol is shown in its most basic way: three parties would like to exchange a one-bit message.
In this example, taken from~\cite{corrigan2013proactively}, each member pairs up with his neighbours.
Both flip a coin and agree on a secret result.
Afterwards, each member \acp{XOR} all results he knows.
The member publishing the message also \acp{XOR} the message would like to publish with the result of the previous \acp{XOR} operation.
Then, the results of the mathematical operations which took place are published and each member \acp{XOR} the messages he receives.
The result is the message that was published anonymously~\cite{corrigan2013proactively}.

In~\cite{krasnova2016footprint}, it is described how the basic \ac{DCN} protocol works with an example, which is also illustrated in Figure~\ref{DCN-6bit}. Three participants want to exchange six-bit messages and every participant has a shared symmetric key (six-bit key) with each other participant. By assuming participant A is the node who wants to send a message in this round, she \acp{XOR} her message with all the keys she shares with the other participants. The result is an output that A sends out to every participant of the \ac{DCN}. The other participants perform the same procedure, but use a zero message instead of a meaningful one. All outputs of all participants are \ac{XOR}ed together to reveal the meaningful message of participant A, because keys are cancelling out (each symmetric key is used twice). This completes a single round of the \ac{DCN}, during which, one participant can transfer (broadcast) one message; in the next round, another participant transmits a message until all participants are done transmitting~\cite{krasnova2016footprint}.

\subsection{The Strength of DCN: Unconditional Unobservability}
\label{subsec:DCN-2-strength}

Since, the construction of \ac{DCN} relies on the `reliable broadcast assumption'~\cite{waidner1989dining}, providing receiver anonymity is easy. As mentioned in Section~\ref{sec:GENERAL OVERVIEW OF ACS}, broadcast-based concepts offer full receiver anonymity. In addition, the reliability of broadcast means the broadcasted message should be received by every receiver unaltered which can be ensured by verifying the integrity~\cite{scholz2007dining}.

In terms of preserving sender anonymity, the designated protocol must be able to eliminate the possibility of identifying the message originator, under any circumstances. The privacy is compromised, if an attacker can learn something about a participant's role from the network. The strongest possible attacker, in this case, is able to observe all communications and can collude with all participants except two honest ones (Colluding means the attacker knows all the exchanged keys and the individual messages of the participants~\cite{scholz2007dining}. In addition, it is worth mentioning that there is no anonymity if all participants are colluding against one single victim~\cite{bauer2017dining}). Nevertheless, in the case of using \ac{DCN} protocol, the attacker still cannot discover the exact role of the honest participants, i.e. acting as a sender or receiver, even if they are only two.

Indeed, for every message, every participant in a network can be in the role of the receiver, sender, none or both~\cite{scholz2007dining}. Therefore, the attacker is not able to distinguish whether honest nodes are communicating, or just being idle, which is consistent with the definition of unobservability.  
Obviously, the assured unobservability set of a single node in the \ac{DCN} protocol is equivalent to the set of honest nodes participating in the protocol and the node has already shared secrets with them~\cite{bauer2017dining}. 

Moreover, it should be noted that, to set up keys before message transmission, every participant exchanges random keys with everyone else and add them to the value he sends on the network~\cite{scholz2007dining}. Therefore, if the participants of a \ac{DCN}, share secrets through an unconditional secure channel, the \ac{DCN} provides unconditional unobservability (or in general privacy). While, if the key exchange is done through a public-key cryptography system, the privacy of a \ac{DCN} degrades to the degree of computationally robustness of the security for the used public-key cryptography~\cite{staudemeyer2019takes}.

Subsequent to keys establishment, players may accomplish their anonymous message transmission in a single broadcast round, with no player-to-player communication~\cite{golle2004dining}. This very attractive and compelling feature of the basic \ac{DCN}, as formulated by Chaum, is called non-interactivity, which is not possible in other privacy-preserving tools like Mix networks based communication protocols~\cite{ren2010survey}.

In the end, it can be guaranteed that the basic privacy offering of \ac{DCN} are much stronger than solutions like \ac{Tor}~\cite{bauer2017dining}, and they guaranteed a higher privacy level by providing provable sender and receiver unobservability without relying on a trusted third party~\cite{ren2010survey}.
\subsection{Main Challenges of DCN}
\label{subsec:DCN-3-challenges}

The listed advantages of \ac{DCN} protocol and achieving to unconditional unobservability are very decisive to choose an appropriate privacy preserving solution; however, they come at the cost of low throughput and higher computation and delay in particular when scaling to many participants~\cite{krasnova2016footprint}. In addition, there are major drawbacks that are large obstacles to the development of protocols based on \ac{DCN}. The following can be considered as the main challenging issues and practical problems of \ac{DCN} adoption and implementation in reality:
\subsubsection{Scheduling (Collision Prevention)}
If two participants send in the same round, their messages collide and become unusable~\cite{krasnova2016footprint}. Even when all participants are honest and adhere to the protocol, there is still no perfect means of enabling them to select distinct rounds in order to transmit their messages in a non-interactive manner~\cite{golle2004dining}. This problem can be avoided by scheduling the rounds or slot reservation in advance. In doing so, each participant needs to know when it is his or her turn to send a message, and it is mandatory to schedule the rounds before any transmission. Therefore, the task of slot reservation or scheduling is to agree on a transmission schedule in a way that each participant knows when to send, but does not learn who is sending in the other slots~\cite{krasnova2016footprint}.

For this reason, many of the standard slot-reservation protocols are not applicable due to compromising anonymity~\cite{krasnova2016footprint}. In addition, the probability of a collision never reaches zero with reservation procedures, and in any case there is some possibility that two (or more) players attempt to transmit messages in the same round (or slot). Hence, in all cases, \ac{DCN} protocols involve collisions (whether of messages or even reservation requests) which mandates retransmissions and causes more cost and delay in delivering messages. The general approaches for slot reservation could be divided into three classes including reservation-map methods, collision-resolution algorithms and secure multi-party computation~\cite{krasnova2016footprint}.
\subsubsection{Disruption and Jamming Protection}   
Besides accidental collisions, a single malicious or dishonest insider – a participant who wants to disrupt the communication – can straightforwardly jam the network by intentionally corrupt or block the transmission of messages from honest participants. These disruptions by malicious participants can prevent the delivery of messages, either by broadcasting invalid messages via tampering bits in encrypted messages of others or even simply by dropping out of the protocol~\cite{golle2004dining}. What makes things worse is that tracing the jamming source, in this case, is challenging due to the privacy guarantee in \ac{DCN} and each participant is as anonymous as any message originator. Hence, the malicious participant can choose to send a message every round or not following the protocol to launch a denial-of-service attack and disrupt the entire \ac{DCN} communications without being identified~\cite{ren2010survey}. Detecting cheating players comes at a cost as well; multiple broadcast rounds, high computational and communications overhead and fault recovery are required~\cite{golle2004dining}. For this reason, \ac{DCN} are known to be vulnerable to disruption attacks (i.e., jamming)~\cite{chaum1988dining}, and many solutions like using trap rounds, relying on commitments or blame mechanisms have been proposed in the literature to find and exclude the disruptor~\cite{barman2020prifi}.
\subsubsection{Churn Handling}
Churn handling means the ability of participants to join or leave the network and is another fundamental \ac{DCN} challenge. A single missing ciphertext in each round prevents the discovery of message. Therefore, unlike other \acp{ACS}, the disconnection of any participant invalidates the current communication, forces re-transmission of the data and leads to global downtime where no one can communicate. Hence, it should be handled intelligently in order to prevent imposing extra overheads in each round~\cite{barman2020prifi}.
\subsubsection{Topology and Scalability}
The initial \ac{DCN} design requires a shared secret between every pair of participants. In this way, the number of nodes in the network to run \ac{DCN} protocol and their topology dictates the number of required shared secret key pairs, latency, overhead and the scalability~\cite{barman2020prifi}. As the public becomes increasingly concerned about threats to personal privacy, the number of anonymity systems’ users is likely to grow and \acp{ACS} must be able to support more users~\cite{edman2009anonymity}. However, since \ac{DCN} requires every user of the network to participate in every round of the protocol, it quickly becomes impractical as the number of users grows~\cite{corrigan2015riposte}. For this reason, scalability is one of the main problems prevent \ac{DCN} to be implemented in current real-world scenarios~\cite{bauer2017dining}.
\section{Review of Existing DCN-based Methods}
\label{subsec:DCN-4-methods}

As stated in the \ac{DCN} challenges, they originally offer unobservability at the cost of high-latency and communication overheads, and they scale poorly~\cite{barman2017prifi}. However, over the years, \acp{DCN} have been re-used in several \acp{ACS} and various improvements have been made to fix the challenges and decrease their cost. This section reviews the efforts in developing methods based on \acp{DCN} and explore how their contributions tried to mitigate the weaknesses and vulnerabilities of \acp{DCN}.
\subsection{Preliminary Studies on DCN}

Chaum himself made the early improvements; a ring topology to decrease the overhead of broadcasting messages and the number of required paired keys~\cite{chaum1988dining}. With this topology, each global broadcast message has to travel twice through the ring in order to be received completely by all members. Further, Chaum pointed trap mechanisms out to address disrupter problems. This way, an honest sender places a trap by sending a random message with a secret key in its reserved slot instead of sending an actual message. Then, if the attacker tries to disrupt the communication in this reserved round, the honest node that placed the trap detects it and signals the other participants~\cite{ren2010survey,bauer2017dining}. Unfortunately, even a computationally limited attacker can forge a trap for an arbitrary slot~\cite{ren2010survey}.

Later, Waidner and Pfitzmann generalise the concept of superposed sending by deciding to take the advantage of Abelian finite group $(F,\bigoplus)$ instead of the \ac{XOR} operation~\cite{waidner1989dining,waidner1989unconditional}. They also propose a multi-round solution to the disruption problem, which is only guaranteed to identify one dishonest player for a given “trap”, but without any chance for fault recovery~\cite{ren2010survey}. In addition, Waidner used additive groups of integers modulo m to improve the original reservation technique. In this case, after broadcasting reservation vectors, participants can sum the positions up to find slots that more than one participant wants to reserve (the addition result is higher than 1). Then, all slots with collisions are skipped, and the \ac{DCN} protocol is executed only in successfully reserved slots. Franck used the Pfitzmann’s collision resolution algorithm again in 2014 in a protocol, which is called \ac{SICTA}. This scheduling protocol operates with multiplication of ciphertexts instead of addition~\cite{franck2008new}.

Additionally, other schemes were proposed to resolve collision~\cite{bos1989detection} and to make \ac{DCN} robust to disruption by employing cryptographic proofs of correctness rather than traps to detect cheating players. However, the bandwidth costs and inefficiencies of the protocol remain high. For instance, in \ac{DCN} variant by von Ahn, Bortz, and Hopper~\cite{von2003k}, participants initially partitioned into autonomous groups, then a secret sharing protocol used to establish secrets and the correctness of pads is proven via a cut-and-choose protocol. The communications complexity of this scheme, in the worst case in the presence of cheating player, rises to $O(n^4)$.

\subsection{Herbivore}

Herbivore is a peer-to-peer scalable \ac{ACS} introduced in 2002 by Goel \emph{et al.}~\cite{goel2003herbivore}. It takes a hierarchical two-level approach. At the lower level, a round protocol governs how bits are sent among the participating nodes, while, at the next level, a global topology control algorithm is employed to divide the network into smaller anonymizing groups~\cite{ren2010survey}. When a new user joins the network, it is assigned to one of many smaller groups of users called cliques. Herbivore guarantees that each clique will have at least $k$ users (between $k$ and $3k$ users), where $k$ is a predetermined constant that describes the degree of anonymity offered by the system. If a clique grows too large, it can be separated into smaller cliques. Similarly, once a clique becomes too small, its users are merged into other cliques. Users within each clique are logically arranged in a star topology, with one user at the centre, and all other nodes communicate via this centre node using \acp{DCN}. Each user in the clique still has a shared key with every other member of the clique. At a higher level, cliques are arranged in a ring topology, allowing inter-clique communication~\cite{edman2009anonymity}.

Further contributions of the Herbivore system design are reservation maps and an enhancement in collision avoidance via optimisation in the size of the scheduling message by allowing some collisions during the reservation cycle depending on the message size (collisions for smaller messages are more likely)~\cite{goel2003herbivore,krasnova2016footprint}. Herbivore attempted to adapt \acp{DCN} into a design that would make them more efficient and suitable for use in low latency, real-time Internet applications~\cite{edman2009anonymity}.
\subsection{Dining Cryptographers Revisited}
The `Dining Cryptographer revisited' paper, presented in 2004~\cite{golle2004dining}, is one of the papers related to jamming detection and proposes asymmetric constructions to detect cheating. It describes the intuition behind \acp{DCN} based on Chaum’s original paper~\cite{chaum1988dining} and reviews the solutions using traps in multi-round protocols such as~\cite{chaum1988dining, waidner1989dining}. The paper shows that catching cheating players with an overwhelming probability comes at the price of higher computation and communication costs. Hence, it proposed two new \ac{DCN} constructions to achieve non-interactivity and high-probability detection in identification of cheating players.

For this purpose, by assuming the presence of a reliable broadcast channel and that all messages have authenticity, a different strategy for pad computation (messages or dummy traffic transmitted by players) has been used and more computationally efficient cryptographic proofs of correctness are employed to proof the pad computation. In these new asymmetric constructions, the players only employ bilinear maps to identify the cheating players with a cost that is linear in the number of participating players, which is reasonable for small sets of players. Moreover, in the case of cheating, full fault recovery is possible with just a single additional broadcast and there is no need for repeating the whole transmission round. However, the issues of collisions are not covered in these constructions and \ac{DCN} considered as a primitive to provide partial throughput~\cite{golle2004dining}.
\subsection{Adding Accountability to \acp{DCN} with Verifiable Shuffles (Dissent, Dissent in Numbers and Verdict)}

Another set of enhancements to \ac{DCN} was made by presenting the secret shuffle methods~\cite{furukawa2001efficient,neff2001verifiable}. In 2007, Studholme and Blake proposed a way to implement secret shuffle based on multi-party computation~\cite{studholme2007multiparty}. In this method, \ac{DCN} participants are organized in a Mix network and use it to transmit their encrypted round reservation vectors. By passing the reservation vectors through the entire Mix, a secretly permuted vector is obtained, so that each participant can recognise his own request only after the permutation is completed. Therefore, the participant nodes can derive their corresponding reserved round number from the position of their requests within the secretly shuffled vector without any possibility to find the round owners by others~\cite{krasnova2016footprint}.

Later, Franck used this idea to derive a fully verifiable variant of \ac{DCN}~\cite{franck2008new}. Then in 2010, Corrigan-Gibbs and Ford built an accountable anonymous messaging protocol upon Verifiable shuffles and called it \acf{Dissent}~\cite{corrigan2010dissent}. This protocol was designed to be used for sending a message anonymously in a small-distributed group to a single recipient or the whole group.

\begin{figure}
	\centering
	\includegraphics[width=0.5\linewidth]{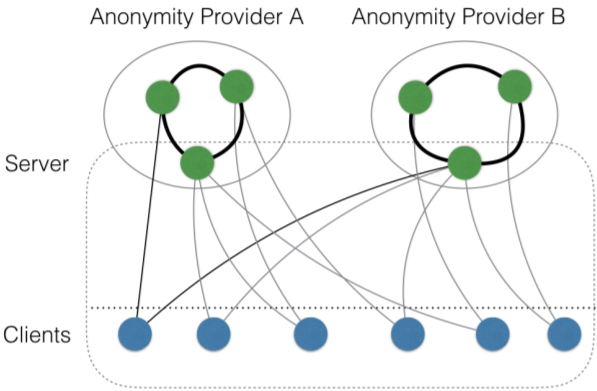}
	\caption{The Client-Server architecture of \ac{Dissent} (adopted from~\cite{wolinsky2012dissent}). \label{Dissent-Architecture}}
\end{figure}

To achieve this goal, the topology in \ac{Dissent} differs with the complete decentralized approach of the \ac{DCN}; \ac{Dissent} uses a client-server architecture and communications always happens among client-server and server-server, but never between two clients directly (Figure~\ref{Dissent-Architecture}). Hence, each client node only shares keys with the servers, but not with other clients~\cite{bauer2017dining}.

The communication protocol in \ac{Dissent} is implemented as follows; first, each client creates a secret pseudonym key \textit{k} and sends it encrypted to a server.
The servers receive and decrypt the keys, then, shuffle them in a random and secret manner to create a vector with the secret permutation of all clients’ received keys.
This vector is sent to the clients, then, each client can determine his own key within the shuffled key vector without being able to discover the position of other clients’ keys.
In this way, the secret shuffling works as a scheduling algorithm and one slot is assigned for each key included in the secret shuffled vector, which is reserved for the corresponding key owner.

Next, for each communication round, according to the assigned slot number, each client can determine if he can actively send his payload, or he has to participate passively by sending a message with zeros as payload.
Afterwards, the client creates pseudo-random strings from the secret keys he shared with every server and combines those strings with \ac{XOR}, then sends the output to at least one server.
The servers synchronize messages from each other (by ignoring messages from the same client through the pseudonym keys) and continue the protocol if the number of participating clients in the round is above a certain threshold (to not threaten the anonymity of clients sending in the current round).
Finally, each client receives the output of this round, which is signed by every server.
Then, each client can extract the round message after verifying the signatures\footnote{The signatures are generated and verified using a public key cryptosystem. Each server uses its private key to sign a message and clients verify the message by using the server's public key in the verification step. In Dissent's implementation, 1024-bit RSA-OAEP is used for this purpose~\cite{corrigan2010dissent}.}~\cite{corrigan2010dissent,bauer2017dining}.

The approaches used in \ac{Dissent}’s design has added several advantages to it. First, \ac{Dissent} offers participants the ability to send variable-sized messages by announcing the length of their intended message instead of sending equal-sized messages in original \ac{DCN}. The length-dissemination phase scales linearly in the number of group members~\cite{modinger2021arbitrary}.

The next positive point of \ac{Dissent} is that it allows servers to leave out clients that were leaving the network or clients that are too slow during each communication round. In Herbivore, the ring topology or traditional \ac{DCN}, the global result of each round relies on all participants’ committed messages and the slowest client determines the latency and bandwidth characteristics of the network. However, a fixed receiving window is used in \ac{Dissent} to tackle this bottleneck to overall performance; clients, who send their messages too late or do not send anything at all, are left out for that round. In addition, a threshold for minimum number of participating nodes, who have to participate in a given round, is considered to avoid the possibility for attackers to isolate currently sending client. If the threshold is not met, the servers will abort the round, or they increase the receiving windows~\cite{staudemeyer2019takes}.

As a final point, each node in \ac{Dissent} shares secrets with each server. If there is at least one honest server, the anonymity of a client cannot easily be compromised. In this implementation, the client does not have to know which one of the servers is classified as the honest server, which is commonly referred to as the “any-trust model”~\cite{wolinsky2012scalable}.

In brief, by using \ac{DCN} and verifiable shuffle algorithms, \ac{Dissent} provides integrity, anonymity guaranty and supports variable-length messages~\cite{kotzanikolaou2017broadcast}. In addition, it holds members accountable, by ensuring that no malicious member can abuse his strong anonymity to disrupt the groups' operation~\cite{corrigan2010dissent}. A prototype of \ac{Dissent} protocol is implemented in the C++ programming language with the Qt-Framework. In practice, normal \ac{TCP} traffic can be tunneled through the \acs{SOCKS} server provided by \ac{Dissent} and therefore normal application level protocol can transparently use the \ac{DCN} implementation~\cite{bauer2017dining}.

However, \ac{Dissent} has limitations of course, and it is not intended for large scale, `open-access' anonymous messaging or file sharing. \ac{Dissent} works for the cases in which very few clients share very small messages and suffers from poor performance at scale. Moreover, since this protocol is designed to ensure resilience to any traffic analysis, its imposed overhead grows linearly with the size of the anonymity set scale~\cite{akhoondi2015scalable,corrigan2010dissent}. Besides, \ac{Dissent}’s accountability properties assume closed groups and are ineffective if a malicious member can leave and re-join the group under a new (public) identity after expulsion. The serialized shuffle protocol also imposes a per-round start-up delay that makes \ac{Dissent} impractical for latency-sensitive applications~\cite{corrigan2010dissent}.

Later, the same group of authors proposed an improved implementation of \ac{Dissent} protocol in 2012. \ac{Dissent} in Numbers~\cite{wolinsky2012dissent} is an extension of the original \ac{Dissent}. The protocol uses a few powerful core servers as anonymity providers in a round-based multiphase protocol, to make it accessible for a large number of clients~\cite{modinger2021arbitrary}.

In addition, a verifiable \ac{DCN} was implemented in 2013 under the name Verdict~\cite{corrigan2013proactively}. The advantage of Verdict is that it allows switching between traditional \ac{DCN} and verifiable \ac{DCN}, depending on the presence of disruption and for scheduling, it uses a similar approach as \ac{Dissent}~\cite{corrigan2010dissent} and~\cite{studholme2007multiparty,wolinsky2012dissent}. Additionally, Verdict uses zero-knowledge proofs to proactively exclude misbehaving users before jamming the communication. The proactive exclusion of insider disruption attacks relieves the system from the need to trace a disruptor after the attack. In contrast, Verdict relies on public key cryptography for message encryption, which increases the computation cost in relative to traditional \ac{DCN}. Verdict is suitable for low-latency communications for small groups of users. However, similar to other approaches based on the verifiable shuffle, it does not scale well.
\subsection{Riposte}
\label{subsec:4.4.5Riposte}
A Riposte~\cite{corrigan2015riposte} deployment consists of a few servers, which collectively maintain a shared write-private database and a large number of clients that are allowed to write into it. To post a message, which should be proceeded in regular time intervals, called epochs, a client generates a write request to encode his message and the row index at which he wants to write. Then, the client splits this write request into many shares via secret sharing and sends one share to each Riposte server. A coalition of servers smaller than a pre-specified threshold cannot learn anything about the contents of the write request or write location.

During each epoch time, the servers collect write requests and apply them to their local state. When the servers agree that the epoch has ended, they publish the aggregation of the write requests they received during the epoch to reveal the plaintexts represented by the write requests.

When using a Riposte as a platform for anonymous message broadcasting, the messages are limited to be long enough as database row size; hence, there is a limit for uploading a message. In addition, at the end of each epoch, anyone can recover the set of posts uploaded, so the identity of the entire set of clients who posted during this epoch is known, but no one can link a client to a post. A particular client’s anonymity set consists of all the honest clients who submitted write requests to the servers during each epoch. Thus, each time epoch must be long enough to ensure that many honest clients are able to participate. Therefore, the definition of what constitutes an epoch is a crucial decision for the level of offered anonymity.

In comparison with \ac{Dissent} systems, which also use partially trusted distributed servers to provide anonymity guarantee, \ac{Dissent} requires a weaker trust assumption than Riposte. At least one of the \ac{Dissent} servers must be honest, and it has any-trust model, whereas, for a Riposte system at least three non-colluding servers are required, and it has a three-server protocol. On the opposite side, when \ac{Dissent} clients want to send an \textit{l}-bit message, they must send the whole message length to all servers, but Riposte clients split their write request into a number of shares, and they have to send the shares in fewer bits to servers, which leads to bandwidth-efficiency by Riposte.
\subsection{Footprint Scheduling}

In~\cite{krasnova2016footprint}, a new reservation-map method called footprint scheduling was proposed to address the ‘slot reservation’ issue. Footprint scheduling modifies the original map-reservation algorithm described in the Chaum’s \ac{DCN}~\cite{chaum1988dining}. This scheduling uses footprints in \textit{B} bits ($B > 1$ bit) instead of 1-bit per slot in the reservation vector to decrease the likelihood of undetected collisions in the transmission phase, which occurs when an odd number of players attempt to reserve the same slot. In this case, although the number of bits to represent each slot is multiplied by the B factor, the number of slots in the reservation vector can be decreased.

To reserve a slot in footprint scheduling, players should change the corresponding bits of a slot to a random value. If each slot in the final reservation vector, which results from \ac{XOR}ing of all participants’ individual vectors, contains a footprint of a participant, the slot is reserved successfully. In contrast, if participants cannot find their original footprints in the round result, then the corresponding participants detect the collision and due to multiple scheduling rounds in a cycle, players can try again to choose another slot. Additionally, skipping non-reserved slots at the end of the scheduling and the possibility to reserve multiple slots which can hide the number of actively sending users, are other improvements of the footprint scheduling method.

A publicly available simulation~\cite{DCSimulator2015} was used to define optimum values for the footprint scheduling parameters (the number of bits per slot, number of slots and number of scheduling rounds per scheduling cycle) based on the scheduling overhead imposed on each participant for slot reservation. The simulation results for scenarios with three different activity rates show that footprint scheduling yields excellent results, particularly in very dynamic networks with a frequently changing set of participants and frequently changing activity rate. Therefore, using the footprint approach will reduce the probability of undetected collisions in the reservation vector; participants can negotiate for communication slots without losing anonymity, while at the same time, the number of actively sending users will be hidden.
\subsection{Dining Cryptographers Group}

Dining Cryptographers groups refer to the protocols where only a part of the whole network participates in the execution of \ac{DC} protocol. This idea is being used in different state-of-the-art protocols such as \ac{Dissent} variations~\cite{corrigan2010dissent,wolinsky2012dissent,corrigan2013proactively} and k-anonymous groups~\cite{von2003k}. Performing \ac{DC} within the groups with a smaller number of participants provides efficiency in communication with strong privacy guarantee. However, the non-cooperating participants in a protocol based on \ac{DC} might force the true originator to step up which leads to jeopardising its anonymity. This problem in protocols based on \ac{DC} groups such as~\cite{modinger2018flexible} creates additional risks. The common approach to address this issue is to punish and exclude the misbehaving nodes from the group.
A better alternative is to incentivize nodes to participate as in the Shared-Dining approach~\cite{modinger2021shared}, designed in a way that messages can only be read when enough participants cooperate to cross a threshold.

Shared-Dining introduces a combination of Shamir’s secret sharing and classical \ac{DCN} to provide a manageable performance impact on dissemination while enforcing the anonymity guarantees of the protocol throughout the network. In this approach, the broadcasting of a message to all participants in the original \ac{DC} is replaced by the transmission of \textit{n} distinct parts. Therefore, in the first phase called Split, the message parts are created by splitting the original message into \textit{n} shares using a $(n,k)$ Shamir’s secret sharing technique. Then, in Distribute phase as the second phase of the protocol, each part is transmitted simultaneously during a modified \ac{DC} round, resulting in each participant ending up with a single share of the message. In the third phase, Broadcast and Combine, all group members broadcast their received share through the network, allowing any recipient of \textit{k} shares to reconstruct the message, enforcing anonymity.

If at least $k$ participants broadcast their message, every recipient can decode the original message. When exactly $k-1$ participants broadcast, only non-broadcasting participants of the group can decode the message, as they possess the last share required to decrypt the message. If $k-2$ or fewer group members broadcast their shares, no one can decode the message, thus preventing privacy breaches for the originator.

The shared-Dining approach is designed to address privacy requirements for financial information. In this regard, a proof-of-concept prototype of Shared-Dining is implemented in Java~\cite{SharedDininggithub2020} and its performance in terms of latency and throughput rates are investigated according to different system parameters. The anonymous transmission of transaction data for blockchains in peer-to-peer networks is selected as an evaluation scenario and the results show throughput rates between 10 and 100 kB/s~\cite{modinger2021shared}.
\subsection{PriFi}
PriFi~\cite{barman2020prifi} is another \ac{DCN} based anonymous communication protocol which assures connected users on the \ac{LAN}/\ac{WLAN} to be indistinguishable from other users. This protocol is similar to a low-latency proxy service (e.g., a \ac{VPN} or SOCKS tunnel) working within a \ac{LAN}, creating tunnels between clients and the PriFi relay (e.g., the \ac{LAN}’s router) and these tunnels protect honest client’s traffic from eavesdropping attacks.

The main technical contribution of PriFi is a low-latency 3-layer architecture, which removes an important latency bottleneck seen in Mix networks and eliminates the need for costly client-server communication, while adding to the security of all PriFi clients. The clients only `stream' ciphertexts to the relay and this design dramatically reduces the latency experienced by the clients. The PriFi system goals are anonymity, traffic-analysis resistance, low-latency, accountability and scalability. Additionally, server does not need to be trusted, i.e., security properties hold in case of compromise. 

The PriFi communication protocol can be deployed to existing infrastructures with minimal changes. Consider a set of \textit{n} \emph{clients}, $\{C_1, \ldots, C_{n}\}$, which are connected within an organizational network through a \emph{relay}, \textit{R}, which acts as a gateway and connects the \ac{LAN} to the Internet. The relay can process regular network traffic in addition to running the anonymity protocol (PriFi software). Furthermore, on the Internet, there is a small set of \textit{m} servers called \emph{guards}, $\{S_1, \ldots, S_{m}\}$, whose role is to assist the relay in the anonymization process. These guards could be maintained by independent third parties or sold as a `privacy service' by companies, and it is preferable to be distributed around the world, across different jurisdictions to maximize diversity and collective trustworthiness.

The PriFi protocol is jointly executed by clients, guards and the relay and proceeds in time slots to allow an \textit{l}-bit anonymous message transmission within each slot according to a defined schedule. The protocol starts with a \emph{Setup} phase to establish a schedule and share secrets for a predetermined timespan (e.g. 10 minutes) which is called epoch. The configuration (i.e. shared secrets and schedule) does not change during an epoch and when its time expires or due to network churn, a new epoch will be created by re-executing of the \emph{Setup} phase. During \emph{Setup} phase, each client ($C_i$) authenticates itself to the relay using its individual long-term public keys, generates a fresh ephemeral key-pairs, then runs authenticated Diffie-Hellman key exchange protocol with each guard ($S_j$) to agree on a shared secret ($r_{ij}$) which is used later to compute the \ac{DCN}’s ciphertexts.

The clients need to know how they should participate in the protocol.
For this purpose, the relay prepares a vector of n client’s ephemeral public keys and sends it sequentially to all guards in order to perform a verifiable shuffle on the vector.
Finally, after the finishing of last server shuffling, relay broadcasts the resulted random permutation of the vector to all clients.
At the end of scheduling, each client uses its own ephemeral private key to recognise the corresponding pseudonym key in the vector to find its allocated slot.

After \emph{Setup}, all nodes continuously run the \emph{Anonymize} phase. At each time slot, all the clients and guards participate in a \ac{DCN} protocol. Each guard seeds a \ac{PRG} function with all of its shared secrets with clients and \ac{XOR} all these \textit{N} values to compute one \textit{l}-bit pseudo-random message and sends it to the relay. On the other side, all clients, except the slot owner, perform likewise and generate pseudo-random numbers by using the same \ac{PRG} functions for all shared secrets with guards and \ac{XOR} all of them to compute one \textit{l}-bit pseudo-random message to send to the relay. The client owning the time-slot ($C_i$) additionally includes its upstream message(s) - $m_i$ - in the computation. The upstream message is one or more \ac{IP} packet(s) without source address, up to a total length \textit{l}. If the slot owner has nothing to transmit, it sets $m_i$  to $0$ in \textit{l}-bit.

Once the relay receives the ciphertexts from all clients and guards ($n + m$ messages), it \ac{XOR}s them together to obtain $m_k$. The values of each \ac{PRG}($r_{ij}$), $ i \epsilon  \{ 1 \ldots n \}, j \epsilon \{ 1 \ldots m \} $, appears twice in the computation and cancel out, hence $m_k = m_i$, if the protocol is executed correctly. Finally, if $m_k$ is a full \ac{IP} packet, the relay replaces the null source \ac{IP} in the header by its own (just like in \ac{NAT}) and forwards it to its destination. If it is a partial packet, the relay buffers it and completes it during the next schedule. Besides, by receiving an answer to an anonymous message sent in some time-slot, the relay encrypts it under the (anonymous) slot-owner’s public key, then broadcasts the ciphertext to all clients. In addition, client churn can be handled in background either as a graceful churn or as abrupt disconnection.

In the threat model of PriFi, all nodes including clients, guards and relay are honest if they follow the protocol faithfully and does not collude or leak sensitive information to any other node. The relay might be malicious and actively tries to de-anonymize honest users but considered trusted for availability, means it will not perform actions that affect the availability of PriFi communications such as delaying, corrupting or deleting clients’ messages. In addition, clients can be malicious (controlled by an adversary), but at least two honest clients at all times are required; otherwise, the anonymization will be meaningless. The guards are all highly available and follow the any-trust model (clients do not need to know which one). Thus, the PriFi protects an honest user's traffic among all honest user's traffic and suggested further ideas to hide global/aggregate communication volumes or time series of packets. However, the proposed solutions are not perfect against intersection attacks and could just make them harder.   

In terms of security and practical considerations, the PriFi protocol provides techniques for protection against disruption attacks by malicious clients and equivocation attacks by relays that try to de-anonymize clients. An open-source prototype of PriFi was evaluated by Barman and Wolinsky with realistic datasets.
The experiments results are publicly available for further investigation~\cite{Barman2020PrifiGithub,barman2020prifiexperiments}.
\subsection{Verifiable \acp{DCN} by Adding Commitments and Zero-Knowledge Proofs}
For a long time, \acp{DCN} were considered unpractical for real-world applications because of the tremendous communication and computation overhead they introduce, in particular for handling malicious participants who disrupt protocol~\cite{Franck2021FastECC}. The advances in cryptographic techniques, such as commitment schemes and \acfp{ZKP}, provide a great opportunity to reduce the communicational cost of modern \acp{DCN} and a possibility to detect misbehaving participants more efficiently than before. This led researchers to re-assess the \ac{DCN}-based solutions and consider a more fundamental role for \ac{DCN} in future communication~\cite{Franck2014practical}.

In 2015, Franck extended the \ac{DC} scheme with the Pedersen commitments to provide Zero-Knowledge verification and unconditional anonymity at the same time~\cite{Franck2014practical}. The Pedersen commitments computationally bind participants to their secret keys, and then they could be used to construct \acp{ZKP} about the retransmission of data. A \ac{ZKP} allows a prover to convince a verifier that he knows a witness, which verifies a given statement without revealing the witness or giving the verifier any other information~\cite{Franck2014practical}. This verifiable \ac{DC} scheme does not require any kind of reservation phase prior to the message transmission; hence, it shows a significant improvement over the reservation based techniques~\cite{Franck2014verifiable}.

Later in 2021, Franck and his colleagues introduced a library specifically designed to efficiently implement the cryptographic primitives they proposed. The X64ECC is a self-contained library for \ac{ECC} developed from scratch to fulfil all public-key operations needed by modern \acp{DCN}: key exchange, digital signatures, Pedersen commitments, and \ac{ZKP}~\cite{Franck2021FastECC}.

The use of \ac{ECC} in the library implementation allows keeping the cryptography as compact and as efficient as possible. Furthermore, X64ECC supports three different levels of security, which can be chosen independently for each of the four high-level functionalities. This makes X64ECC easy to use for the implementation of \acp{DCN} with arbitrary message sizes, and trade-offs between the cryptographic strength and throughput are possible. Additionally, the arithmetic functions of the X64ECC are parameterized to provide a high level of flexibility and scalability. Also, compiler intrinsics are used to speed up performance-critical arithmetic operations~\cite{Franck2021FastECC}.
\subsection{Arbitrary Length k-Anonymous DC-based Communication}
In the most recent research effort in 2021, Mödinger \emph{et al.} highlighted the lack of privacy for blockchain systems in transactions’ dissemination within peer-to-peer networks connecting all participants and took the advantage of \ac{DCN} based protocols to address this need~\cite{modinger2021arbitrary}.
In their work, a number of developed network-layer protocols are reviewed, and a design for a privacy-preserving protocol is proposed with strong privacy guarantees. The main idea of this design is derived from two previous works on \acp{DCN}:
\begin{enumerate}
	\item The k-anonymous message-transmission protocol (which is earlier mentioned as a broadcast based anonymous method in Section~\ref{sec:GENERAL OVERVIEW OF ACS}) by von Ahn \emph{et al.}~\cite{von2003k}. This sender- and receiver-k-anonymous protocol is realised to mitigate the scalability weakness of \acp{DCN}. For this purpose, participants are assigned to several disjoint groups with only \textit{k} members and a message is first shared anonymously within a local group providing k-anonymity. Then, all participants of this group send the message to all members of the target group. The participants also create commitments on all messages and broadcast them to the group to provide a blame protocol to identify malicious actors in case of misbehaviour detection. In addition, participants are allowed to reserve more than one slot per round. As a result, the k-anonymous protocol provides fairness and robustness at a higher overhead than basic \ac{DCN} in malicious environments.
	\item 3P3 design~\cite{modinger20203p3}. In 2020, the same group of authors, Mödinger \emph{et al.}, addressed the major limitation of predetermined message size and extended the k-anonymous message transmission protocol by proposing a design to support arbitrary-length message transmission. The 3P3 design consists of two consecutive phases, which are each built on a \ac{DCN} protocol. In the initial phase, the participants anonymously announce the size of their messages while trying to reserve a slot. For this purpose, each participant shares a vector, which includes the length of his message in his randomly selected slot and sets the remaining slots to zeros. The protocol then merges the vectors of all participants using secure multi-party computation. During the next phase, participants will follow the \ac{DCN} protocol in successfully reserved slots to disseminate the actual messages. To do this, every participant prepares a message with a length equal to the sum of the sizes announced for this slot during the initial round. If the slot number and the aggregated announced length of message for this slot are equal with the participant’s request in the initial phase, hence, this slot is properly reserved for him. Therefore, he should include his actual message; otherwise, he sends a zero message in the specified length. This protocol assumed single group topology, so the final inter-group transmissions of von Ahn \emph{et al.} protocol are not required.
\end{enumerate}    
The 3P3 design is extended and its realisation within a real-world use-case is provided in~\cite{modinger2021arbitrary}. In addition, since this protocol has massive overhead to secure its operations in a malicious environment, an unsecured variant (without creating commitments) is proposed to severely reduce the overhead in environments where maliciousness is the exception, i.e., in normal operations. This construction results in a more complex protocol state, it starts with the unsecured variant and will switch to the secure variation once a likely attack is detected. Thus, the performance of the extended 3P3 is optimised by reducing the overhead for the most common cases. A prototype implementation of 3P3 in C++ is used in the simulation to evaluate the expected performance of the extended protocol in blockchain applications and results show the protocol provides enough throughput for most applications by the secured and unsecured variants~\cite{Hess20203p3}. The fully secure version is applicable for highly security-relevant applications such as blockchain transactions, while, the version only using the secure version as a fallback mechanism can easily be used for many less critical text-based applications~\cite{modinger2021arbitrary}.
\subsection{Discussion and Open Issues of Dining Cryptographers Networks}
\label{subsec:DCN-5-discussion}
\begin{figure}
	\includegraphics[width=\textwidth, height=\textheight, keepaspectratio]{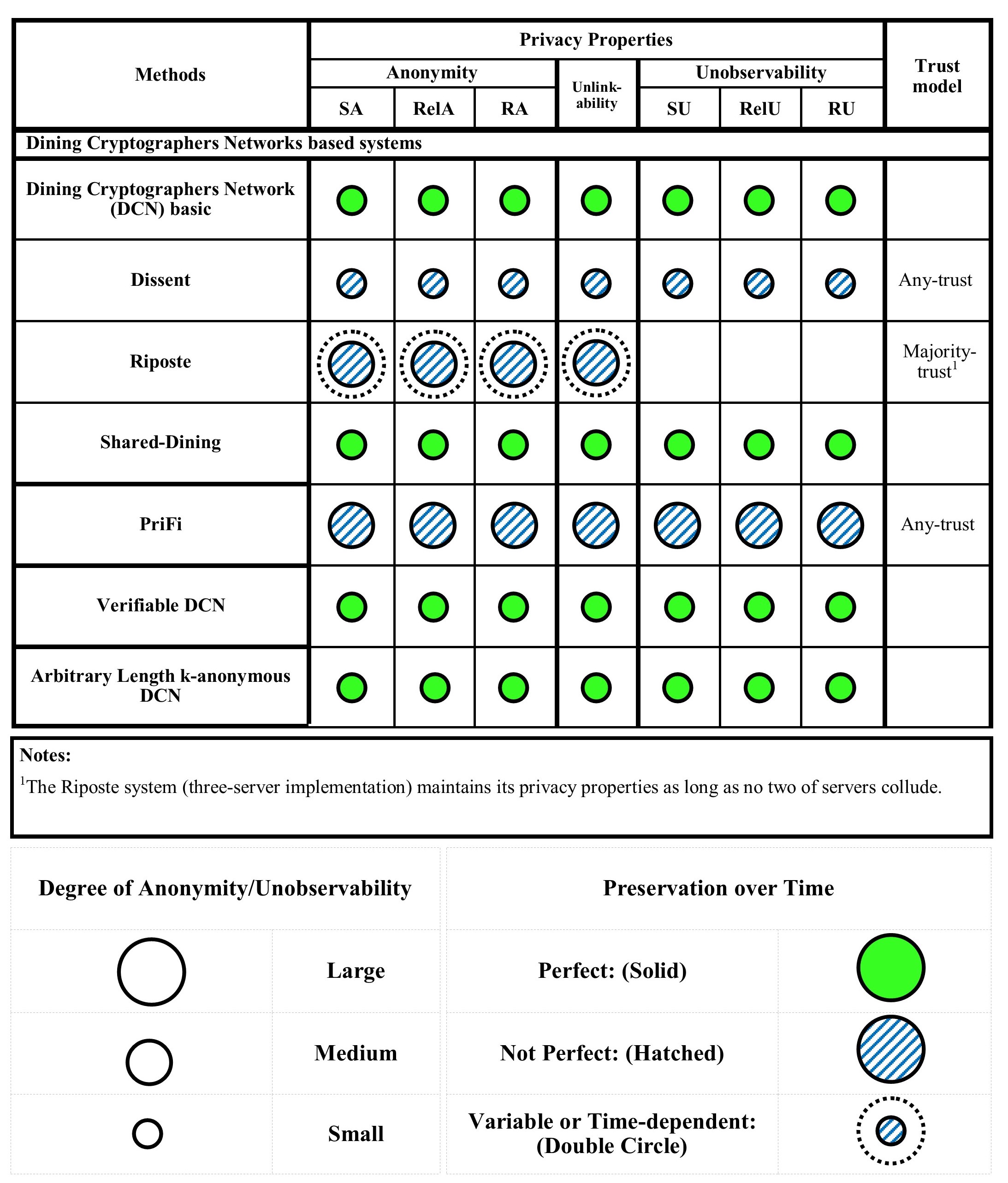}
	\caption{Comparison of main \acs{DCN}-based \acp{ACS} from the privacy properties perspective.\label{fig:privacy-matrix-dcn}}
\end{figure}

The anonymity offered by original \acp{DCN} is information-theoretically secure, which makes \acp{DCN} unique among other solutions. The \acp{DCN}’ security cannot be broken given unlimited time and computation power~\cite{chaum1988dining}. Hence, they are the perfect choice to provide unconditional privacy under any circumstances in current communications through the Internet. 
As discussed in this section, many efforts have been made up until now to make \acp{DCN} more practical by degrading their unconditional privacy to provide more applicable methods with reduced computation and communication overhead.
Figure~\ref{fig:privacy-matrix-dcn} illustrates a privacy map to compare main DCN-based ACSs regarding their offered privacy properties.

The privacy map of \ac{DCN}-based \acp{ACS} clearly shows that these methods have no problem in terms of privacy and are able to provide unobservability.
However, the performance of these methods drastically drops when they are scaled up to be used by a large number of participants.
Indeed, their imposed computation and communication overhead in association with latency have made them quite inefficient and infeasible in practice.
For instance, the basic \ac{DCN}~\cite{chaum1988dining} is almost impractical for groups with more than 10 participants (due to its overhead for disruption protection).
The Shared-Dining~\cite{modinger2021shared} is designed to implement \ac{DCN} in groups, and the results of other methods, including Arbitrary length k-anonymous~\cite{modinger2021arbitrary} and verifiable \ac{DCN}~\cite{Franck2014verifiable,Franck2021FastECC}, also show that they have too much overhead when used by large number of participants and are not applicable.
This problem is reflected by small-sized circles in the privacy map. 

Considering the performance and overhead issues that hinder the implementation of scalable and practical \acp{DCN}, the challenges of \acp{DCN} are addressed in several ways. For instance, there have been enhancements to the original \ac{DCN} to support the transmission of variable-length messages. Moreover, due to the major impact of scheduling algorithms on the fairness, bandwidth utilisation and the perceived latency of the clients, several methods have been proposed to reduce the probability of collision and the amount of computation in the setup phases. However, by scheduling in advance, participants have one or more slots per round, which force them to buffer messages. Still, all active nodes in the network must participate in every round, and they should transmit a message. Even in some protocols, messages must be encrypted before the transmission, which makes things worse. Nevertheless, if a node wants to be inactive for a while in order to avoid this overhead, it has to leave and re-join the network. This change in the network members consequently imposes additional overhead on the remaining participants to establish fresh ephemeral key pairs with others and perform setup and scheduling phases again. Also, the network will be not available for message transmission during the required actions to handle this churn. Therefore, \ac{DCN}-based methods need more investigation and contribution in this regard.

The next decisive factor is the network topology and architecture. A two-tier architecture, clients and servers, provides higher scalability and reduces the number of required shared keys since each client no longer needs to exchange secrets with all other client. In contrast, the need of relaying messages by servers during message transmission will have a significant negative impact on latency. It requires several client-to-server and server-to-server messages to ensure integrity, accountability and the ability of handling churn.

Furthermore, even if the high overhead is tolerated in a particular scenario, the jamming and disruption attacks should not be overlooked, due to their impact on the overall network performance. Nevertheless, the methods proposed to solve the disruption problem are time-consuming and always computationally intensive. The advances in cryptographic primitives such as commitments and \acp{ZKP} help detect misbehaving participants efficiently with fewer computations; however, the feasibility of using them in real scenarios needs more investigation. Therefore, it might be preferable to use a \ac{DCN}-based method, which is able to switch between two different modes.
One to be used in environments where disruption is an exception and requires reduced overhead, and the other is a secure or resilient mode, which employs commitments and robust proof of correctness to discover disruptors for a stronger privacy guarantee.

Additionally, regarding the privacy-support classification of the \acp{ACS}, as described in section~\ref{sec:acs_privacy_map}, the privacy map provided in Figure~\ref{fig:privacy-matrix-dcn} foregrounds that \ac{DCN}-based systems are placed in the unobservability-support class. In addition, the investigation of \ac{DCN} variations mostly highlights that except for recent implementations such as PriFi~\cite{barman2020prifi}, Verifiable \ac{DCN}~\cite{Franck2014verifiable,Franck2021FastECC} and Arbitrary length k-anonymous~\cite{modinger2021arbitrary}, the legacy methods are inappropriate for latency-sensitive applications. 

In the end, despite the definite necessity of privacy preservation for users on the Internet; however, the applicability of \ac{DCN}-based methods in constrained environments has rarely been explored.
The recent innovative solutions and adaptations applied to modern \ac{DCN} variations have broken down the barriers in its implementation.
Some of the most promising innovations are energy-efficient cryptographic primitives, highly scalable architecture designs with reduced latency, and the flexibility of supporting multiple modes of a protocol to provide different levels of privacy and resiliency.
Therefore, if a novel variant of \ac{DCN}-based \ac{ACS} utilises a combination of these recent innovations, it might eventually be able to facilitate the integration of \acp{DCN} in daily real-world scenarios, including the equipment within resource-constrained environments such as \ac{IoT} devices.
In conclusion, \ac{DCN}-based methods with efficient overhead are worth to be considered; hence, further evaluation of \ac{DCN}-based methods in \ac{IoT} testbeds and simulations using realistic datasets is recommended.
To facilitate this evaluation, the detailed analysis of de-anonymising techniques could also help system developers realise the security threats and detect possible vulnerabilities of anonymous communications in a better manner by identifying more attack models.
\section{Conclusion}
\label{sec:conclusion}

Although data encryption algorithms protect the content of communications from unauthorised access, traffic analysis methods can be used to extract valuable information about communicating parties.
\acfp{ACS} are able to additionally provide strong privacy properties, like anonymity and ideally unobservability. 
Hence, the growing trend of privacy-sensitive data collection, for instance, via \ac{IoT} devices, and the increasing usage of wireless communications in data transmission require us to rethink to applying the privacy protections offered by \acp{ACS}.

These systems are proposed to prevent traffic-analysis attacks by making the traffic of the network’s users indiscernible, and this article was devoted to a comprehensive survey of \acp{ACS} with a focus on \ac{DCN}-based methods, including the latest contributions.
In addition, this article provides a common ground privacy terminology to define privacy properties.
Then, we investigate the \acp{ACS} from the privacy perspective and visualised their alignment with the terminology.
It is highlighted that modern \acp{ACS}, in particular, \ac{DCN}-based methods offer better privacy protections. 
However, their intrinsically introduce significant computation and communications overhead and are less scalable, which often makes them initially appear impractical for real-world scenarios.

Reduction of overhead becomes the next impetus to design and develop practical, low-latency and robust \acp{ACS}. 
In this direction, \ac{DCN}-based methods with guaranteed unconditional unobservability and provable traffic-analysis resistance seem to be the most appropriate option to choose because they can offer the highest guarantees.
The analysed enhancements, such as variable-length message transmission support, commitment techniques or using two-tier architectures in the more recent variants of \ac{DCN}-based methods have reduced latency, computation and communication overhead and offer higher scalability compared to the original \ac{DCN} protocol. 
Still, the \ac{DCN}-based methods are limitedly realised in real-world use cases such as the constrained environments or for blockchain’s transaction dissemination privacy, which has revealed the necessity for further improvements. 
However, the latest advancements highlight again the strength of \ac{DCN}-based methods (and also other types of \acp{ACS}) for privacy preservation but make these methods more operational in practice.

Thus, we expect a reinforcement in the research on those types of \acp{ACS}s that bring the most suitable privacy guarantees to the application and hope that they will finally become woven into future communication networks to intrinsically protect our privacy. 

\section*{Acknowledgments}

The authors would like to express their gratitude to Professor Jorge Cuellar for his constructive feedback, particularly on the literature review section.
The collaboration on this article was established during the research stay of the corresponding author at the chair of IT-Security at the University of Passau.
Hence, the corresponding author would also like to acknowledge Professor Joachim Posegga and Dr. Mona Ghassemian for their support in making his research visit happen and for their valuable comments.
The research stay was (partially) funded by DAAD under grant no. 57440917.
\appendix
\label{appendices}

\section{Abbreviations}
\label{Appendix-A-Terminology}

\printacronyms[template=supertabular,heading=none,preamble={A complete list of appeared abbreviated terms in this article and their correspondig definitions are provided below.\\}]

\section{Background Knowledge}
\label{sec:BACKGROUND KNOWLEDGE}

To have a common understanding of the properties used to describe anonymous communications, we must introduce the commonly used terms from the literature in a harmonized way to avoid any confusion and to establish a common ground for later usage.
We start by presenting networking terms in Section~\ref{sec:background:network-terms}.
Then, in Section~\ref{sec:background:cia}, the well-known security properties are presented to describe the foundational requirements of a secure information system.
Afterwards, in Section~\ref{sec:background:adversarial-model}, we lay the foundation to characterise an adversary, against whom an \ac{ACS} seeks to defend.
Finally, in Section~\ref{sec:background:attacks}, a brief overview of traffic analysis and de-anonymization attacks is given.
\subsection{Network Terminology}
\label{sec:background:network-terms}

The communications that should be hidden take part between multiple participants which form a network.
There are different terms that are used to describe the participants of a network and the actions occurring in it.
In order to make clear the networking terminology, which is used throughout the article, these terms are explained here briefly.

\textbf{Sender.} A networking device acts as a sender whenever it sends a signal to the network (e.g. in order to communicate with another device)~\cite[pp.~35-36]{Schiller2003}.

\textbf{Recipient.} In difference from being a sender, a device is a recipient whenever it receives a signal from the network~\cite[pp.~35-36]{Schiller2003}.

\textbf{Packet.} Data transmission over a network happens using so-called `datagrams'.
A datagram may be too large to be transmitted over a network. Then, it has to be fragmented into multiple smaller packets~\cite[p.~8]{ISI1981}.

\textbf{Node.} A network node is an element which takes part in communication.
A node can either be an endpoint (a sender or recipient) or a point which receives a packet and redistributes it to another one~\cite[p.~473]{Schiller2003}.

\textbf{Metadata.} Metadata is, as the name indicates, data about data.
This term is widely used for describing data that is generated when data exchanges occur.
If packets are transmitted between network nodes, there is also data exchanged besides the payload, e.g. the \acp{MACAddress} of the involved nodes.
This data is then called metadata~\cite[pp.~1-2]{Gartner2016}.

\textbf{Internet of Things (IoT).} The term Internet of Things is now more and more broadly used; however, it is hard to find a common definition or understanding of what \ac{IoT} actually encompasses~\cite{wortmann2015internet} and manifold definitions are presented within the research community.
For instance, the \acf{ITU} defines the Internet of Things as `a global infrastructure for the Information Society, enabling advanced services by interconnecting (physical and virtual) things based on, existing and evolving, interoperable information and communication technologies'~\cite[p.~1]{UniInternationalTelecommunication2012}.

The basic idea of the \ac{IoT} concept is the ubiquitous presence of a variety of things or objects around us, which are able to interact with each other and cooperate with their neighbours to reach common goals~\cite{ATZORI2010IoT}.
In this way, an \ac{IoT} system logically can be depicted as a collection of smart devices that interact on a collaborative basis to fulfil a common goal~\cite{sicari2015security}. 
\subsection{Security Properties}
\label{sec:background:cia}

The main goals of information security are confidentiality, integrity and availability~\cite[pp.~2-4]{Bishop2004}. 
Regarding communication, we define them as properties following the \ac{ISO}\footnote{Note, \ac{ISO} maintains a free terminology database at \url{www.iso.org/obp}.} as follows:

\textbf{Confidentiality}. The `property that information is not made available or disclosed to unauthorised individuals, entities or processes'~\cite[def. 3.36.4]{isots21089}.
Thus, confidentiality protected information is only accessible to the people or systems which were allowed to access it.
Confidentiality can be ensured by using access control mechanisms, e.g. encryption and giving a decryption key only to authorised people or systems.

\textbf{Integrity.} Defined as the property that allows to proof that the message content has not been altered, deliberately or accidentally in unauthorised manner during transmission~\cite[def. 3.46 \& 3.79]{isots21089}. 
This allows maintaining the correctness and trustworthiness of data, as unauthorised third parties are not able to change the data without detection.

\textbf{Availability.} The `property of being accessible and useable upon demand by an authorised entity'~\cite[def. 3.27.1]{isots21089}, individual or process. 
It means a legitimate user or system can always access a system or a resource without suffering restrictions.

In addition to maintaining integrity, we often need to explicitly make sure that received information is from a certain origin, e.g. in communication, the receiver would like to ensure that a message originated from a certain sender.
By intuition the authentication of the origin is thought to be included in integrity, but not clearly by definition~\cite{Poehls_Contingency_2013}, thus we explicitly define it for later use in this contribution:

\textbf{Authenticity of data.} A positive output of the `process of corroboration that the source of data received is as claimed'~\cite[def. 3.23.3]{isots21089} indicates authenticity. 
Thus making sure that when we require authentication of messages the origin of the data or message can be verified.
\subsection{Adversary Model}
\label{sec:background:adversarial-model}
Due to the wide range of attacks applicable on \acp{ACS}, modelling realistic capabilities of the adversary can provide better insight to assess the resiliency of various \acp{ACS} against attacks.
Adversaries are mostly defined according to their goals, i.e. breaking a privacy or security property, and their strengths; the following properties as suggested by Raymonds~\cite{raymond2001traffic} can be used in combination to describe the strength of an adversary~\cite{edman2009anonymity}:

\textbf{Capability.} Defines the ability of an adversary to monitor or manipulate the network traffic.
A \textit{passive adversary} is able to monitor and record the traffic on the network links.
This also applies to metadata about network flows.
Whereas, an \textit{active adversary} has all capabilities of a passive attacker and is also able to manipulate network traffic.
This can happen by either controlling one or more network links or by operating a node in the network.

\textbf{Visibility.} Determines how much of the network is passively monitored or actively manipulated by the adversary.
A \textit{global adversary} is able to access and observe all communication lines within a network~\cite{staudemeyer2019takes}.
In contrast, a \textit{partial} (or local) \textit{adversary} is only able to monitor a subset of links or nodes of the network.

\textbf{Mobility.} Categorises the adversary depending on the ability to select a specific subset of a network.
An \textit{adaptive adversary} can choose subsets of the network to monitor, while a \textit{static adversary} is not able to change the observable subset of the network at will.

\textbf{Participation.} Characterises adversaries based on their engagement in the network protocol.
An \textit{internal adversary} is one who participates in the anonymity network’s protocol as a node; the adversary can be a client (node at endpoint) or perhaps operate a piece of the network’s infrastructure by running a server in the network.
An \textit{external adversary}, on the other hand, does not participate in the anonymity network nor in its protocol.
Instead, the adversary compromises the communication medium used by nodes (i.e., their network links) in order to monitor or manipulate their network traffic.

It is a prudent practice in information security to try to defend against a worst-case scenario.
The adversary model most often assumed in the literature on \acp{ACS} is a passive global adversary (who can access and observe the whole network)~\cite{edman2009anonymity}.
This could be made worse by assuming also an internal and adaptive, as well as an active adversary. 
\subsection{Traffic Analysis and De-anonymization Attacks}
\label{sec:background:attacks}

Traffic analysis attacks are inherently not detectable as they occur~\cite{staudemeyer2019takes}.
These attacks ignore the content of messages and instead try to obtain as much information as possible from only network traffic metadata, such as packet arrival time and message lengths.
The attackers passively collect metadata about the messages or traffic flows and try to correlate senders and recipients~\cite{edman2009anonymity}.
In the same way that attackers can use traffic analysis techniques to compromise the anonymity of users in a system, system designers also can use these techniques to find the vulnerabilities within their designs~\cite{staudemeyer2005attacker}.
Attacks based on traffic analysis have been the subject of several studies~\cite{hashemi2018fingerprinting, fernandez2012survey}.
Hence, a brief introduction to the most significant traffic analysis attacks -- that have been applied to \acp{ACS} -- are presented in the following. 

\textbf{Website fingerprinting.} Encrypted communication still reveals who is communicating with whom and how much data is transmitted through the network~\cite{hintz2002fingerprinting}.
To start a fingerprinting attack, an attacker connects to websites and records the generated metadata and traffic.
Then, the attacker analyses this metadata (packets’ length and quantity) and uses a supervised classifier to build a fingerprint of what the website’s response looks like when it is fetched via an encrypted connection~\cite{fernandez2012survey,edman2009anonymity}.
Afterwards, the attacker can compare the observed traffic pattern of target users against the stored fingerprint database to figure out which websites the observed users are visiting~\cite{juarez2014critical, erdin2012anonymous}.
Active, passive, and semi-passive are three classes of fingerprinting techniques~\cite{kohno2005remote}.
To mitigate this attack vector, most \acp{ACS} split messages into equal-length packets.

\textbf{Timing attacks.} A passive global adversary, who is able to observe connections entering and exiting the anonymity network, observes the duration of communication between nodes.
Then, the attacker is maybe able to correlate incoming and outgoing packets through timing analysis~\cite{staudemeyer2019takes,edman2009anonymity,diaz2010impact}.
The adversary uses the patterns of packet inter-arrival times to link the network participation's patterns~\cite{clarke2001freenet}.

\textbf{Disclosure attacks.} When users engage in repeated or persistent communications, their frequent communication partners may eventually be uncovered just by observing the edges of the network and correlating the activity at the two ends of the communication~\cite{danezis2008survey}.
These attacks exploit the fact that a sender’s correspondents will appear more frequently when that sender is active~\cite{diaz2022mix}.
Therefore, the adversary observes multiple sets of recipients for consecutive message transmissions by that user.
Each of these sets contains precisely one communication partner of the sender.
In a long run, the adversary refines these sets by intersecting observed new recipient sets with previous sets and continues to do this until he reduces all sets to only a single element, thus de-anonymizing each correspondent of the sender~\cite{danezis2008survey,fernandez2012survey,edman2009anonymity}.
The disclosure attacks were first presented in~\cite{berthold2001disadvantages} and have also been referred to as intersection attacks in the literature~\cite{danezis2004statistical}.
Rather than mounting exact disclosure attacks that precisely identify users’ communication partners, an improved statistical variant of these attacks is also presented in~\cite{danezis2003statistical, danezis2004statistical, mathewson2004practical} to effectively reveal probable communication partners.
These kinds of attacks, in fact, explore fundamental limitations for any systems that select trusted parties, at random, from a larger set of potentially corrupt nodes to provide security~\cite{danezis2008survey}.
No efficient method for absolutely preventing intersection attacks has been found so far.
However, inserting dummy traffic is a measure often proposed to reduce the effectiveness of such attacks~\cite{edman2009anonymity}.

The traffic analysis attacks applied in Tor networks~\cite{juarez2014critical} - the most popular overlay network to provide anonymous communication by redirecting traffic with the largest anonymity set-, cellular LTE communications~\cite{rupprecht2019breaking} and several feasibility studies in various environments, such as smart home use-cases by utilising network traffic rates~\cite{hashemi2018fingerprinting, apthorpe2017smart} and~\cite{staudemeyer2018road}, are examples in this regard.

\section{Detailed Overview of \acp{ACS}}
\label{AppendixC:Detailed_Overview}

A summary of the major \acp{ACS} methods from the overall performance aspects is presented in Table~\ref{tab:acs-overall-comparision}.
In this table, the main affecting factors in the overall performance of a method including computational and communication overheads, latency, disadvantages, main features and their possible attacks are listed. 
The computations cost, transmission overhead and latency of various \acp{ACS} are evaluated according to the efforts required for different phases of each method such as setup, transmission and recovery from the failures or disruptions. Moreover, the likelihood of conducting successful traffic analysis attacks for each method is also represented in table and \faCheckCircle~symbols in different sizes have been used for this purpose.

\renewcommand{\arraystretch}{1}
\setlist[itemize]{leftmargin=*,label=$\bullet$,noitemsep,partopsep=0pt,topsep=0pt,parsep=0pt,nosep}
\tablecaption{Detailed overview of \acp{ACS} from the performance aspect.\label{tab:acs-overall-comparision}}
\tiny
\tablehead{\hline
	\hline
	\textbf{ACS}			&	\textbf{Computation Cost}	&	\textbf{Transmission overhead}	&	\textbf{Latency}	&	\textbf{Disadvantages}	&	\textbf{Main Features}	&	\textbf{Desired Applications}	&	\multicolumn{2}{l}{\textbf{Traffic Analysis Attacks}}\\
	\hline
	\hline}
\begin{landscape}
\begin{supertabular}{p{1.8cm}p{1.8cm}p{1.8cm}p{1.8cm}p{4.2cm}p{4.8cm}p{2cm}p{0.1cm}p{1.8cm}}
	\shrinkheight{2cm}
VPN, Proxy	&	Low	&	No Overhead	&	Low	&	\begin{itemize}
	\item Single point of failure
	\item Fully trusted party (server)
\end{itemize}	&	Simple	&	Supports most network protocols	&	\raisebox{-\height}{\normalsize \faCheckCircle[regular]}	&	\begin{itemize}
	\item Timing
	\item Fingerprinting
\end{itemize}\\
\hline
\multicolumn{9}{c}{\textbf{Mix-Networks Systems}}\\
\hline
Basic Mix networks~\cite{chaum1981untraceable}	&	High	&	Low	&	High	&	\begin{itemize}
	\item Anonymity and overheads are topology-dependent
	\item Poor performance
	\item Undesirable for real-time applications
\end{itemize}	&	Re-ordering and transferring messages cryptographically	&	Delay-tolerant-Applications	&	\raisebox{-0.5\height}{\normalsize \faCheckCircle[regular]}	&	Timing\\
Loopix~\cite{piotrowska2017loopix}	&	High	&	High	&	\begin{itemize}
	\item Low (without delay)
	\item High (added delay at Mixes)
\end{itemize}	&	\begin{itemize}
	\item Third-party anonymity/unobservability
	\item Message processing and packaging is the most computationally expensive part
\end{itemize}	&	\begin{itemize}
	\item Obfuscate timing (via independent poisson mixing delay)
	\item Tunable latency/anonymity trade-off
	\item Tunable real/cover traffic load
	\item Continuous (does not operate in synchronized rounds)
	\item Supports delivering messages to offline users
\end{itemize}	&	\begin{itemize}
	\item Private emails
	\item Instant messaging
\end{itemize}	&	\multicolumn{2}{l}{N/A}\\
cMix~\cite{chaum2017cmix}	&	\begin{itemize}
	\item Setup: Medium
	\item Precomputation: Zero
	\item Real-time: Low
\end{itemize} 	&	\begin{itemize}
	\item Setup: Medium
	\item Precomputation: Low
	\item Real-time: Low
\end{itemize} 	&	\begin{itemize}
	\item Setup: Medium
	\item Precomputation: Medium
	\item Real-time: Medium
\end{itemize}	&	User activity at any given time is known	&	\begin{itemize}
	\item Precomputation based
	\item No public key encryption at run time
	\item Cascade architecture
	\item Multi-party group homomorphic cryptographic system
	\item Resistance to traffic analysis and intersection attacks
	\item Linear scalability
\end{itemize}	&	Low latency applications with lightweight clients	&	\multicolumn{2}{l}{N/A}\\
Nym~\cite{diaz2021nym}	&	N/A	&	N/A	&	N/A	&\begin{itemize}
	\item Security analysis is not available
	\item implementation details are not available
\end{itemize} 	&	\begin{itemize}
	\item Generic - Incentivized - Decentralized and stratified infrastructure
	\item Continuous-time
	\item Using loops of cover traffic
	\item Better privacy and performance trade-offs when scales up to support multiple services and applications
\end{itemize}	&	Ranging from instant messaging to cryptocurrency transactions	&	\multicolumn{2}{l}{N/A}\\
\hline
\multicolumn{9}{c}{\textbf{Onion Routing Based Systems}}\\
\hline
TOR (Curcuit-based Onion Routing)~\cite{syverson2004tor}	&	High	&	Medium	&	Low	&	Does not perform permutation of messages	&	\begin{itemize}
	\item Centralized (any)trusted design
	\item Low-latency
	\item Circuit-based
\end{itemize}	&	generic	&	\raisebox{-1.75\height}{\normalsize \faCheckCircle[regular]}	&	\begin{itemize}
	\item Correlation attacks
	\item Timing
	\item Fingerprinting
	\item Disclosure attacks
\end{itemize}\\
I2P~\cite{timpanaro2012i2p,timpanaro2011monitoring}	&	Medium	&	High	&High	&	\begin{itemize}
	\item Does not hide running I2P and is publicly known
	\item Aim is defending against local network adversaries
\end{itemize}	&
\begin{itemize}
	\item Decentralized with no trusted parties
	\item Message-oriented (packet-switched)
	\item Peer-to-peer overlay network
	\item Encrypted unidirectional tunnels
	\item Stores routing and contact information in distributed hash tables (DHT)
\end{itemize}	&	\begin{itemize}
	\item Web hosting
	\item Web browsing
	\item Email and chat
	\item File sharing
\end{itemize}	&	\raisebox{-0.5\height}{\normalsize \faCheckCircle[regular]}	&	De-anonymizing and disclosure attacks\\
Crowds~\cite{reiter1998crowds}	&	Small	&	Small	&	Medium	&	No protection against global adversaries	&	Crowds members cannot identify message originator	&	Web browsing	&	\raisebox{-0.3\height}{\normalsize \faCheckCircle[regular]}	&	Timing attacks\\
Vuvuzela~\cite{van2015vuvuzela}	&	High	&	\begin{itemize}
	\item Conversion: Medium
	\item Dialing: High
\end{itemize}	&	\begin{itemize}
	\item Conversion: Medium
	\item Dialing: High
\end{itemize}	&	\begin{itemize}
	\item High bandwidth cost (almost 30GB over a month)
	\item Operates in rounds
	\item Users can  send and receive only one message per round
	\item Offline users might miss messages
\end{itemize}	&	\begin{itemize}
	\item Scalable with linear cost
	\item Two protocols (dialing and point-to-point conversation)
	\item Using differential privacy
	\item Preserve fixed message sizes and rate
\end{itemize} 	&	Private messaging	&	\multicolumn{2}{l}{N/A}\\
\hline
\multicolumn{9}{c}{\textbf{Broadcast/Multicast-based Systems and MPC-Based Systems}}\\
\hline
BAR~\cite{kotzanikolaou2017broadcast}	& \begin{itemize}
	\item Registration: High
	\item Communication: High
\end{itemize}	& \begin{itemize}
	\item Registration: High
	\item Communication: Medium
\end{itemize}	& \begin{itemize}
	\item Registration: High
	\item Communication: Medium
\end{itemize}	&	High bandwidth and latency cost	&	\begin{itemize}
	\item Trade-off between bandwidth and latency with selective Broadcast mechanism
	\item Filtering mechanism to distinguish between noise and unobservable bilateral communication
\end{itemize}	&	Anonymous broadcasting	&	\multicolumn{2}{l}{N/A}\\
MCMix~\cite{alexopoulos2017mcmix}	&	\begin{itemize}
	\item Registration: Low
	\item Dialing: Medium
	\item Conversation: Medium
\end{itemize} &	Low	&	\begin{itemize}
	\item Registration: Medium
	\item Dialing: High
	\item Conversation: High
\end{itemize}	& Each user can send and receive one message per conversation round	&	\begin{itemize}
	\item Unobservable bilateral communication
	\item Completely hides metadata
	\item Using MPC and oblivious sorting algorithm for efficient implementation
	\item Proceeds in rounds via two types of protocols (Dialing and Conversation)
\end{itemize}	&	Point-to-point messaging	& \multicolumn{2}{l}{N/A}\\
\hline 
\multicolumn{9}{c}{\textbf{DCN-based Systems}}\\
\hline
Basic \ac{DCN}~\cite{chaum1988dining}	&	Medium	&	High	&	High	&	\begin{itemize}
	\item Disruptions
	\item Only supports transmission of one-bit messages per round
\end{itemize}	&	Information-theoretically secure	&	& &	\\
Dissent~\cite{corrigan2010dissent,wolinsky2012dissent}	& Medium	&	High	&	High	&	\begin{itemize}
	\item Linear increase of overhead with anonymity set size
	\item Not intended for large scale
	\item Per round start-up delay due to serialized shuffle protocol
\end{itemize}	&	\begin{itemize}
	\item Client-server architecture
	\item Scheduling via secret shuffling
	\item Support variable-size message transmissions
	\item Minimum participating users threshold (minimum anonymity set)
	\item Leaves out slow users
\end{itemize}	&	\begin{itemize}
	\item Latency tolerant messaging
	\item File sharing
\end{itemize}	&	\raisebox{-0.3\height}{\footnotesize \faCheckCircle[regular]}	&	Intersection Attacks\\
Riposte~\cite{corrigan2015riposte}	&	Medium	&	High	&	High	&	\begin{itemize}
	\item Trust model (three non-colluding servers)
	\item Long epochs to ensure enough user participation
	\item Uploading limit for message sizes
	\item Identity of active users in each epoch is known
\end{itemize}	&	\begin{itemize}
	\item Secret sharing of write request
	\item Bandwidth-efficiency (transmission of message shares in fewer bits)
\end{itemize}	&	 Anonymous message broadcasting	&	\raisebox{-0.3\height}{\footnotesize \faCheckCircle[regular]}	&	Intersection attacks\\
Shared-Dining~\cite{modinger2021shared}	&	High	&	High	&	High	&	\begin{itemize}
	\item Supports group with small number of participants
	\item Fixed-length messages
\end{itemize}	&	\begin{itemize}	
	\item Combination of secret sharing with classical \ac{DCN}
	\item Incentivize nodes to participate in the protocol
	\item Prevents privacy breach by considereing a threshold for participating users in a round
\end{itemize}	&	 Applications with high privacy requirements, e.g. financial systems	&	\multicolumn{2}{l}{N/A}\\
PriFi~\cite{barman2017prifi,barman2020prifi}	&	\begin{itemize}
	\item Setup: High
	\item Anonymize: Medium
\end{itemize}	&	\begin{itemize}
	\item Setup: Medium
	\item Anonymize: High
\end{itemize}	&	\begin{itemize}
	\item Setup: Medium
	\item Anonymize: Low
\end{itemize}	&	\begin{itemize}
	\item Fix-length messages
	\item High overhead for disruption attacks
\end{itemize}	&	
\begin{itemize}
	\item 3-layer low-latency architecture
	\item Keeps packets inside usual local low-latency path
	\item Deployable to existing infrastructures with minimal changes
\end{itemize}	&	 Organizational communication networks	&	\raisebox{-0.3\height}{\tiny \faCheckCircle[regular]}	&	Intersection attacks\\
Verifiable DCN~\cite{Franck2014verifiable,Franck2021FastECC}	&	Medium	&	Medium	&	Medium	&	& \begin{itemize}
	\item Development of self contained ECC library for all public key operations
	\item Trade-off between cryptographic strength and throughput
	\item Arbitrary-length messages
	\item Efficient misbehaving user detection
	\item No need for reservation phase
\end{itemize}	&			& \multicolumn{2}{l}{N/A}\\
Arbitrary-Length k-Anonymous DCN~\cite{modinger2021arbitrary}	&	\begin{itemize}
	\item Unsecured: Low
	\item Secured: High
\end{itemize}	&	\begin{itemize}
	\item Unsecured: Medium
	\item Secured: High
\end{itemize}	& \begin{itemize}
	\item Unsecured: Low
	\item Secured: High
\end{itemize}	&	\begin{itemize}
	\item Massive overhead to secure operations
	\item Only single group transmissions are implemented in the current version
\end{itemize}	&	\begin{itemize}
	\item Two variant and reducing the overhead for the most common cases
	\item Secured variant is available as a fallback mechanism
	\item Arbitrary-length messages
\end{itemize}	&	 From highly security relevant applications to less critical applications	&	\multicolumn{2}{l}{N/A}\\
\end{supertabular}
\end{landscape}
\normalsize

%
%

\AtNextBibliography{\scriptsize}

\printbibliography[title={References}]

@InProceedings{Poehls_Contingency_2013,
  author     = {Henrich C. P{\"o}hls},
  title      = {{Contingency Revisited: Secure Construction and Legal Implications of Verifiably Weak Integrity}},
  booktitle  = {{IFIP WG 11.11 International Conference on Trust Management (IFIPTM'13)}},
  year       = {2013},
  volume     = {401},
  series     = {IFIPAICT},
  pages      = {136--150},
  month      = {Jun.},
  publisher  = {Springer},
  url        = {http://dx.doi.org/10.1007/978-3-642-38323-6_10}
}

@misc{isots21089,
  author = {{International Organization for Standardization (ISO)}},
  title = {{ISO/TS 21089: Health informatics --- Trusted end-to-end information flows}},
  year = {2018}
}

@Article{cranor2016towards,
  author    = {Cranor, Lorrie and Rabin, Tal and Shmatikov, Vitaly and Vadhan, Salil and Weitzner, Daniel},
  title     = {Towards a privacy research roadmap for the computing community},
  journal   = {arXiv preprint arXiv:1604.03160},
  year      = {2016},
  copyright = {arXiv.org perpetual, non-exclusive license},
  doi       = {10.48550/arXiv.1604.03160},
  publisher = {arXiv}
}

@Article{mendes2017privacy,
  author    = {Mendes, Ricardo and Vilela, Jo{\~a}o P},
  title     = {Privacy-preserving data mining: methods, metrics, and applications},
  journal   = {IEEE Access},
  year      = {2017},
  volume    = {5},
  pages     = {10562--10582},
  doi       = {10.1109/access.2017.2706947},
  publisher = {IEEE}
}

@InProceedings{pohls2014rerum,
  author       = {P{\"o}hls, Henrich C and Angelakis, Vangelis and Suppan, Santiago and Fischer, Kai and Oikonomou, George and Tragos, Elias Z and Rodriguez, Rodrigo Diaz and Mouroutis, Theodoros},
  title        = {{RERUM}: Building a reliable {IoT} upon privacy-and security-enabled smart objects},
  booktitle    = {Wireless Communications and Networking Conference Workshops (WCNCW'14)},
  year         = {2014},
  pages        = {122--127},
  month        = {apr},
  publisher    = {IEEE},
  doi          = {10.1109/WCNCW.2014.6934872}
}

@Article{cook2018using,
  author    = {Cook, Diane J and Duncan, Glen and Sprint, Gina and Fritz, Roschelle L},
  title     = {Using smart city technology to make healthcare smarter},
  journal   = {Proceedings of the IEEE},
  year      = {2018},
  volume    = {106},
  number    = {4},
  pages     = {708--722},
  month     = {apr},
  doi       = {10.1109/jproc.2017.2787688},
  publisher = {IEEE}
}

@Article{edman2009anonymity,
  author    = {Edman, Matthew and Yener, B{\"u}lent},
  title     = {On anonymity in an electronic society: A survey of anonymous communication systems},
  journal   = {ACM Computing Surveys (CSUR)},
  year      = {2009},
  volume    = {42},
  number    = {1},
  pages     = {1--35},
  month     = {dec},
  doi       = {10.1145/1592451.1592456},
  publisher = {ACM},
}

@InCollection{tragos2015securing,
  author    = {Elias Z. Tragos and Henrich C. Pöhls and Ralf C. Staudemeyer and Daniel Slamanig and Adam Kapovits and Santiago Suppan and Alexandros Fragkiadakis and Gianmarco Baldini and Ricardo Neisse and Peter Langendörfer and Zoya Dyka and Christian Wittke},
  title     = {Securing the internet of things—security and privacy in a hyperconnected world},
  booktitle = {Building the Hyperconnected Society- Internet of Things Research and Innovation Value Chains, Ecosystems and Markets},
  publisher = {River Publishers},
  year      = {2015},
  pages     = {189--219},
  month     = {sep},
  doi       = {10.1201/9781003337454-6}
}

@Article{ren2010survey,
  author    = {Ren, Jian and Wu, Jie},
  title     = {Survey on anonymous communications in computer networks},
  journal   = {Elsevier Computer Communications},
  year      = {2010},
  volume    = {33},
  number    = {4},
  pages     = {420--431},
  month     = {mar},
  doi       = {10.1016/j.comcom.2009.11.009},
  publisher = {Elsevier},
}

@InProceedings{Syverson1997,
  author    = {Syverson, P.F. and Goldschlag, D.M. and Reed, M.G.},
  title     = {Anonymous connections and onion routing},
  booktitle = {Symposium on Security and Privacy (S\&P'97)},
  year      = {1997},
  pages     = {44-54},
  publisher = {IEEE},
  doi       = {10.1109/SECPRI.1997.601314},
}

@InProceedings{rupprecht2019breaking,
  author       = {Rupprecht, David and Kohls, Katharina and Holz, Thorsten and P{\"o}pper, Christina},
  title        = {Breaking {LTE} on layer two},
  booktitle    = {Symposium on Security and Privacy (S\&P'19)},
  year         = {2019},
  pages        = {1121--1136},
  month        = {may},
  publisher    = {IEEE},
  doi          = {10.1109/sp.2019.00006}
}

@InProceedings{staudemeyer2018road,
  author       = {Staudemeyer, Ralf C and P{\"o}hls, Henrich C and W{\'o}jcik, Marcin},
  title        = {The road to privacy in IoT: beyond encryption and signatures, towards unobservable communication},
  booktitle    = {19th International Symposium on A World of Wireless, Mobile and Multimedia Networks (WoWMoM'18)},
  year         = {2018},
  pages        = {14--20},
  month        = {jun},
  publisher    = {IEEE},
  doi          = {10.1109/WoWMoM.2018.8449779}
}

@Article{staudemeyer2019takes,
  author    = {Staudemeyer, Ralf C and P{\"o}hls, Henrich C and W{\'o}jcik, Marcin},
  title     = {What it takes to boost Internet of Things privacy beyond encryption with unobservable communication: a survey and lessons learned from the first implementation of DC-net},
  journal   = {Springer Journal of Reliable Intelligent Environments},
  year      = {2019},
  volume    = {5},
  number    = {1},
  pages     = {41--64},
  month     = {feb},
  doi       = {10.1007/s40860-019-00075-0},
  publisher = {Springer},
}

@Article{barman2017prifi,
  author    = {Barman, Ludovic and Dacosta, Italo and Zamani, Mahdi and Zhai, Ennan and Pyrgelis, Apostolos and Ford, Bryan and Hubaux, Jean-Pierre and Feigenbaum, Joan},
  title     = {{PriFi}: Low-latency anonymity for organizational networks},
  journal   = {arXiv preprint arXiv:1710.10237},
  year      = {2017},
  copyright = {arXiv.org perpetual, non-exclusive license},
  doi       = {10.48550/ARXIV.1710.10237},
  publisher = {arXiv}
}

@InProceedings{barman2020prifi,
  author    = {Barman, Ludovic and Dacosta, Italo and Zamani, Mahdi and Zhai, Ennan and Pyrgelis, Apostolos and Ford, Bryan and Feigenbaum, Joan and Hubaux, Jean-Pierre},
  title     = {{PriFi}: Low-latency anonymity for organizational networks},
  booktitle = {Privacy Enhancing Technologies Symposium (PETS'20)},
  year      = {2020},
  number    = {4},
  pages     = {24--47},
  month     = {aug},
  publisher = {de Gruyter},
  doi       = {10.2478/popets-2020-0061},
  journal   = {Proceedings on Privacy Enhancing Technologies (PETS'20)},
}

@InProceedings{raymond2001traffic,
  author    = {Raymond, Jean-Fran{\c{c}}ois},
  title     = {Traffic analysis: Protocols, attacks, design issues, and open problems},
  booktitle = {Designing privacy enhancing technologies (LNCS)},
  year      = {2001},
  pages     = {10--29},
  publisher = {Springer},
  doi       = {10.1007/3-540-44702-4_2},
}

@book{danezis2007introducing,
  title={Introducing traffic analysis},
  author={Danezis, George and Clayton, Richard},
  year={2007},
  publisher={Auerbach Publications, Boca Raton, FL}
}

@inproceedings{staudemeyer2005attacker,
  title={Attacker models, traffic analysis and privacy threats in {IP} networks},
  author={Staudemeyer, RC and Umuhoza, D and Omlin, CW},
  booktitle={12th International Conference on Telecommunications (ICT’05)},
  year={2005}
}

@InProceedings{juarez2014critical,
  author    = {Juarez, Marc and Afroz, Sadia and Acar, Gunes and Diaz, Claudia and Greenstadt, Rachel},
  title     = {A critical evaluation of website fingerprinting attacks},
  booktitle = {Conference on Computer and Communications Security (CCS'14)},
  year      = {2014},
  pages     = {263--274},
  month     = {nov},
  publisher = {ACM},
  doi       = {10.1145/2660267.2660368}
}

@Article{kotzanikolaou2017broadcast,
  author    = {Kotzanikolaou, Panayiotis and Chatzisofroniou, George and Burmester, Mike},
  title     = {Broadcast anonymous routing {(BAR)}: scalable real-time anonymous communication},
  journal   = {Springer International Journal of Information Security},
  year      = {2017},
  volume    = {16},
  number    = {3},
  pages     = {313--326},
  month     = {feb},
  doi       = {10.1007/s10207-016-0318-0},
  publisher = {Springer},
}

@Article{chaum1981untraceable,
  author    = {Chaum, David L},
  title     = {Untraceable electronic mail, return addresses, and digital pseudonyms},
  journal   = {Communications of the ACM},
  year      = {1981},
  volume    = {24},
  number    = {2},
  pages     = {84--90},
  month     = {feb},
  doi       = {10.1145/358549.358563},
  publisher = {ACM}
}

@Article{chaum1988dining,
  author    = {Chaum, David},
  title     = {The dining cryptographers problem: Unconditional sender and recipient untraceability},
  journal   = {Springer Journal of Cryptology},
  year      = {1988},
  volume    = {1},
  number    = {1},
  pages     = {65--75},
  month     = {jan},
  doi       = {10.1007/bf00206326},
  publisher = {Springer},
}

@techreport{dingledine2004tor,
  title={Tor: The second-generation onion router},
  author={Dingledine, Roger and Mathewson, Nick and Syverson, Paul},
  year={2004},
  institution={Naval Research Lab Washington DC}
}

@InProceedings{nguyen2003breaking,
  author       = {Nguyen, Lan and Safavi-Naini, Rei},
  title        = {Breaking and mending resilient mix-nets},
  booktitle    = {International Workshop on Privacy Enhancing Technologies},
  year         = {2003},
  pages        = {66--80},
  publisher    = {Springer},
  doi          = {10.1007/978-3-540-40956-4_5}
}

@InProceedings{pfitzmann1994breaking,
  author    = {Pfitzmann, Birgit},
  title     = {Breaking an efficient anonymous channel},
  booktitle = {Advances in Cryptology (EUROCRYPT'94)},
  year      = {1994},
  pages     = {332--340},
  publisher = {Springer},
  doi       = {10.1007/bfb0053448},
}

@InProceedings{bauer2017dining,
  author       = {Bauer, Johannes and Staudemeyer, Ralf C},
  title        = {From Dining Cryptographers to dining things: Unobservable communication in the {IoT}},
  booktitle    = {22nd International Workshop on Computer Aided Modeling and Design of Communication Links and Networks (CAMAD'17)},
  year         = {2017},
  pages        = {1--7},
  month        = {jun},
  publisher    = {IEEE},
  doi          = {10.1109/CAMAD.2017.8031529}
}

@Article{modinger2021arbitrary,
  author    = {M{\"o}dinger, David and He{\ss}, Alexander and Hauck, Franz J},
  title     = {Arbitrary Length k-Anonymous Dining-Cryptographers Communication},
  journal   = {arXiv preprint arXiv:2103.17091},
  year      = {2021},
  copyright = {arXiv.org perpetual, non-exclusive license},
  doi       = {10.48550/ARXIV.2103.17091},
  publisher = {arXiv}
}

@Misc{pfitzmann2010terminology,
  author       = {Pfitzmann, Andreas and Hansen, Marit},
  title        = {A terminology for talking about privacy by data minimization: Anonymity, unlinkability, undetectability, unobservability, pseudonymity, and identity management},
  howpublished = {online},
  year         = {2010},
  note         = {\url{http://dud.inf.tu-dresden.de/literatur/Anon_Terminology_v0.34.pdf}},
  publisher    = {Dresden, Germany},
}

@mastersthesis{erdin2012anonymous,
  title={Anonymous Communication Networks: Usage Analysis and Attack Mechanisms},
  author={Erdin, Esra},
  year={2012},
  school={University of Nevada, Reno}
}

@techreport{danezis2008survey,
  title={A survey of anonymous communication channels},
  author={Danezis, George and Diaz, Claudia},
  year={2008},
  institution={Technical Report MSR-TR-2008-35, Microsoft Research}
}

@InProceedings{goldschlag1996hiding,
  author       = {Goldschlag, David M and Reed, Michael G and Syverson, Paul F},
  title        = {Hiding routing information},
  booktitle    = {International Workshop on Information Hiding (IH'96)},
  year         = {1996},
  pages        = {137--150},
  publisher    = {Springer},
  doi          = {10.1007/3-540-61996-8_37}
}

@Article{reed1998anonymous,
  author    = {Reed, Michael G and Syverson, Paul F and Goldschlag, David M},
  title     = {Anonymous connections and onion routing},
  journal   = {IEEE Journal on Selected areas in Communications},
  year      = {1998},
  volume    = {16},
  number    = {4},
  pages     = {482--494},
  month     = {may},
  doi       = {10.1109/49.668972},
  publisher = {IEEE},
}

@Article{goldschlag1999onion,
  author    = {Goldschlag, David and Reed, Michael and Syverson, Paul},
  title     = {Onion routing},
  journal   = {Communications of the ACM},
  year      = {1999},
  volume    = {42},
  number    = {2},
  pages     = {39--41},
  month     = {feb},
  doi       = {10.1145/293411.293443},
  publisher = {ACM}
}

@InProceedings{syverson2004tor,
  author    = {Syverson, Paul and Dingledine, Roger and Mathewson, Nick},
  title     = {Tor: The second generation onion router},
  booktitle = {26th USENIX Security Symposium (USENIX Security '04)},
  year      = {2004},
  pages     = {303--320}
}

@Article{levine2002hordes,
  author    = {Levine, Brian Neil and Shields, Clay},
  title     = {Hordes: a multicast based protocol for anonymity},
  journal   = {Journal of Computer Security (IOS Press)},
  year      = {2002},
  volume    = {10},
  number    = {3},
  pages     = {213--240},
  month     = {jul},
  doi       = {10.3233/jcs-2002-10302},
  publisher = {IOS Press},
}

@InProceedings{akhoondi2012lastor,
  author    = {Akhoondi, Masoud and Yu, Curtis and Madhyastha, Harsha V},
  title     = {{LASTor}: A low-latency AS-aware Tor client},
  booktitle = {Symposium on Security and Privacy (S\&P'12)},
  year      = {2012},
  pages     = {476--490},
  month     = {may},
  publisher = {IEEE},
  doi       = {10.1109/sp.2012.35},
}

@InProceedings{mclachlan2009scalable,
  author    = {McLachlan, Jon and Tran, Andrew and Hopper, Nicholas and Kim, Yongdae},
  title     = {Scalable onion routing with torsk},
  booktitle = {16th Conference on Computer and Communications Security (CCS'09)},
  year      = {2009},
  pages     = {590--599},
  publisher = {ACM},
  doi       = {http://dx.doi.org/10.1145/1653662.1653733}
}

@InProceedings{chen2015hornet,
  author    = {Chen, Chen and Asoni, Daniele E and Barrera, David and Danezis, George and Perrig, Adrain},
  title     = {{HORNET}: High-speed onion routing at the network layer},
  booktitle = {22nd Conference on Computer and Communications Security (CCS'15)},
  year      = {2015},
  pages     = {1441--1454},
  month     = {oct},
  publisher = {ACM},
  doi       = {10.1145/2810103.2813628}
}

@InProceedings{shi2021non,
  author       = {Shi, Elaine and Wu, Ke},
  title        = {Non-interactive anonymous router},
  booktitle    = {Annual International Conference on the Theory and Applications of Cryptographic Techniques},
  year         = {2021},
  pages        = {489--520},
  publisher    = {Springer},
  doi          = {10.1007/978-3-030-77883-5_17}
}

@InProceedings{mokhtar2013rac,
  author       = {Mokhtar, Sonia Ben and Berthou, Gautier and Diarra, Amadou and Qu{\'e}ma, Vivien and Shoker, Ali},
  title        = {Rac: A freerider-resilient, scalable, anonymous communication protocol},
  booktitle    = {33rd International Conference on Distributed Computing Systems (ICDCS'13)},
  year         = {2013},
  pages        = {520--529},
  month        = {jul},
  publisher    = {IEEE},
  doi          = {10.1109/ICDCS.2013.52}
}

@InProceedings{van2015vuvuzela,
  author    = {Van Den Hooff, Jelle and Lazar, David and Zaharia, Matei and Zeldovich, Nickolai},
  title     = {Vuvuzela: Scalable private messaging resistant to traffic analysis},
  booktitle = {25th Symposium on Operating Systems Principles (SOSP'15)},
  year      = {2015},
  pages     = {137--152},
  month     = {oct},
  publisher = {ACM},
  doi       = {10.1145/2815400.2815417},
}

@InProceedings{angel2016unobservable,
  author    = {Angel, Sebastian and Setty, Srinath},
  title     = {Unobservable communication over fully untrusted infrastructure},
  booktitle = {12th USENIX Symposium on Operating Systems Design and Implementation (OSDI'16)},
  year      = {2016},
  pages     = {551--569}
}

@InProceedings{tyagi2017stadium,
  author    = {Tyagi, Nirvan and Gilad, Yossi and Leung, Derek and Zaharia, Matei and Zeldovich, Nickolai},
  title     = {Stadium: A distributed metadata-private messaging system},
  booktitle = {26th Symposium on Operating Systems Principles (SOSP'17)},
  year      = {2017},
  pages     = {423--440},
  month     = {oct},
  publisher = {ACM},
  doi       = {10.1145/3132747.3132783},
}

@InProceedings{lazar2018karaoke,
  author    = {Lazar, David and Gilad, Yossi and Zeldovich, Nickolai},
  title     = {Karaoke: Distributed private messaging immune to passive traffic analysis},
  booktitle = {13th USENIX Symposium on Operating Systems Design and Implementation (OSDI'18)},
  year      = {2018},
  pages     = {711--725}
}

@Article{dwork2014algorithmic,
  author    = {Dwork, Cynthia and Roth, Aaron},
  title     = {The algorithmic foundations of differential privacy},
  journal   = {Found. Trends Theor. Comput. Sci. (Now Publishers)},
  year      = {2014},
  volume    = {9},
  number    = {3-4},
  pages     = {211--407},
  doi       = {10.1561/0400000042},
  publisher = {Now Publishers},
}

@phdthesis{akhoondi2015scalable,
  title={Scalable Techniques for Security and Anonymity in Distributed Systems},
  author={Akhoondi, Masoud},
  year={2015},
  school={UC Riverside}
}

@InProceedings{johnson2013users,
  author    = {Johnson, Aaron and Wacek, Chris and Jansen, Rob and Sherr, Micah and Syverson, Paul},
  title     = {Users get routed: Traffic correlation on Tor by realistic adversaries},
  booktitle = {Conference on Computer and Communications Security (CCS'13)},
  year      = {2013},
  pages     = {337--348},
  publisher = {ACM},
  doi       = {10.1145/2508859.2516651},
}

@Article{wright2004predecessor,
  author    = {Wright, Matthew K and Adler, Micah and Levine, Brian Neil and Shields, Clay},
  title     = {The predecessor attack: An analysis of a threat to anonymous communications systems},
  journal   = {ACM Transactions on Information and System Security (TISSEC'04)},
  year      = {2004},
  volume    = {7},
  number    = {4},
  pages     = {489--522},
  month     = {nov},
  doi       = {10.1145/1042031.1042032},
  publisher = {ACM},
}

@InProceedings{murdoch2005low,
  author       = {Murdoch, Steven J and Danezis, George},
  title        = {Low-cost traffic analysis of Tor},
  booktitle    = {Symposium on Security and Privacy (S\&P'05)},
  year         = {2005},
  pages        = {183--195},
  publisher    = {IEEE},
  doi          = {10.1109/sp.2005.12}
}

@InProceedings{bauer2007low,
  author    = {Bauer, Kevin and McCoy, Damon and Grunwald, Dirk and Kohno, Tadayoshi and Sicker, Douglas},
  title     = {Low-resource routing attacks against Tor},
  booktitle = {Workshop on Privacy in the Electronic Society (WPES'07)},
  year      = {2007},
  pages     = {11--20},
  month     = {oct},
  publisher = {ACM},
  doi       = {10.1145/1314333.1314336},
}

@Article{hopper2010much,
  author    = {Hopper, Nicholas and Vasserman, Eugene Y and Chan-Tin, Eric},
  title     = {How much anonymity does network latency leak?},
  journal   = {ACM Transactions on Information and System Security (TISSEC'10)},
  year      = {2010},
  volume    = {13},
  number    = {2},
  pages     = {1--28},
  month     = {feb},
  doi       = {10.1145/1698750.1698753},
  publisher = {ACM},
}

@InProceedings{mittal2011stealthy,
  author    = {Mittal, Prateek and Khurshid, Ahmed and Juen, Joshua and Caesar, Matthew and Borisov, Nikita},
  title     = {Stealthy traffic analysis of low-latency anonymous communication using throughput fingerprinting},
  booktitle = {18th Conference on Computer and Communications Security (CCS'11)},
  year      = {2011},
  pages     = {215--226},
  month     = {oct},
  publisher = {ACM},
  doi       = {10.1145/2046707.2046732},
}

@InProceedings{cai2012touching,
  author    = {Cai, Xiang and Zhang, Xin Cheng and Joshi, Brijesh and Johnson, Rob},
  title     = {Touching from a distance: Website fingerprinting attacks and defenses},
  booktitle = {19th Conference on Computer and Communications Security (CCS'12)},
  year      = {2012},
  pages     = {605--616},
  publisher = {ACM},
  doi       = {10.1145/2382196.2382260},
}

@InProceedings{kwon2015circuit,
  author    = {Kwon, Albert and AlSabah, Mashael and Lazar, David and Dacier, Marc and Devadas, Srinivas},
  title     = {Circuit fingerprinting attacks: Passive deanonymization of Tor hidden services},
  booktitle = {24th USENIX Security Symposium (USENIX Security '15)},
  year      = {2015},
  pages     = {287--302}
}

@InProceedings{panchenko2011website,
  author    = {Panchenko, Andriy and Niessen, Lukas and Zinnen, Andreas and Engel, Thomas},
  title     = {Website fingerprinting in onion routing based anonymization networks},
  booktitle = {10th Workshop on Privacy in the Electronic Society (WPES'11)},
  year      = {2011},
  pages     = {103--114},
  month     = {oct},
  publisher = {ACM},
  doi       = {10.1145/2046556.2046570},
}

@InProceedings{wang2013improved,
  author    = {Wang, Tao and Goldberg, Ian},
  title     = {Improved website fingerprinting on tor},
  booktitle = {12th Workshop on Privacy in the Electronic Society (WPES'13)},
  year      = {2013},
  pages     = {201--212},
  month     = {nov},
  publisher = {ACM},
  doi       = {10.1145/2517840.2517851},
}

@InProceedings{chakravarty2014effectiveness,
  author       = {Chakravarty, Sambuddho and Barbera, Marco V and Portokalidis, Georgios and Polychronakis, Michalis and Keromytis, Angelos D},
  title        = {On the effectiveness of traffic analysis against anonymity networks using flow records},
  booktitle    = {International Conference on Passive and Active Network Measurement (PAM'14)},
  year         = {2014},
  pages        = {247--257},
  publisher    = {Springer},
  doi          = {10.1007/978-3-319-04918-2_24}
}

@Article{lu2019survey,
  author    = {Lu, Tianbo and Du, Zeyu and Wang, Z Jane},
  title     = {A survey on measuring anonymity in anonymous communication systems},
  journal   = {IEEE Access},
  year      = {2019},
  volume    = {7},
  pages     = {70584--70609},
  doi       = {10.1109/access.2019.2919322},
  publisher = {IEEE}
}

@InCollection{staudemeyer2017security,
  author    = {Staudemeyer, Ralf C and P{\"o}hls, Henrich C and Watson, Bruce W},
  title     = {Security and Privacy for the {Internet of Things} Communication in the SmartCity},
  booktitle = {Designing, Developing, and Facilitating Smart Cities},
  publisher = {Springer},
  year      = {2017},
  pages     = {109--137},
  month     = {dec},
  doi       = {10.1007/978-3-319-44924-1_7}
}

@Article{srinivasan2002p5,
  author    = {Rob Sherwood and Bobby Bhattacharjee and Aravind Srinivasan},
  title     = {P5: A protocol for scalable anonymous communication},
  journal   = {Journal of Computer Security ({IOS} Press)},
  year      = {2005},
  volume    = {13},
  number    = {6},
  pages     = {839--876},
  month     = {dec},
  doi       = {10.3233/jcs-2005-13602},
  publisher = {{IOS} Press},
}

@InProceedings{von2003k,
  author    = {Von Ahn, Luis and Bortz, Andrew and Hopper, Nicholas J},
  title     = {K-anonymous message transmission},
  booktitle = {10th Conference on Computer and Communications Security (CCS'03)},
  year      = {2003},
  pages     = {122--130},
  month     = {oct},
  publisher = {ACM},
  doi       = {10.1145/948109.948128}
}

@InProceedings{perng2006m2,
  author       = {Perng, Ginger and Reiter, Michael K and Wang, Chenxi},
  title        = {M2: Multicasting mixes for efficient and anonymous communication},
  booktitle    = {26th International Conference on Distributed Computing Systems (ICDCS'06)},
  year         = {2006},
  pages        = {59--59},
  publisher    = {IEEE},
  doi          = {10.1109/icdcs.2006.53}
}

@InProceedings{xiao2006design,
  author       = {Xiao, Li and Liu, Xiaomei and Gu, Wenjun and Xuan, Dong and Liu, Yunhao},
  title        = {A design of overlay anonymous multicast protocol},
  booktitle    = {20th International Parallel \& Distributed Processing Symposium (IPDPS'06)},
  year         = {2006},
  pages        = {10pp},
  publisher    = {IEEE},
  doi          = {10.1109/ipdps.2006.1639286}
}

@misc{BARgithub2016,
  author = {George Chatzisofroniou, Panayiotis Kotzanikolaou},
  title = {Broadcast Anonymous Routing - A scalable system for efficient anonymous communications},
  year = {2016},
  publisher = {GitHub},
  journal = {GitHub repository},
  howpublished = "\url{https://github.com/sophron/BAR}",
  note = "[Online; accessed 10-April-2022]"
}

@InProceedings{chor1995private,
  author       = {Chor, Benny and Goldreich, Oded and Kushilevitz, Eyal and Sudan, Madhu},
  title        = {Private information retrieval},
  booktitle    = {36th Annual Foundations of Computer Science},
  year         = {1995},
  pages        = {41--50},
  publisher    = {IEEE},
  doi          = {10.1109/sfcs.1995.492461}
}

@InProceedings{Kwon2015riffle,
  author    = {Albert Kwon and David Lazar and Srinivas Devadas and Bryan Ford},
  title     = {Riffle: An efficient communication system with strong anonymity},
  booktitle = {Privacy Enhancing Technologies Symposium (PETS'16)},
  year      = {2015},
  volume    = {2016},
  number    = {2},
  pages     = {115--134},
  publisher = {de Gruyter},
  doi       = {doi:10.1515/popets-2016-0008},
  journal   = {Privacy Enhancing Technologies Symposium (PETS'16)},
  url       = {https://doi.org/10.1515/popets-2016-0008},
}

@InProceedings{kwon2017atom,
  author    = {Kwon, Albert and Corrigan-Gibbs, Henry and Devadas, Srinivas and Ford, Bryan},
  title     = {Atom: Horizontally scaling strong anonymity},
  booktitle = {26th Symposium on Operating Systems Principles (SOSP'17)},
  year      = {2017},
  pages     = {406--422},
  month     = {oct},
  publisher = {ACM},
  doi       = {10.1145/3132747.3132755},
}

@InProceedings{von2018management,
  author       = {Von Maltitz, Marcel and Smarzly, Stefan and Kinkelin, Holger and Carle, Georg},
  title        = {A management framework for secure multiparty computation in dynamic environments},
  booktitle    = {Network Operations and Management Symposium (NOMS'18)},
  year         = {2018},
  pages        = {1--7},
  month        = {apr},
  publisher    = {IEEE},
  doi          = {10.1109/noms.2018.8406322}
}

@mastersthesis{franck2008new,
  title={New directions for dining cryptographers},
  author={Franck, Christian},
  year={2008},
  school={University of Luxembourg}
}

@InProceedings{kissner2004private,
  author       = {Kissner, Lea and Oprea, Alina and Reiter, Michael K and Song, Dawn and Yang, Ke},
  title        = {Private keyword-based push and pull with applications to anonymous communication},
  booktitle    = {International Conference on Applied Cryptography and Network Security (ACNS'04)},
  year         = {2004},
  pages        = {16--30},
  publisher    = {Springer},
  doi          = {10.1007/978-3-540-24852-1_2}
}

@InProceedings{aguilar2016xpir,
  author    = {Aguilar-Melchor, Carlos and Barrier, Joris and Fousse, Laurent and Killijian, Marc-Olivier},
  title     = {{XPIR}: Private information retrieval for everyone},
  booktitle = {Privacy Enhancing Technologies Symposium (PETS'16)},
  year      = {2016},
  volume    = {2},
  pages     = {155--174},
  month     = {dec},
  publisher = {de Gruyter},
  doi       = {10.1515/popets-2016-0010},
}

@Article{kelly2011exploring,
  author    = {Kelly, Douglas and Raines, Richard and Baldwin, Rusty and Grimaila, Michael and Mullins, Barry},
  title     = {Exploring extant and emerging issues in anonymous networks: A taxonomy and survey of protocols and metrics},
  journal   = {IEEE Communications Surveys \& Tutorials},
  year      = {2011},
  volume    = {14},
  number    = {2},
  pages     = {579--606},
  doi       = {10.1109/surv.2011.042011.00080},
  publisher = {IEEE},
}

@InProceedings{wang2019survey,
  author       = {Wang, Shunye and Du, Yanhui and Lu, Tianliang and Wu, Jing and Wang, Tengfei},
  title        = {A survey of anonymous communication methods in {Internet of Things}},
  booktitle    = {9th International Conference on Electronics Information and Emergency Communication (ICEIEC'19)},
  year         = {2019},
  pages        = {627--633},
  month        = {jul},
  publisher    = {IEEE},
  doi          = {10.1109/iceiec.2019.8784538}
}

@inproceedings{scholz2007dining,
  title={Dining Cryptographers. The Protocol},
  author={Scholz, Immanuel},
  booktitle={24th Chaos Communication Congress},
  year={2007}
}

@Article{chaum1985security,
  author    = {Chaum, David},
  title     = {Security without identification: Transaction systems to make big brother obsolete},
  journal   = {Communications of the ACM},
  year      = {1985},
  volume    = {28},
  number    = {10},
  pages     = {1030--1044},
  month     = {oct},
  doi       = {10.1145/4372.4373},
  publisher = {ACM}
}

@Article{pfitzmann1987networks,
  author    = {Pfitzmann, Andreas and Waidner, Michael},
  title     = {Networks without user observability},
  journal   = {Elsevier Computers \& Security},
  year      = {1987},
  volume    = {6},
  number    = {2},
  pages     = {158--166},
  month     = {apr},
  doi       = {10.1016/0167-4048(87)90087-3},
  publisher = {Elsevier},
}

@InProceedings{golle2004dining,
  author       = {Golle, Philippe and Juels, Ari},
  title        = {Dining cryptographers revisited},
  booktitle    = {Advances in Cryptology (EUROCRYPT'04)},
  year         = {2004},
  pages        = {456--473},
  publisher    = {Springer},
  doi          = {10.1007/978-3-540-24676-3_27}
}

@InProceedings{krasnova2016footprint,
  author    = {Krasnova, Anna and Neikes, Moritz and Schwabe, Peter},
  title     = {Footprint scheduling for dining-cryptographer networks},
  booktitle = {Financial Cryptography and Data Security (FC'16)},
  year      = {2016},
  pages     = {385--402},
  publisher = {Springer},
  doi       = {10.1007/978-3-662-54970-4_23},
}

@InProceedings{waidner1989dining,
  author    = {Waidner, Michael and Pfitzmann, Birgit and others},
  title     = {The dining cryptographers in the disco: Unconditional sender and recipient untraceability with computationally secure serviceability},
  booktitle = {Advances in Cryptology (EUROCRYPT'89)},
  year      = {1989},
  volume    = {434},
  pages     = {690},
  publisher = {Springer},
  doi       = {10.1007/3-540-46885-4_69},
  journal   = {Advances in Cryptology (EUROCRYPT'89)},
}

@InProceedings{corrigan2015riposte,
  author       = {Corrigan-Gibbs, Henry and Boneh, Dan and Mazi{\`e}res, David},
  title        = {Riposte: An anonymous messaging system handling millions of users},
  booktitle    = {Symposium on Security and Privacy (S\&P'15)},
  year         = {2015},
  pages        = {321--338},
  month        = {may},
  publisher    = {IEEE},
  doi          = {10.1109/sp.2015.27}
}

@InProceedings{waidner1989unconditional,
  author    = {Waidner, Michael},
  title     = {Unconditional sender and recipient untraceability in spite of active attacks},
  booktitle = {Workshop on the Theory and Application of of Cryptographic Techniques (EUROCRYPT'89)},
  year      = {1989},
  pages     = {302--319},
  publisher = {Springer},
  doi       = {10.1007/3-540-46885-4_32},
}

@InProceedings{bos1989detection,
  author       = {Bos, Jurjen and Boer, Bert den},
  title        = {Detection of disrupters in the DC protocol},
  booktitle    = {Workshop on the Theory and Application of of Cryptographic Techniques (EUROCRYPT'89)},
  year         = {1989},
  pages        = {320--327},
  publisher    = {Springer},
  doi          = {10.1007/3-540-46885-4_33}
}

@techreport{goel2003herbivore,
  title={Herbivore: A scalable and efficient protocol for anonymous communication},
  author={Goel, Sharad and Robson, Mark and Polte, Milo and Sirer, Emin},
  year={2003},
  institution={Cornell University}
}

@InProceedings{furukawa2001efficient,
  author       = {Furukawa, Jun and Sako, Kazue},
  title        = {An efficient scheme for proving a shuffle},
  booktitle    = {Advances in Cryptology (CRYPTO'01)},
  year         = {2001},
  pages        = {368--387},
  publisher    = {Springer},
  doi          = {10.1007/3-540-44647-8_22}
}

@InProceedings{neff2001verifiable,
  author    = {Neff, C Andrew},
  title     = {A verifiable secret shuffle and its application to e-voting},
  booktitle = {8th Conference on Computer and Communications Security (CCS'01)},
  year      = {2001},
  pages     = {116--125},
  month     = {nov},
  publisher = {ACM},
  doi       = {10.1145/501983.502000}
}

@article{studholme2007multiparty,
  title={Multiparty computation to generate secret permutations},
  author={Studholme, Chris and Blake, Ian},
  journal={Cryptology ePrint Archive},
  year={2007}
}

@InProceedings{corrigan2010dissent,
  author    = {Corrigan-Gibbs, Henry and Ford, Bryan},
  title     = {Dissent: accountable anonymous group messaging},
  booktitle = {17th Conference on Computer and Communications Security (CCS'10)},
  year      = {2010},
  pages     = {340--350},
  month     = {oct},
  publisher = {ACM},
  doi       = {10.1145/1866307.1866346},
}

@InProceedings{wolinsky2012dissent,
  author    = {Wolinsky, David Isaac and Corrigan-Gibbs, Henry and Ford, Bryan and Johnson, Aaron},
  title     = {Dissent in numbers: Making strong anonymity scale},
  booktitle = {10th USENIX Symposium on Operating Systems Design and Implementation (OSDI'12)},
  year      = {2012},
  pages     = {179--182}
}

@techreport{wolinsky2012scalable,
  title={Scalable anonymous group communication in the anytrust model},
  author={Wolinsky, David I and Corrigan-Gibbs, Henry and Ford, Bryan and Johnson, Aaron},
  year={2012},
  institution={NAVAL RESEARCH LAB WASHINGTON DC}
}

@InProceedings{corrigan2013proactively,
  author    = {Corrigan-Gibbs, Henry and Wolinsky, David Isaac and Ford, Bryan},
  title     = {Proactively accountable anonymous messaging in verdict},
  booktitle = {22nd USENIX Security Symposium (USENIX Security '13)},
  year      = {2013},
  pages     = {147--162}
}

@misc{DCSimulator2015,
  author = {Moritz Neikes},
  title = {{DCnet-simulator - Simulator to explore and compare various scheduling algorithms for DC-networks}},
  year = {2015},
  publisher = {GitHub},
  journal = {GitHub repository},
  howpublished = "\url{https://github.com/25A0/DCnet-simulator}",
  note = "[Online; accessed 10-April-2022]"
}

@misc{Barman2020PrifiGithub,
  author = {Ludovic Barman, David Wolinsky, Bryan Ford},
  title = {{PriFi}, a low-latency, local-area anonymous communication network},
  year = {2020},
  publisher = {GitHub},
  journal = {GitHub repository},
  howpublished = "\url{https://github.com/dedis/prifi}",
  note = "[Online; accessed 10-April-2022]"
}

@misc{barman2020prifiexperiments,
  author = {Ludovic Barman},
  title = {{PriFi experiments - Raw data and Plots for PriFi}},
  year = {2020},
  publisher = {GitHub},
  journal = {GitHub repository},
  howpublished = "\url{https://github.com/lbarman/prifi-experiments}",
  note = "[Online; accessed 10-April-2022]"
}

@InProceedings{modinger20203p3,
  author       = {M{\"o}dinger, David and Hauck, Franz J},
  title        = {{3P3}: strong flexible privacy for broadcasts},
  booktitle    = {19th International Conference on Trust, Security and Privacy in Computing and Communications (TrustCom'20)},
  year         = {2020},
  pages        = {1630--1637},
  month        = {dec},
  publisher    = {IEEE},
  doi          = {10.1109/trustcom50675.2020.00225}
}

@misc{Hess20203p3,
  author = {Alexander Hess, David Mödinger},
  title = {{3P3} Evaluation},
  year = {2020},
  publisher = {GitHub},
  journal = {GitHub repository},
  howpublished = "\url{https://github.com/vs-uulm/3p3-evaluation}",
  note = "[Online; accessed 10-April-2022]"
}

@inproceedings{conrad2014survey,
  title={A Survey on {Tor and I2P}},
  author={Conrad, Bernd and Shirazi, Fatemeh},
  booktitle={9th International Conference on Internet Monitoring and Protection (ICIMP2014)},
  pages={22--28},
  year={2014}
}

@InProceedings{timpanaro2012i2p,
  author       = {Timpanaro, Juan Pablo and Chrisment, Isabelle and Festor, Olivier},
  title        = {{I2P’s} usage characterization},
  booktitle    = {International Workshop on Traffic Monitoring and Analysis (TMA'12)},
  year         = {2012},
  pages        = {48--51},
  publisher    = {Springer},
  doi          = {10.1007/978-3-642-28534-9_5}
}

@phdthesis{timpanaro2011monitoring,
  title={Monitoring the I2P network},
  author={Timpanaro, Juan Pablo and Isabelle, Chrisment and Olivier, Festor},
  year={2011},
  school={Inria}
}

@InProceedings{clarke2001freenet,
  author    = {Clarke, Ian and Sandberg, Oskar and Wiley, Brandon and Hong, Theodore W},
  title     = {Freenet: A distributed anonymous information storage and retrieval system},
  booktitle = {Designing Privacy Enhancing Technologies (LNCS)},
  year      = {2001},
  pages     = {46--66},
  publisher = {Springer},
  doi       = {10.1007/3-540-44702-4_4},
}

@InProceedings{dingledine2001free,
  author    = {Dingledine, Roger and Freedman, Michael J and Molnar, David},
  title     = {The free haven project: Distributed anonymous storage service},
  booktitle = {Designing Privacy Enhancing Technologies (LNCS)},
  year      = {2001},
  pages     = {67--95},
  publisher = {Springer},
  doi       = {10.1007/3-540-44702-4_5},
}

@InProceedings{bennett2003gap,
  author       = {Bennett, Krista and Grothoff, Christian},
  title        = {{GAP}--practical anonymous networking},
  booktitle    = {International Workshop on Privacy Enhancing Technologies (PETS'03)},
  year         = {2003},
  pages        = {141--160},
  publisher    = {Springer},
  doi          = {10.1007/978-3-540-40956-4_10}
}

@InProceedings{grothoff2003gnunet,
  author    = {Grothoff, Christian and Bennett, Krista and Matheny, Blake},
  title     = {{GNUnet}: a secure peer-to-peer framework},
  booktitle = {4th Annual Information Security Symposium (CERIAS'03)},
  year      = {2003},
  pages     = {1},
  doi       = {10.5555/2793149.2793165}
}

@phdthesis{sergeev2013network,
  title={Network coding for anonymous broadcast},
  author={Sergeev, Ivan A},
  year={2013},
  school={Massachusetts Institute of Technology}
}

@phdthesis{javani2021privacy,
  title={Privacy-Preserving Protocols with Oblivious Transfer and Mix Networks},
  author={Javani, Farid},
  year={2021},
  school={University of Maryland, Baltimore County}
}

@InCollection{Schoenmakers2005,
  author    = {Schoenmakers, Berry},
  title     = {Oblivious Transfer},
  booktitle = {Encyclopedia of Cryptography and Security},
  publisher = {Springer},
  year      = {2005},
  editor    = {van Tilborg, Henk C. A.},
  pages     = {445--446},
  address   = {Boston, MA},
  isbn      = {978-0-387-23483-0},
  doi       = {10.1007/0-387-23483-7_285},
  url       = {https://doi.org/10.1007/0-387-23483-7_285},
}

@InProceedings{brassard1987all,
  author    = {Brassard, G and Crepeau, C and Robert, JM},
  title     = {All-or-Nothing Disclosure of Secrets},
  booktitle = {Advances in Cryptology (CRYPTO'86)},
  year      = {1987},
  volume    = {263},
  pages     = {234--238},
  publisher = {Springer},
  doi       = {10.1007/3-540-47721-7_17},
  journal   = {Lecture Notes in Computer Science (LNCS)},
}

@InProceedings{gupta2015make,
  author    = {Gupta, Vaibhav and Vineeth, Tatipathri S and Aggarwal, Vaibhav},
  title     = {Make Your Query Anonymous With {Oblivious Transfer}},
  booktitle = {6th International Conference on Computer and Communication Technology (ICCCT'15)},
  year      = {2015},
  pages     = {345--349},
  doi       = {10.1145/2818567.2818671}
}

@Article{even1985randomized,
  author    = {Even, Shimon and Goldreich, Oded and Lempel, Abraham},
  title     = {A randomized protocol for signing contracts},
  journal   = {Communications of the ACM},
  year      = {1985},
  volume    = {28},
  number    = {6},
  pages     = {637--647},
  month     = {jun},
  doi       = {10.1145/3812.3818},
  publisher = {ACM}
}

@InProceedings{chaum2017cmix,
  author       = {Chaum, David and Das, Debajyoti and Javani, Farid and Kate, Aniket and Krasnova, Anna and Ruiter, Joeri De and Sherman, Alan T},
  title        = {{cMix}: Mixing with minimal real-time asymmetric cryptographic operations},
  booktitle    = {International Conference on Applied Cryptography and Network Security (ACNS'17)},
  year         = {2017},
  pages        = {557--578},
  publisher    = {Springer},
  doi          = {10.1007/978-3-319-61204-1_28}
}

@Misc{diaz2022mix,
  author    = {Diaz, Claudia},
  title     = {Mix Networks},
  year      = {2022},
  booktitle = {Encyclopedia of Cryptography, Security and Privacy},
  doi       = {10.1007/978-3-642-27739-9_1754-1},
  pages     = {1--5},
  publisher = {Springer}
}

@article{diaz2021nym,
  title={The Nym Network: The Next Generation of Privacy Infrastructure},
  author={Diaz, Claudia and Halpin, Harry and Kiayias, Aggelos},
  journal={White Paper, version},
  volume={1},
  year={2021}
}

@InProceedings{guirat2020mixim,
  author    = {Guirat, Iness Ben and Gosain, Devashish and Diaz, Claudia},
  title     = {{MiXiM}: A general purpose simulator for mixnet},
  booktitle = {Privacy Enhancing Technologies Symposium -- HotPETs Workshop},
  year      = {2020},
  publisher = {de Gruyter},
}

@InProceedings{guirat2021mixim,
  author    = {Ben Guirat, Iness and Gosain, Devashish and Diaz, Claudia},
  title     = {MiXiM: Mixnet Design Decisions and Empirical Evaluation},
  booktitle = {20th Workshop on Privacy in the Electronic Society (WPES'21)},
  year      = {2021},
  pages     = {33–37},
  month     = {nov},
  publisher = {ACM},
  doi       = {10.1145/3463676.3485613},
  isbn      = {9781450385275},
  numpages  = {5},
}

@InProceedings{diaz2010impact,
  author       = {Diaz, Claudia and Murdoch, Steven J and Troncoso, Carmela},
  title        = {Impact of network topology on anonymity and overhead in low-latency anonymity networks},
  booktitle    = {Privacy Enhancing Technologies Symposium (PETS'10)},
  year         = {2010},
  pages        = {184--201},
  publisher    = {Springer},
  doi          = {10.1007/978-3-642-14527-8_11}
}

@InProceedings{piotrowska2017loopix,
  author    = {Piotrowska, Ania M and Hayes, Jamie and Elahi, Tariq and Meiser, Sebastian and Danezis, George},
  title     = {The loopix anonymity system},
  booktitle = {26th USENIX Security Symposium (USENIX Security '17)},
  year      = {2017},
  pages     = {1199--1216}
}

@Misc{Elixxir2022,
  title        = {Elixxir - a decentralized blockchain platform},
  howpublished = {\url{https://boostylabs.com/cases/elixxir}},
  month        = jul,
  year         = {2021},
  note         = {[Online; accessed 4-May-2022]},
  publisher    = {Boostylabs}
}

@techreport{RABIN1981exchange,
  title={How to exchange secrets by oblivious transfer},
  author={Rabin, Michael O},
  year={1981},
  institution={Technical Report TR-81, Aiken Computation Laboratory, Harvard University}
}

@Article{naor2005computationally,
  author    = {Naor, Moni and Pinkas, Benny},
  title     = {Computationally secure oblivious transfer},
  journal   = {Springer Journal of Cryptology},
  year      = {2005},
  volume    = {18},
  number    = {1},
  pages     = {1--35},
  month     = {oct},
  doi       = {10.1007/s00145-004-0102-6},
  publisher = {Springer},
}

@Article{javani2021aot,
  author    = {Javani, Farid and Sherman, Alan T},
  title     = {{AOT}: Anonymization by Oblivious Transfer},
  journal   = {arXiv preprint arXiv:2105.10794},
  year      = {2021},
  copyright = {Creative Commons Attribution Non Commercial Share Alike 4.0 International},
  doi       = {10.48550/ARXIV.2105.10794},
  publisher = {arXiv}
}

@Article{mueller2000anonymous,
  author    = {Mueller-Quade, Joern and Imai, Hideki},
  title     = {Anonymous oblivious transfer},
  journal   = {arXiv preprint arXiv:cs/0011004},
  year      = {2000},
  copyright = {Assumed arXiv.org perpetual, non-exclusive license to distribute this article for submissions made before January 2004},
  doi       = {10.48550/ARXIV.cs/0011004},
  publisher = {arXiv},
}

@Article{evans2018pragmatic,
  author    = {Evans, David and Kolesnikov, Vladimir and Rosulek, Mike and others},
  title     = {A pragmatic introduction to secure multi-party computation},
  journal   = {Foundations and Trends in Privacy and Security (Now Publishers)},
  year      = {2018},
  volume    = {2},
  number    = {2-3},
  pages     = {70--246},
  doi       = {10.1561/3300000019},
  publisher = {Now Publishers, Inc.},
}

@InProceedings{alexopoulos2017mcmix,
  author    = {Alexopoulos, Nikolaos and Kiayias, Aggelos and Talviste, Riivo and Zacharias, Thomas},
  title     = {{MCMix}: Anonymous Messaging via Secure Multiparty Computation},
  booktitle = {26th USENIX Security Symposium (USENIX Security '17)},
  year      = {2017},
  pages     = {1217--1234}
}

@InProceedings{barak2018end,
  author    = {Barak, Assi and Hirt, Martin and Koskas, Lior and Lindell, Yehuda},
  title     = {An end-to-end system for large scale {P2P} MPC-as-a-service and low-bandwidth MPC for weak participants},
  booktitle = {Conference on Computer and Communications Security (CCS'18)},
  year      = {2018},
  pages     = {695--712},
  month     = {oct},
  publisher = {ACM},
  doi       = {10.1145/3243734.3243801}
}

@InProceedings{lu2019honeybadgermpc,
  author    = {Lu, Donghang and Yurek, Thomas and Kulshreshtha, Samarth and Govind, Rahul and Kate, Aniket and Miller, Andrew},
  title     = {{HoneyBadgerMPC and AsynchroMix: Practical asynchronous MPC and its application to anonymous communication}},
  booktitle = {Conference on Computer and Communications Security (CCS'19)},
  year      = {2019},
  pages     = {887--903},
  month     = {nov},
  publisher = {ACM},
  doi       = {10.1145/3319535.3354238}
}

@Article{gilad2019metadata,
  author    = {Gilad, Yossi},
  title     = {Metadata-private communication for the 99\%},
  journal   = {Communications of the ACM},
  year      = {2019},
  volume    = {62},
  number    = {9},
  pages     = {86--93},
  month     = {aug},
  doi       = {10.1145/3338537},
  publisher = {ACM}
}

@techreport{fernandez2012survey,
  title={A survey and comparison of anonymous communication systems: Anonymity and security},
  author={Fern{\'a}ndez Franco, Leticia},
  year={2012},
  institution={Universitat Oberta de Catalunya}
}

@InProceedings{modinger2018flexible,
  author       = {M{\"o}dinger, David and Kopp, Henning and Kargl, Frank and Hauck, Franz J},
  title        = {A flexible network approach to privacy of blockchain transactions},
  booktitle    = {38th International Conference on Distributed Computing Systems (ICDCS'18)},
  year         = {2018},
  pages        = {1486--1491},
  month        = {jul},
  publisher    = {IEEE},
  doi          = {10.1109/icdcs.2018.00153}
}

@InProceedings{modinger2021shared,
  author       = {M{\"o}dinger, David and Dispan, Juri and Hauck, Franz J},
  title        = {Shared-Dining: Broadcasting Secret Shares Using Dining-Cryptographers Groups},
  booktitle    = {International Conference on Distributed Applications and Interoperable Systems (DAIS'21)},
  year         = {2021},
  pages        = {83--98},
  publisher    = {Springer},
  doi          = {10.1007/978-3-030-78198-9_6}
}

@Misc{SharedDininggithub2020,
  author       = {{Juri (gutjuri)}},
  title        = {threshold dining cryptographers simulation},
  howpublished = {\url{https://github.com/vs-uulm/thc-in-dc-simulation}},
  year         = {2020},
  note         = {[Online; accessed 31-may-2022]},
  journal      = {GitHub repository},
  publisher    = {GitHub},
}

@Article{Franck2014verifiable,
  author    = {Christian Franck},
  title     = {Dining Cryptographers with 0.924 Verifiable Collision Resolution},
  journal   = {arXiv preprint arXiv:1402.1732},
  year      = {2014},
  copyright = {arXiv.org perpetual, non-exclusive license},
  doi       = {10.48550/ARXIV.1402.1732},
  publisher = {arXiv},
}

@Article{Franck2014practical,
  author    = {Christian Franck and Jeroen van de Graaf},
  title     = {Dining Cryptographers are Practical},
  journal   = {arXiv preprint arXiv:1402.2269},
  year      = {2014},
  copyright = {arXiv.org perpetual, non-exclusive license},
  doi       = {10.48550/ARXIV.1402.2269},
  publisher = {arXiv},
  url       = {http://arxiv.org/abs/1402.2269}
}

@InProceedings{Franck2021FastECC,
  author    = {Dupont, Briag and Franck, Christian and Gro{\ss}sch{\"a}dl, Johann},
  title     = {Fast and Flexible Elliptic Curve Cryptography for Dining Cryptographers Networks},
  booktitle = {Mobile, Secure, and Programmable Networking (MSPN'21)},
  year      = {2021},
  pages     = {89--109},
  publisher = {Springer},
  doi       = {10.1007/978-3-030-67550-9_7},
  isbn      = {978-3-030-67550-9},
}

@misc{HoneyBadgerMPC2020,
  title = {{HoneyBadgerMPC - Robust MPC-based confidentiality layer for blockchains}},
  year = {2020},
  publisher = {GitHub},
  journal = {GitHub repository},
  howpublished = "\url{https://github.com/initc3/HoneyBadgerMPC}",
  note = "[Online; accessed 2-june-2022]"
  }

@Book{Bishop2004,
  title     = {{Introduction to Computer Security}},
  publisher = {Addison-Wesley},
  year      = {2004},
  author    = {Matt Bishop},
  isbn      = {0-321-24744-2}
}

@Book{Schiller2003,
  title     = {{Mobile Communications}},
  publisher = {Addison-Wesley},
  year      = {2003},
  author    = {Jochen Schiller},
  edition   = {2},
  isbn      = {0-321-12381-6}
}

@TechReport{ISI1981,
  author       = {Postel, Jon},
  title        = {{RFC0791. Internet Protocol}},
  institution  = {{University of Southern California}},
  year         = {1981},
  month        = {sep},
  doi          = {10.17487/RFC0791},
  organization = {{University of Southern California}},
  publisher    = {{RFC} Editor},
  url          = {https://www.rfc-editor.org/rfc/pdfrfc/rfc791.txt.pdf}
}

@Book{Gartner2016,
  title     = {{Metadata. Shaping Knowledge from Antiquity to the Semantic Web}},
  publisher = {Springer Nature},
  year      = {2016},
  author    = {Richard Gartner},
  doi       = {10.1007/978-3-319-40893-4}
}

@Article{UniInternationalTelecommunication2012,
  author  = {{Sector, ITU Telecommunication Standardization}},
  title   = {Recommendation ITU-T Y. 2060: Overview of the Internet of things},
  journal = {Series Y: Global information infrastructure, internet protocol aspects and next-generation networks-Frameworks and functional architecture models},
  year    = {2012},
  url     = {https://www.itu.int/rec/dologin_pub.asp?lang=e&id=T-REC-Y.2060-201206-I!!PDF-E&type=items},
}

@InProceedings{Shields2000,
  author    = {Shields, Clay and Levine, Brian Neil},
  title     = {A Protocol for Anonymous Communication over the Internet},
  booktitle = {7th Conference on Computer and Communications Security (CCS'00)},
  year      = {2000},
  pages     = {33–42},
  publisher = {ACM},
  doi       = {10.1145/352600.352607},
  isbn      = {1581132034},
  location  = {Athens, Greece},
  numpages  = {10},
}

@Book{Aalst2016,
  title     = {{Process Mining}},
  publisher = {Springer},
  year      = {2016},
  author    = {Wil van der Aalst},
  doi       = {10.1007/978-3-662-49851-4},
}

@InProceedings{Serjantov2003,
  author    = {Serjantov, Andrei and Sewell, Peter},
  title     = {Passive Attack Analysis for Connection-Based Anonymity Systems},
  booktitle = {European Symposium on Research in Computer Security (ESORICS'03)},
  year      = {2003},
  editor    = {Snekkenes, Einar and Gollmann, Dieter},
  pages     = {116--131},
  publisher = {Springer},
  doi       = {10.1007/978-3-540-39650-5_7},
  isbn      = {978-3-540-39650-5},
}

@Article{Pfitzmann2000,
  author  = {Pfitzmann, Andreas and Hansen, Marit},
  title   = {Anonymity, Unobservability, and Pseudonymity — A Proposal for Terminology},
  journal = {Designing Privacy Enhancing Technologies (LNCS)},
  year    = {2000},
  volume  = {2009},
  pages   = {1-9},
  month   = {01},
  doi     = {10.1007/3-540-44702-4_1},
  isbn    = {978-3-540-41724-8},
}

@Book{delfs2015introduction,
  title     = {Introduction to Cryptography: Principles and Applications},
  publisher = {Springer},
  year      = {2015},
  author    = {Delfs, H. and Knebl, H.},
  series    = {Information Security and Cryptography},
  edition   = {3rd},
  isbn      = {9783662479742},
  url       = {https://books.google.de/books?id=KwmkCgAAQBAJ},
}

@mastersthesis{hashemi2018fingerprinting,
  title={Fingerprinting the Smart Home: Detection of Smart Assistants Based on Network Activity},
  author={Hashemi, Arshan},
  year={2018},
  school={Department of Electrical Engineering, Faculty of California Polytechnic State University}
}

@InProceedings{hintz2002fingerprinting,
  author       = {Hintz, Andrew},
  title        = {Fingerprinting websites using traffic analysis},
  booktitle    = {International Workshop on Privacy Enhancing Technologies (PETS'02)},
  year         = {2002},
  pages        = {171--178},
  publisher    = {Springer},
  doi          = {10.1007/3-540-36467-6_13}
}

@Article{kohno2005remote,
  author    = {Kohno, Tadayoshi and Broido, Andre and Claffy, Kimberly C},
  title     = {Remote physical device fingerprinting},
  journal   = {IEEE Transactions on Dependable and Secure Computing (TDSC)},
  year      = {2005},
  volume    = {2},
  number    = {2},
  pages     = {93--108},
  month     = {feb},
  doi       = {10.1109/tdsc.2005.26},
  publisher = {IEEE},
}

@InProceedings{berthold2001disadvantages,
  author    = {Berthold, Oliver and Pfitzmann, Andreas and Standtke, Ronny},
  title     = {The disadvantages of free MIX routes and how to overcome them},
  booktitle = {Designing Privacy Enhancing Technologies (LNCS)},
  year      = {2001},
  pages     = {30--45},
  publisher = {Springer},
  doi       = {10.1007/3-540-44702-4_3},
}

@InProceedings{danezis2004statistical,
  author       = {Danezis, George and Serjantov, Andrei},
  title        = {Statistical disclosure or intersection attacks on anonymity systems},
  booktitle    = {International Workshop on Information Hiding (IH'04)},
  year         = {2004},
  pages        = {293--308},
  publisher    = {Springer},
  doi          = {10.1007/978-3-540-30114-1_21}
}

@InCollection{danezis2003statistical,
  author       = {Danezis, George},
  title        = {Statistical disclosure attacks},
  booktitle    = {Security and Privacy in the Age of Uncertainty},
  publisher    = {Springer},
  year         = {2003},
  pages        = {421--426},
  doi          = {10.1007/978-0-387-35691-4_40}
}

@InProceedings{mathewson2004practical,
  author       = {Mathewson, Nick and Dingledine, Roger},
  title        = {Practical traffic analysis: Extending and resisting statistical disclosure},
  booktitle    = {International Workshop on Privacy Enhancing Technologies (PETS'04)},
  year         = {2004},
  pages        = {17--34},
  publisher    = {Springer},
  doi          = {10.1007/11423409_2}
}

@Article{apthorpe2017smart,
  author    = {Apthorpe, Noah and Reisman, Dillon and Feamster, Nick},
  title     = {A smart home is no castle: Privacy vulnerabilities of encrypted iot traffic},
  journal   = {arXiv preprint arXiv:1705.06805},
  year      = {2017},
  copyright = {arXiv.org perpetual, non-exclusive license},
  doi       = {10.48550/arXiv.1705.06805},
  publisher = {arXiv}
}

@misc{Panoramix,
  author = {{The European Commission}},
  title = {{PANORAMIX:} Privacy and Accountability in Networks via Optimized Randomized Mix-nets},
  year = {2019},
  publisher = {The Community Research and Development Information Service {(CORDIS)}},
  howpublished = "\url{https://cordis.europa.eu/project/id/653497}",
  note = "[Online; accessed 26-June-2022]"
}

@Misc{RACE,
  author       = {{Defense Advanced Research Projects Agency (DARPA)}},
  title        = {Resilient Anonymous Communication for Everyone {(RACE)}},
  howpublished = {\url{https://www.darpa.mil/program/resilient-anonymous-communication-for-everyone\#:~:text=The\%20Resilient\%20Anonymous\%20Communication\%20for,any\%20participant\%20in\%20the\%20system}},
  year         = {2018},
  note         = {[Online; accessed 26-June-2022]},
  publisher    = {{DARPA}},
}

@Book{Galuba2009Overlay,
  title     = {Overlay Network},
  publisher = {Springer},
  year      = {2009},
  author    = {Galuba, Wojciech and Girdzijauskas, Sarunas},
  editor    = {LIU, LING and {\"O}ZSU, M. TAMER},
  address   = {Boston, MA},
  isbn      = {978-0-387-39940-9},
  booktitle = {Encyclopedia of Database Systems},
  doi       = {10.1007/978-0-387-39940-9_1231},
  url       = {https://doi.org/10.1007/978-0-387-39940-9_1231},
}

@Article{ATZORI2010IoT,
  author    = {Luigi Atzori and Antonio Iera and Giacomo Morabito},
  title     = {The Internet of Things: A survey},
  journal   = {Elsevier Computer Networks},
  year      = {2010},
  volume    = {54},
  number    = {15},
  pages     = {2787-2805},
  month     = {oct},
  issn      = {1389-1286},
  doi       = {10.1016/j.comnet.2010.05.010},
  publisher = {Elsevier},
  url       = {https://www.sciencedirect.com/science/article/pii/S1389128610001568},
}

@Article{sicari2015security,
  author    = {Sicari, Sabrina and Rizzardi, Alessandra and Grieco, Luigi Alfredo and Coen-Porisini, Alberto},
  title     = {Security, privacy and trust in Internet of Things: The road ahead},
  journal   = {Elsevier Computer Networks},
  year      = {2015},
  volume    = {76},
  pages     = {146--164},
  month     = {jan},
  doi       = {10.1016/j.comnet.2014.11.008},
  publisher = {Elsevier},
}

@Article{wortmann2015internet,
  author    = {Wortmann, Felix and Fl{\"u}chter, Kristina},
  title     = {Internet of Things},
  journal   = {Springer Business \& Information Systems Engineering},
  year      = {2015},
  volume    = {57},
  number    = {3},
  pages     = {221--224},
  month     = {mar},
  doi       = {10.1007/s12599-015-0383-3},
  publisher = {Springer},
}

@misc{isoiec27551,
  author = {{International Organization for Standardization (ISO)}},
  title = {{ISO/IEC 27551: Information security, cybersecurity and privacy protection — Requirements for attribute-based unlinkable entity authentication}},
  year = {2021}
}

@Article{reiter1998crowds,
  author     = {Reiter, Michael K. and Rubin, Aviel D.},
  title      = {Crowds: Anonymity for web transactions},
  journal    = {ACM Transactions on Information and System Security (TISSEC'98)},
  year       = {1998},
  volume     = {1},
  number     = {1},
  pages      = {66--92},
  month      = {nov},
  issn       = {1094-9224},
  doi        = {10.1145/290163.290168},
  issue_date = {Nov. 1998},
  numpages   = {27},
  publisher  = {ACM},
}

\vfill

\pagebreak

\begin{wrapfigure}{l}{0.3\textwidth}
	\centering
	\includegraphics[width=1in,height=1.25in,clip,keepaspectratio]{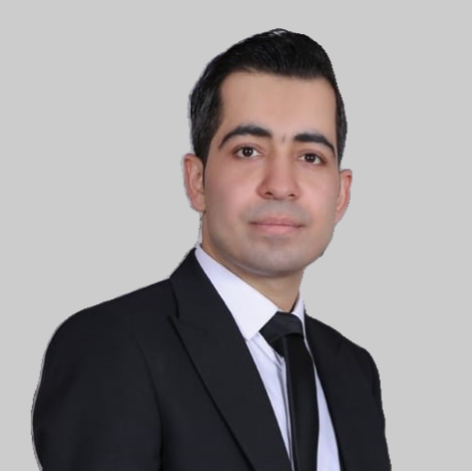}
\end{wrapfigure}

\noindent Mohsen Shirali received his M.Sc. in computer architecture from Shahid Beheshti University (SBU) and his B.Sc. in Software Engineering from Islamic Azad University-Tehran North Branch with distinction in 2013 and 2010, respectively. Currently, he is a PhD candidate at SBU, and he works on using the Internet of Things in the healthcare domain by considering privacy and energy efficiency.
	
\noindent He worked at SBU and Ahwaz Sama University as a lecturer and a visiting researcher in the chair of IT-Security at the Passau Institute of Digital Security (PIDS) (2019-2020). His research stay at the University of Passau was funded by two scholarship awards by the DAAD and Iran MSRT. His research interest includes Internet of Things, security and privacy, anonymous communication and process mining.\bigskip
\begin{wrapfigure}{r}{0.3\textwidth}
	\centering
	\includegraphics[width=1in,height=1.25in,clip,keepaspectratio]{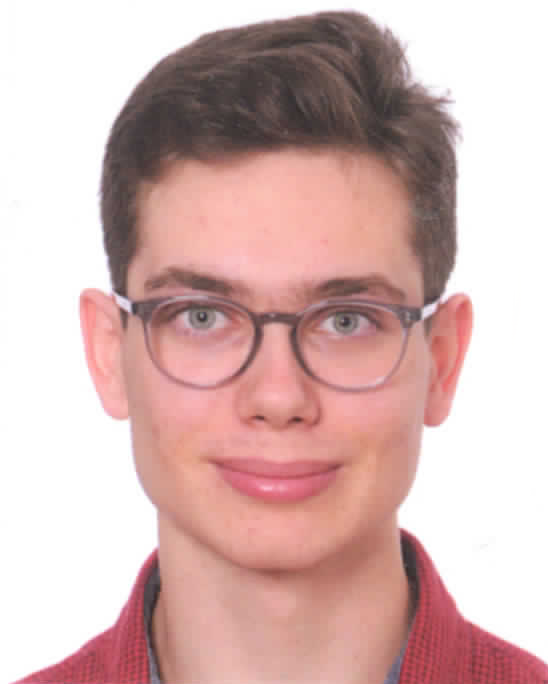}
\end{wrapfigure}

\bigskip
\begin{spacing}{1.8}
	\noindent Tobias Tefke studies Applied Computer Science at Schmalkalden University of Applied Sciences, Germany. He received his B.Sc. in Computer Science from Schmalkalden University of Applied Sciences in 2022. His research interests are IT security and privacy, particularly on mobile and constrained devices. He enjoys software development.	
\end{spacing}
\bigskip

\begin{wrapfigure}{l}{0.3\textwidth}
	\centering
	\includegraphics[width=1in,height=1.25in,clip,keepaspectratio]{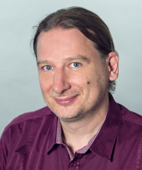}
\end{wrapfigure}

\bigskip
\begin{spacing}{1.2}
	\noindent Ralf C. Staudemeyer is Professor for IT-Security at the Faculty of Computer Science of the Schmalkalden University of Applied Sciences in Germany.  He holds a Ph.D. and a German Diploma in Computer Science (Diplom-Informatiker). He is an designated expert in IT security \& privacy, computer networks, and deep learning. He lecturered at several academic institutions in Germany, South Africa and the Fiji Islands. He was recipient of various research fellowships and supported EU research projects as a coordinator, work-package leader and researcher. He is author of more than 30 publications.
\end{spacing}
\bigskip

\begin{wrapfigure}{r}{0.3\textwidth}
	\centering
	\includegraphics[width=1in,height=1.25in,clip,keepaspectratio]{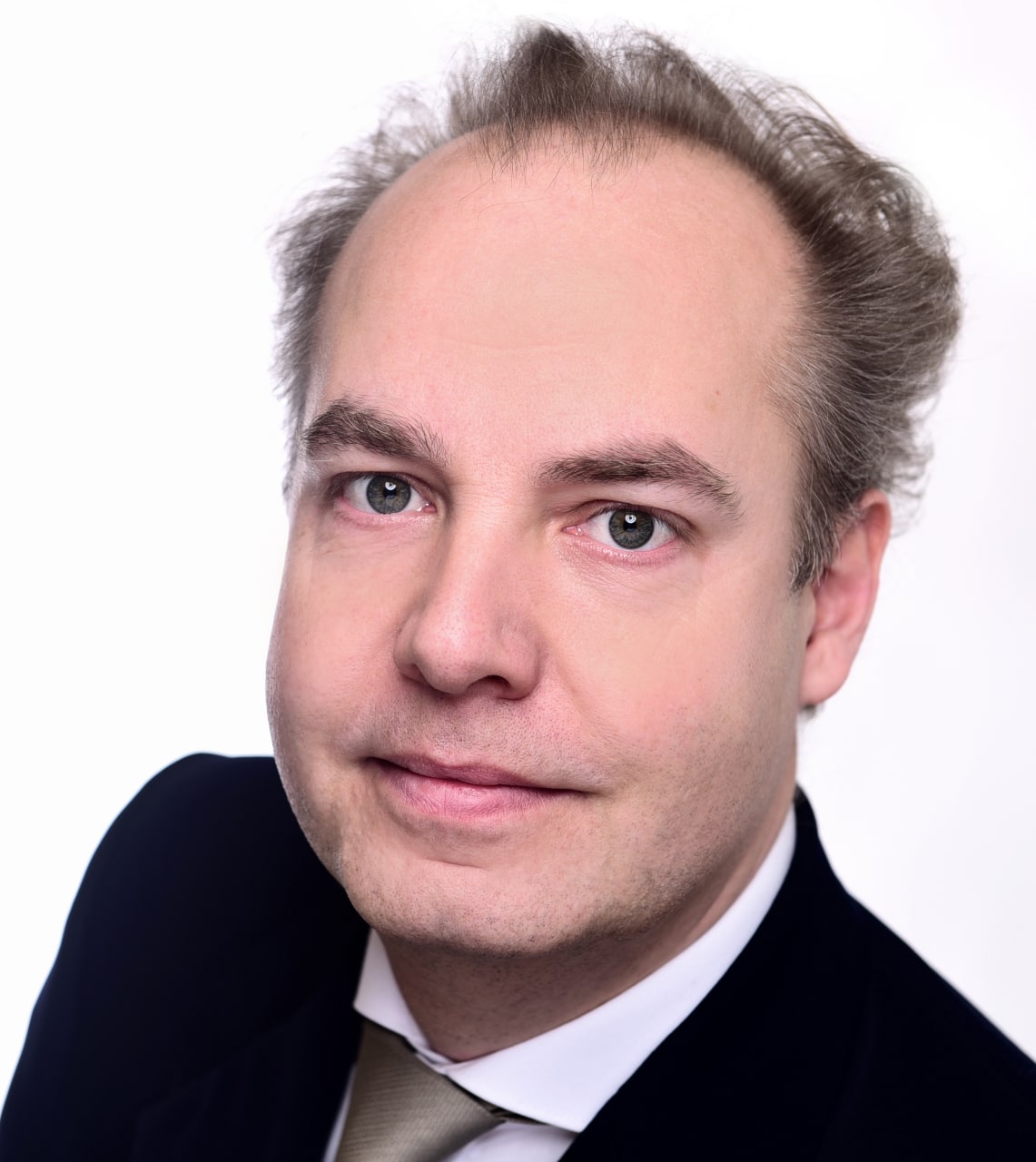}
\end{wrapfigure}

\begin{spacing}{1.2}
	\noindent Henrich C. P{\"o}hls is a Senior Researcher and Lecturer on Secure AI Systems and other topics at University of Passau, Germany. He received a Diploma degree in Computer Science from University of Hamburg (DE), a Master's degree in Information Security from RHUL (UK) and a Ph.D. degree from University of Passau (DE). He researches IoT-, AI- and Cloud-security with a focus on cryptographic integrity protection mechanisms. He serves on several security conferences' committees and is an appointed expert of the German national standardisation body (DIN) in the areas of Information Security (SC27), Artificial Intelligence (SC42); he is an editor of several ISO/IEC standards, e.g. ISO/IEC 23264-1 on redaction of authentic data.
\end{spacing}

\end{document}